\documentclass[aps,preprint,preprintnumbers,superscriptaddress,amsmath,amssymb,floatfix]{revtex4}
\usepackage{graphicx}
\newcommand{\sss}{\scriptscriptstyle}
\begin{document}
\preprint{BNL-HET-03/9}
\preprint{FSU-HEP-2003-0503}
\preprint{UB-HET-03/02}
\preprint{hep-ph/0305087}
\title{Associated Higgs production with top quarks \\
at the Large Hadron Collider: NLO QCD corrections}
\author{S.~Dawson}
\email{dawson@quark.phy.bnl.gov}
\affiliation{Physics Department, Brookhaven National Laboratory,
Upton, NY 11973-5000, USA}
\author{C.~Jackson}
\email{jackson@hep.fsu.edu}
\affiliation{Physics Department, Florida State University,
Tallahassee, FL 32306-4350, USA}
\author{L.~H.~Orr}
\email{orr@pas.rochester.edu}
\affiliation{Department of Physics $\&$ Astronomy, University of Rochester,
Rochester, NY 14627-0171, USA}
\author{L.~Reina}
\email{reina@hep.fsu.edu}
\affiliation{Physics Department, Florida State University,
Tallahassee, FL 32306-4350, USA}
\author{D.~Wackeroth}
\email{dow@ubpheno.physics.buffalo.edu}
\affiliation{Department of Physics, SUNY at Buffalo,
Buffalo, NY 14260-1500, USA}

\date{\today}

\begin{abstract} 
  We present in detail the calculation of the ${\cal O}(\alpha_s^3)$
  inclusive total cross section for the process $pp\to t\bar{t}h$, in
  the Standard Model, at the CERN Large Hadron Collider with
  center-of-mass energy $\sqrt{s_{\sss H}}\!=\!14$~TeV. The
  calculation is based on the complete set of virtual and real ${\cal
    O}(\alpha_s)$ corrections to the parton level processes $q\bar
  q\to t\bar{t}h$ and $gg\to t\bar{t}h$, as well as the tree level
  processes $(q,\bar{q})g\to t\bar{t}h+(q,\bar{q})$. The virtual
  corrections involve the computation of pentagon diagrams with
  several internal and external massive particles, first encountered
  in this process. The real corrections are computed using both the
  single and the two cutoff phase space slicing method. The
  next-to-leading order QCD corrections significantly reduce the
  renormalization and factorization scale dependence of the Born cross
  section and moderately increase the Born cross section for values of
  the renormalization and factorization scales above $m_t$.
\end{abstract}
%
\maketitle
\section{Introduction}
\label{sec:intro}

One of the critical goals of present and future colliders is the study
of the electroweak symmetry breaking mechanism and the origin of
fermion masses.  If the introduction of one or more Higgs fields is
responsible for the breaking of the electroweak symmetry and for the
generation of fermion masses, then one Higgs boson should be
relatively light.  The present lower bounds on the Higgs boson mass
from direct searches at LEP2 are $M_h\!>\!114.4$~GeV (at $95\%$ CL)
\cite{Lephwg1:2002} for the Standard Model (SM) Higgs boson ($h$), and
$M_{h^0}\!>\!91.0$~GeV and $M_{A^0}\!>\!91.9$~GeV (at $95\%$ CL,
$0.5\!<\!\tan\beta\!<\!2.4$ excluded) \cite{Lephwg2:2001} for the
light scalar ($h^0$) and pseudoscalar ($A^0$) Higgs bosons of the
minimal supersymmetric standard model (MSSM).  At the same time,
global SM fits to electroweak precision data imply $M_{h}<211$~GeV (at
$95\%$ CL) \cite{Lepewwg:2003}, while the MSSM requires the existence
of a scalar Higgs boson lighter than about 130~GeV. The possibility of
a Higgs boson discovery in the mass range near 115-130~GeV thus seems
increasingly likely.

The associated production of a Higgs boson with a $t\bar{t}$ pair can
play a very important role at hadron colliders as has been suggested
by many studies over the past several years
\cite{Marciano:1991qq,cms:1994,atlas:1999,Goldstein:2000bp}.  In
particular, it is an important discovery channel for a SM-like Higgs
boson at the LHC if $M_{h}\!<\!130$~GeV
\cite{atlas:1999,Richter-Was:1999sa,Beneke:2000hk,Drollinger:2001ym}.
Although the event rate is small, the signature is quite distinctive.
Given the statistics expected at the LHC, $pp\to t\bar{t}h$, with
$h\to b\bar{b},\tau^+\tau^-,W^+W^-,\gamma\gamma$ will also be
instrumental to the determination of the couplings of a discovered
Higgs boson, and will in particular give the only handle on a direct
measurement of the top quark Yukawa coupling
\cite{Beneke:2000hk,Zeppenfeld:2000td,Zeppenfeld:2002ng,Belyaev:2002ua,
  Maltoni:2002jr}.

The total cross section for $pp\to t\bar{t}h$ has been known at
tree-level, i.e. at leading order (LO) of QCD, for quite some time
\cite{Kunszt:1984ri,Ng:1984jm}. Next-to-leading order (NLO) QCD
corrections are crucial in order to reduce the dependence of the cross
section on the renormalization and factorization scales.  The
calculation of the total cross section for $pp\to t\bar{t}h$ to ${\cal
  O}(\alpha_s^3)$ has been performed by the Authors of
Refs.~\cite{Beenakker:2001rj,Beenakker:2002nc} and by our group.  The
results of the two independent calculations have been compared and
they are in very good agreement. In Ref.~\cite{Dawson:2002tg}, we
presented our first numerical results for the total inclusive NLO QCD
cross section for $pp\to t\bar{t}h$ at the LHC center of mass energy,
$\sqrt{s_{\sss H}}\!=\!14$~TeV.  Here we provide a detailed
description of the calculation.

At the LHC center-of-mass energy, the dominant subprocess for
$t\bar{t}h$ production is $gg\to t\bar{t}h$, but the other
subprocesses, $q\bar{q}\to t\bar{t}h$ and $(q,\bar q)g\to
t\bar{t}h+(q,\bar q)$, which contribute to the cross section at ${\cal
  O}(\alpha_s^3)$, cannot be neglected and are included in this
calculation.  The NLO QCD corrections to the $q\bar{q}\to t\bar{t}h$
subprocess constitute a gauge invariant subset of the entire NLO QCD
calculation and have been presented in
Refs.~\cite{Reina:2001sf,Reina:2001bc} to which we refer for a
thorough discussion of the results. Here we concentrate on a detailed
description of the calculation of the ${\cal O}(\alpha_s)$ corrections
to the $gg\to t\bar{t}h$ subprocess.  The Feynman diagrams
contributing to $gg\to t\bar{t}h$ at lowest order are shown in
Fig.~\ref{fg:tree_gg}, while the ${\cal O}(\alpha_s)$ virtual and real
corrections are given in Figs.~\ref{fg:self_gg}-\ref{fg:pentagons_gg}
and Fig.~\ref{fg:real_gg}, respectively.

The main challenge in the calculation of the ${\cal O}(\alpha_s)$
virtual corrections comes from the presence of pentagon diagrams with
several massive external and internal particles. The pentagon scalar
and tensor Feynman integrals originating from these diagrams present
either analytical (scalar) or numerical (tensor) challenges. We have
calculated the pentagon scalar integrals as linear combinations of
scalar box integrals using the method of
Ref.~\cite{Bern:1993em,Bern:1994kr}, and cross checked them using the
techniques of Ref.~\cite{Denner:1993kt}.  Pentagon tensor integrals
have been calculated and cross checked in two ways: numerically, by
isolating the numerical instabilities and extrapolating from the
numerically safe to the numerically unsafe region using various
methods; and analytically, by reducing them to a numerically stable
form.  The real corrections have been computed using the phase space
slicing method, in both the double (for a review see, e.g.
\cite{Harris:2001sx}) and single
\cite{Giele:1992vf,Giele:1993dj,Keller:1998tf} cutoff approaches.
Together with the corresponding $q\bar{q}\to t\bar{t}h$ calculation
\cite{Reina:2001sf,Reina:2001bc}, this is the first application of the
single cutoff phase space slicing method to a cross section involving
more than one massive particle in the final state and agreement
between the two cutoff and the single cutoff approaches is a strong
check of the calculation.

The outline of our paper is as follows. In Section~\ref{sec:framework}
we summarize the general structure of the NLO cross section for $pp\to
t\bar{t}h$. In Section~\ref{sec:sigma_lo} we briefly review the case
of the LO cross section for $pp\to t\bar{t}h$, introducing some
fundamental notation. We proceed in Sections~\ref{sec:virtual} and
\ref{sec:real} to present the details of the calculation of both the
virtual and real parts of the NLO QCD corrections to $gg\to
t\bar{t}h$.  Section~\ref{sec:real} also includes a discussion of the
tree level $(q,\bar q)g\to t\bar{t}h+(q,\bar q)$ processes. In
Section~\ref{sec:total} we explicitly show the factorization of the
initial state infrared singularities into the gluon distribution
functions, and finally summarize our results for the NLO inclusive
total cross section for $pp\to t\bar{t}h$ at the LHC in
Eqs.~({\ref{eq:sigma_nlo_ij}) and
  (\ref{eq:sigmatot_gg2})-(\ref{eq:sigmatot_qg1}).  Finally, numerical
  results for the total cross section are presented in
  Section~\ref{sec:results}. We collect most of the technical details,
  including a list of box and pentagon integrals, in a series of
  Appendices.
\section{The calculation: general setup}
\label{sec:framework}

The inclusive total cross section for $pp\to t\bar{t}h$ at ${\cal
  O}(\alpha_s^3)$ can be written as:
\begin{eqnarray}
\label{eq:sigma_nlo}
&&\sigma_{\sss NLO}(pp\to t\bar{t}h)=\nonumber\\
&&\sum_{ij}\frac{1}{1+\delta_{ij}}
\int dx_1 dx_2 \left[{\cal F}_i^p(x_1,\mu) {\cal F}_j^{p}(x_2,\mu)
{\hat \sigma}^{ij}_{\sss NLO}(x_1,x_2,\mu)+(1\leftrightarrow 2)\right]
\,\,\,,
\end{eqnarray}
where ${\cal F}_i^p$ are the NLO parton distribution functions (PDFs)
for parton $i$ in a proton, defined at a generic factorization scale
$\mu_f\!=\!\mu$, and ${\hat \sigma}^{ij}_{\sss NLO}$ is the ${\cal
  O}(\alpha_s^3)$ parton-level total cross section for incoming
partons $i$ and $j$, made of the channels $q\bar{q},gg\to t\bar{t}h$
and $(q,\bar{q})g\to t\bar{t}h(q,\bar{q})$, and renormalized at an
arbitrary scale $\mu_r$ which we also take to be $\mu_r\!=\!\mu$.
Throughout this paper we will always assume the factorization and
renormalization scales to be equal, $\mu_r\!=\!\mu_f\!=\!\mu$, unless
differently specified.  The partonic center-of-mass energy squared,
$s$, is given in terms of the hadronic center-of-mass energy squared,
$s_{\sss H}$, by $s=x_1 x_2 s_{\sss H}$. At the LHC center-of-mass
energy the cross section is dominated by the $gg$ initial state,
although the other contributions cannot be neglected and are included
in this calculation.

We write the NLO parton-level total cross section ${\hat
  \sigma}_{\sss NLO}^{ij}(x_1,x_2,\mu)$ as:
\begin{eqnarray}
\label{eq:sigmahat_nlo}
{\hat\sigma}_{\sss NLO}^{ij}(x_1,x_2,\mu)&=&
\alpha_s^2(\mu)\left\{f_{\sss LO}^{ij}(x_1,x_2)+
\frac{\alpha_s(\mu)}{4\pi}f_{\sss NLO}^{ij}(x_1,x_2,\mu)\right\}
\nonumber\\
&\equiv&{\hat \sigma}_{\sss LO}^{ij}(x_1,x_2,\mu)+
\delta {\hat \sigma}_{\sss NLO}^{ij}(x_1,x_2,\mu)\,\,\,,
\end{eqnarray}
where $\alpha_s(\mu)$ is the strong coupling constant renormalized at
the arbitrary scale $\mu_r\!=\!\mu$, ${\hat\sigma}_{\sss
  LO}^{ij}(x_1,x_2,\mu)$ is the ${\cal O}(\alpha_s^2)$ Born cross
section, and $\delta{\hat\sigma}_{\sss NLO}^{ij}(x_1,x_2,\mu)$
consists of the ${\cal O}(\alpha_s)$ corrections to the Born cross
sections for $gg,q\bar{q}\to t\bar{t}h$ and of the tree level
$(q,\bar{q})g\to t\bar{t}h(q,\bar{q})$ processes, including the
effects of mass factorization (see Section~\ref{sec:total}).
$\delta{\hat\sigma}_{\sss NLO}^{ij}(x_1,x_2,\mu)$ can be written as
the sum of two terms:
\begin{eqnarray}
\label{eq:delta_sigmahat}
\delta{\hat\sigma}_{\sss NLO}^{ij}(x_1,x_2,\mu)&=&
\int d(PS_3) \overline{\sum}|{\cal A}_{virt}(ij\to t\bar{t}h)|^2+
\int d(PS_4)\overline{\sum}|{\cal A}_{real}(ij\to t\bar{t}h+l)|^2
\nonumber \\
&\equiv&\hat{\sigma}^{ij}_{virt}(x_1,x_2,\mu)+
\hat{\sigma}^{ij}_{real}(x_1,x_2,\mu)\,\,\,,
\end{eqnarray}
where $|{\cal A}_{virt}(ij\to t\bar{t}h)|^2$ and $|{\cal
  A}_{real}(ij\to t\bar{t}h+l)|^2$ (for $ij\!=\!q\bar{q},gg$ and
$l\!=\!g$, or $ij\!=\!qg,\bar{q}g$ and $l\!=\!q,\bar{q}$) are
respectively the ${\cal O}(\alpha_s^3)$ terms of the squared matrix
elements for the $ij\rightarrow t\bar th$ and $ij\rightarrow t\bar t
h+l$ processes, and $\overline{\sum}$ indicates that they have been
averaged over the initial state degrees of freedom and summed over the
final state ones.  Moreover, $d(PS_3)$ and $d(PS_4)$ in
Eq.~(\ref{eq:delta_sigmahat}) denote the integration over the
corresponding three and four-particle phase spaces respectively.  The
first term in Eq.~(\ref{eq:delta_sigmahat}) represents the
contribution of the virtual one gluon corrections to $q\bar{q}\to
t\bar{t}h$ and $gg\to t\bar{t}h$, while the second one is due to the
real one gluon and real one quark/antiquark emission, i.e.
$q\bar{q},gg\to t\bar{t}h+g$ and $qg(\bar qg) \to
t\bar{t}h+q(\bar{q})$.

The ${\cal O}(\alpha_s)$ virtual and real corrections to $q\bar{q}\to
t\bar{t}h$ have been discussed in detail in Ref.~\cite{Reina:2001bc},
and will not be repeated here. In the following sections we present
the general structure of the ${\cal O}(\alpha_s)$ virtual and real
corrections to $gg\to t\bar{t}h$. The contribution of the
$(q,\bar{q})g$ initiated process will be considered in
Section~\ref{sec:real}, when dealing with the real part of the ${\cal
  O}(\alpha_s^3)$ cross section.  The results presented in the
following sections have been obtained by two completely independent
calculations, based on a combination of FORM~\cite{Vermaseren:2000nd}
and \emph{Maple} codes in one case, and on the \emph{Mathematica}
based code Tracer~\cite{Jamin:1991dp} in the other.  The matrix
elements squared for the tree level processes $gg\to t\bar{t}h$,
$gg\to t\bar{t}h+g$, and $(q,\bar{q})g\to t\bar{t}h+(q,\bar{q})$ have
been checked with Madgraph \cite{Stelzer:1994ta}. The numerical
results presented in Section~\ref{sec:results} have been obtained with
two independent \emph{Fortran} codes.

Finally, we observe that the scale dependence of the total cross
section at NLO is dictated by renormalization group arguments, and
$f_{\sss NLO}^{ij}(x_1,x_2,\mu)$ in Eq.~(\ref{eq:sigmahat_nlo}) must
be of the form:
\begin{equation}
\label{eq:f_ij_nlo}
f_{\sss NLO}^{ij}(x_1,x_2,\mu)=f_1^{ij}(x_1,x_2)+
\tilde{f}_1^{ij}(x_1,x_2)\ln\left(\frac{\mu^2}{s}\right)\,\,\,,
\end{equation}
with $\tilde{f}_1^{ij}(x_1,x_2)$ given by:
\begin{eqnarray}
\label{eq:mudep_coeff}
\tilde{f}_1^{ij}(x_1,x_2)&=&
2\,\left\{4\pi b_0 f^{ij}_{\sss LO}(x_1,x_2)-
\sum_k\left[\int_\rho^1 dz_1 P_{ik}(z_1)f^{kj}_{\sss LO}(x_1 z_1,x_2)
\right. \right.\nonumber\\
&+&\left.\left.\int_\rho^1 dz_2 P_{jk}(z_2)f^{ik}_{\sss LO}(x_1,x_2z_2) 
\right]\right\}
\,\,\,,
\end{eqnarray}
where $\rho\!=\!(2m_t+M_h)^2/s$, $P_{ij}(z)$ denotes the lowest-order
regulated Altarelli-Parisi splitting function \cite{Altarelli:1977zs}
of parton $i$ into parton $j$, when $j$ carries a fraction $z$ of the
momentum of parton $i$, (see e.g. Section~\ref{sec:real}), and $b_0$
is determined by the one-loop renormalization group evolution of the
strong coupling constant $\alpha_s$:
\begin{equation}
\frac{d\alpha_s(\mu)}{d\ln(\mu^2)}=-b_0\alpha_s^2+{\cal O}(\alpha_s^3)
\,\,\,\,\,,\,\,\,\,\,
b_0=\frac{1}{4\pi}\left(\frac{11}{3}N-\frac{2}{3}n_{lf}\right)\,\,\,,
\end{equation}
with $N=3$, the number of colors, and $n_{lf}\!=\!5$, the number of
light flavors. The origin of the terms in Eq.~(\ref{eq:mudep_coeff})
will become manifest in Sections~\ref{sec:virtual}, \ref{sec:real},
and \ref{sec:total} when we describe in detail the calculation of both
virtual and real ${\cal O}(\alpha_s)$ corrections.
\boldmath
\section{The tree level cross section for $gg\to t\bar{t}h$}
\unboldmath
\label{sec:sigma_lo}

The tree level amplitude for the process
\[
g^A(q_1)+g^B(q_2)\to t(p_t)+\bar{t}(p_t^\prime)+h(p_h) \; ,
\]
where $q_1+q_2=p_t+p_t^\prime +p_h$ and $A,B$ denote the color of the
incoming gluons, is obtained from the three classes of Feynman
diagrams represented in Fig.~\ref{fg:tree_gg}, identified as
$s-$channel, $t-$channel, and $u-$channel diagrams respectively. We
find it convenient to organize the color structure of both the tree
level amplitude and the one-loop virtual amplitude in terms of only
two color factors, one symmetric and one antisymmetric in the color
indices of the initial gluons.  Following this prescription, the tree
level amplitude for $gg\to t\bar{t}h$ can be written as:
\begin{equation}
  \label{eq:amp_tree}
  {\cal A}_0={\cal A}_0^{nab}[T^A,T^B]+{\cal A}_0^{ab}\{T^A,T^B\}\,\,\,,
\end{equation}
where $T^{A,B}=\lambda^{A,B}/2$ in terms of the Gell-Mann matrices
$\lambda^{A,B}~$\footnote{We note that the one-loop virtual amplitude
  can be expressed in terms of the same antisymmetric color factor
  $[T^A,T^B]$ and a symmetric color factor made of $\{T^A,T^B\}$ and
  $\delta^{AB}$.}. ${\cal A}_0^{ab}$ and ${\cal A}_0^{nab}$ correspond
to the terms in the amplitude that are proportional respectively to
the \emph{abelian} (or symmetric) and \emph{non-abelian} (or
antisymmetric) color factors and are explicitly given by:
\begin{equation}
  \label{eq:a0_ab_nab}
  {\cal A}_0^{ab}=\frac{1}{2}({\cal A}_{0,t}+{\cal A}_{0,u})\,\,\,,\,\,\,
  {\cal A}_0^{nab}={\cal A}_{0,s}+\frac{1}{2}({\cal A}_{0,t}-{\cal A}_{0,u})\,\,\,,
\end{equation}
where ${\cal A}_{0,s}$, ${\cal A}_{0,t}$, and ${\cal A}_{0,u}$ are the
amplitudes corresponding to the sum of the $s-$channel, $t-$channel,
and $u-$channel tree level diagrams in Fig.~\ref{fg:tree_gg}. ${\cal
  A}_{0,s}$, ${\cal A}_{0,t}$, and ${\cal A}_{0,u}$ are given
explicitly in Appendix~\ref{sec:app_tree_level}.

Due to the \emph{orthogonality} between symmetric and antisymmetric
color factors, the tree level amplitude squared takes the very simple
form:
\begin{equation}
\label{eq:a0_square}
\overline{\sum}|{\cal A}_0|^2=
\overline{\sum}\left[\frac{N}{2}(N^2-1)\left(
|{\cal A}_0^{nab}|^2+|{\cal A}_0^{ab}|^2\right)-
\frac{1}{N}(N^2-1)|{\cal A}_0^{ab}|^2
\right]\,\,\,,
\end{equation}
from which we can derive the LO partonic cross section, upon
integration over the final state phase space:
\begin{equation}
{\hat \sigma}_{\sss LO}^{ij}(x_1,x_2,\mu)=
\int d(PS_3) \overline{\sum}|{\cal A}_0|^2(x_1,x_2,\mu)\,\,\,,
\end{equation}
where the dependence of $|{\cal A}_0|^2$ on $x_1$ and $x_2$ (through
$s\!=\!x_1x_2s_H$) and on the renormalization scale $\mu$ (through
$\alpha_s(\mu))$) has been made explicit.

When averaging over the polarization states of the initial gluons, the
polarization sum of the gluon polarization vectors,
$\epsilon_{\mu}(q_1,\lambda_1)$ and $\epsilon_{\nu}(q_2,\lambda_2)$,
has to be performed in such a way that only the physical (transverse)
polarization states of the gluons contribute to the matrix element
squared. We adopt the general prescription:
\begin{equation}
\sum_{\lambda_i=1,2}\epsilon_{\mu}(q_i,\lambda_i) 
\epsilon_{\nu}^*(q_i,\lambda_i)=
-g_{\mu\nu}+\frac{n_{i\mu}q_{i\nu}+q_{i\mu}n_{i\nu}}{n_i\cdot q_i}-
\frac{n_i^2q_{i\mu}q_{j\nu}}{(n_i\cdot q_i)^2}\,\,\,,
\end{equation}
where $i\!=\!1,2$ and the arbitrary vectors $n_i$ have to satisfy the
relations:
\begin{equation}
n_i^\mu\sum_{\lambda_i=1,2} \epsilon_{\mu}(q_i,\lambda_i) 
\epsilon_{\nu}^*(q_i,\lambda_i)=0\,\,\,\,,\,\,\,\,
n_i^\nu\sum_{\lambda_i=1,2} \epsilon_{\mu}(q_i,\lambda_i) 
\epsilon_{\nu}^*(q_i,\lambda_i)=0\,\,\,,
\end{equation}
together with $n_i^2\!\neq\!0$ and $n_1\!\neq\! n_2$ We choose
$n_1\!=\!q_2$ and $n_2\!=\!q_1$, such that:
\begin{equation}
\sum_{\lambda_i=1,2}\epsilon_{\mu}(q_i,\lambda_i) 
\epsilon_{\nu}^*(q_i,\lambda_i)=
-g_{\mu\nu} +2 \frac{q_{1\mu} q_{2\nu}+q_{2\mu} q_{1\nu}}{s}\,\,\,.
\end{equation}
Finally, the entire calculation is performed using Feynman gauge for
both internal and external gluons.
\begin{figure}[t]
\begin{center}
\includegraphics[scale=0.85]{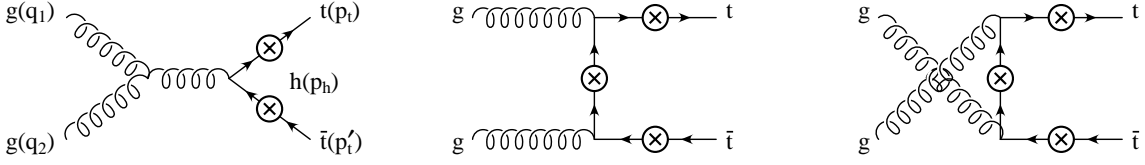} 
\caption[]{Feynman diagrams contributing to the tree level process
  $gg\to t\bar{t}h$. The circled crosses indicate all possible
  insertions of the final state Higgs boson leg, each insertion
  corresponding to a different diagram.}
\label{fg:tree_gg}
\end{center}
\end{figure}
\boldmath
\section{NLO virtual QCD corrections to $gg\to t\bar{t}h$: 
 the $\hat{\sigma}_{virt}^{gg}$ cross section.}
\unboldmath
\label{sec:virtual}

The ${\cal O}(\alpha_s)$ virtual corrections to the $gg\to t\bar{t}h$
tree level process consist of the self-energy, vertex, box, and
pentagon diagrams illustrated in
Figs.~\ref{fg:self_gg}-\ref{fg:pentagons_gg}. The ${\cal
  O}(\alpha_s^3)$ contribution to the virtual amplitude squared of
Eq.~(\ref{eq:delta_sigmahat}) can then be written as:
\begin{equation}
\label{eq:amp2_virt_gen}
\overline{\sum}|{\cal A}_{virt}(gg\to t\bar th)|^2=
\sum_{D_{i,j}}\overline{\sum}\left({\cal A}_0 {\cal A}_{D_{i,j}}^*+
{\cal A}_0^* {\cal A}_{D_{i,j}}\right)=
\sum_{D_{i,j}}\overline{\sum}2\,{\cal R}e\left({\cal A}_0 
{\cal A}_{D_{i,j}}^*\right)\,\,\,,
\end{equation}
where ${\cal A}_0$ is the tree level amplitude given in
Eq.~(\ref{eq:amp_tree}), while ${\cal A}_{D_{i,j}}$ denotes the
amplitude for a class of virtual diagrams that only differ by the
insertion of the final state Higgs boson leg, i.e. $D_{i,j}=\sum_k
D_{i,j}^{(k)}$ with $D_i\!=\!S_i,V_i,B_i,P_i$, $j=s,t,u$, and $k$
running over all possible Higgs boson insertions, as illustrated in
Figs.~\ref{fg:self_gg}-\ref{fg:pentagons_gg}.
\begin{figure}[t]
\begin{center}
\includegraphics[scale=0.85]{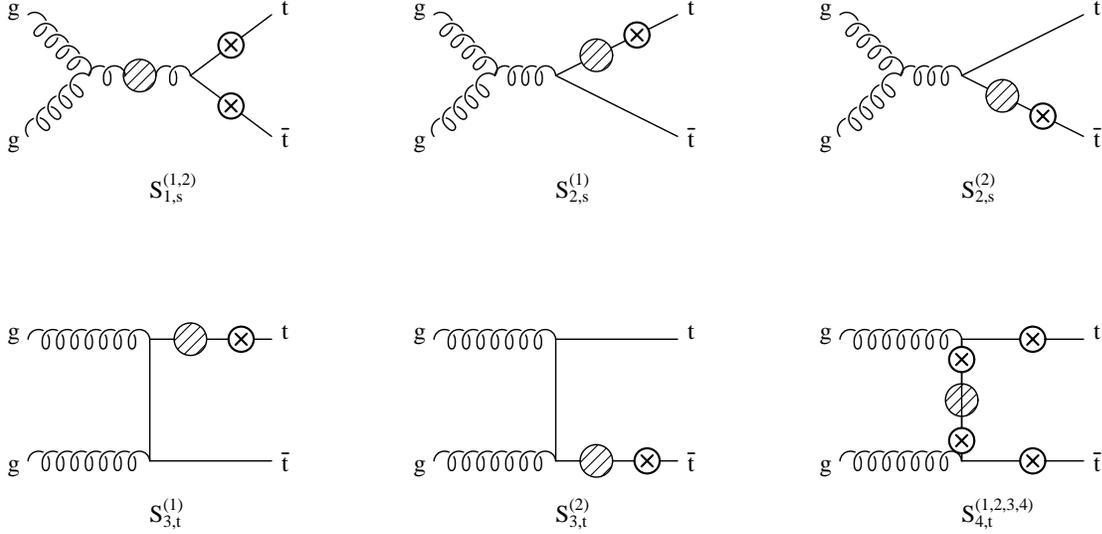} 
\caption[]{${\cal O}(\alpha_s)$ virtual corrections to $gg\to
  t\bar{t}h$: self-energy diagrams. The shaded blobs denote standard
  one-loop QCD corrections to the gluon and top quark propagators
  respectively. The circled crosses denote all possible insertions of
  the final state Higgs boson leg, each insertion corresponding to a
  different diagram. All $t$-channel diagrams (labeled as
  $S_{i,t}^{(j)}$) have corresponding $u$-channel diagrams.}
\label{fg:self_gg}
\end{center}
\end{figure}
\begin{figure}[t]
\begin{center}
\includegraphics[scale=0.85]{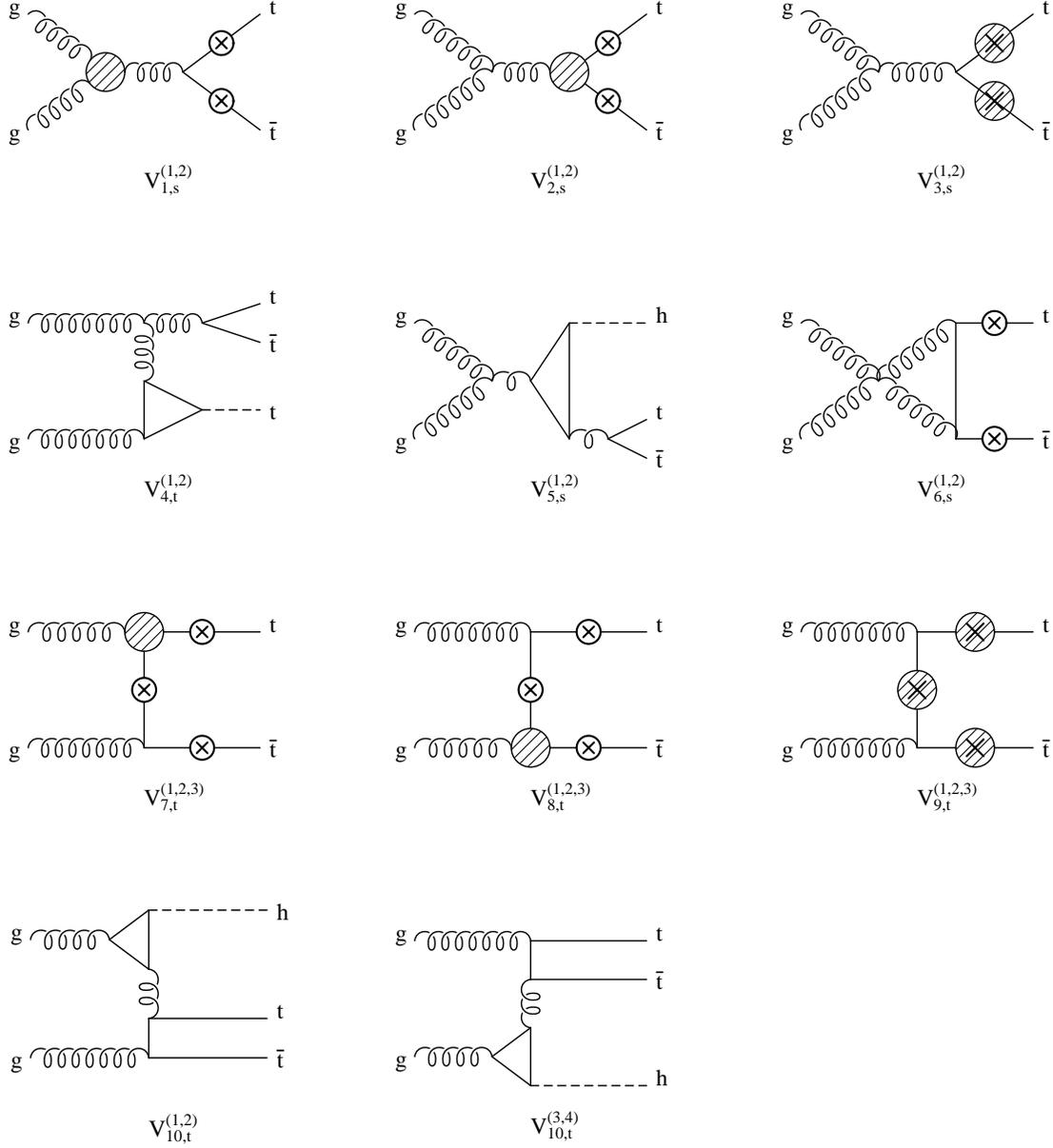} 
\caption[]{${\cal O}(\alpha_s)$ virtual corrections to $gg\to
  t\bar{t}h$: vertex diagrams. The shaded blobs denote standard
  one-loop QCD corrections to the $ggg$, $gt\bar{t}$, or $ht\bar{t}$
  vertices respectively. The circled crosses denote all possible
  insertions of the final Higgs boson leg, each insertion
  corresponding to a different diagram. Diagrams with a closed fermion
  loop have to be counted twice, once for each orientation of the loop
  fermion line.  All $t$-channel diagrams (labeled as $V_{i,t}^{(j)}$)
  have corresponding $u$-channel diagrams.}
\label{fg:vertices_gg}
\end{center}
\end{figure}
\begin{figure}[t]
\begin{center}
\includegraphics[scale=0.85]{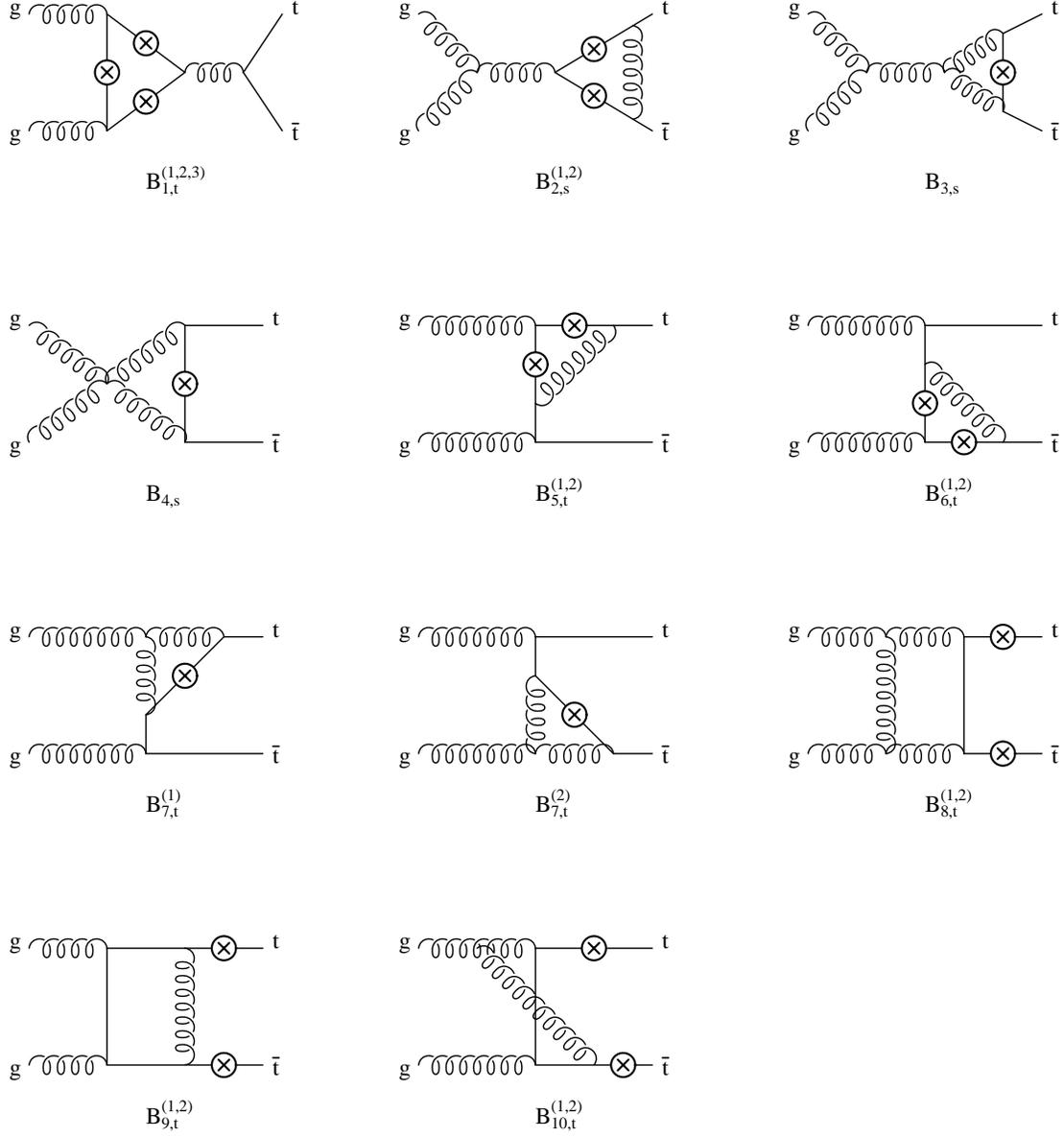} 
\caption[]{${\cal O}(\alpha_s)$ virtual corrections to $gg\to
  t\bar{t}h$: box diagrams. The circled crosses denote all possible
  insertions of the final Higgs boson leg, each insertion
  corresponding to a different diagram.  Diagrams with a closed
  fermion loop have to be counted twice, once for each orientation of
  the loop fermion line. All $t$-channel diagrams (labeled as
  $B_{i,t}^{(j)}$) have corresponding $u$-channel diagrams.}
\label{fg:boxes_gg}
\end{center}
\end{figure}
\begin{figure}[t]
\begin{center}
\includegraphics[scale=0.85]{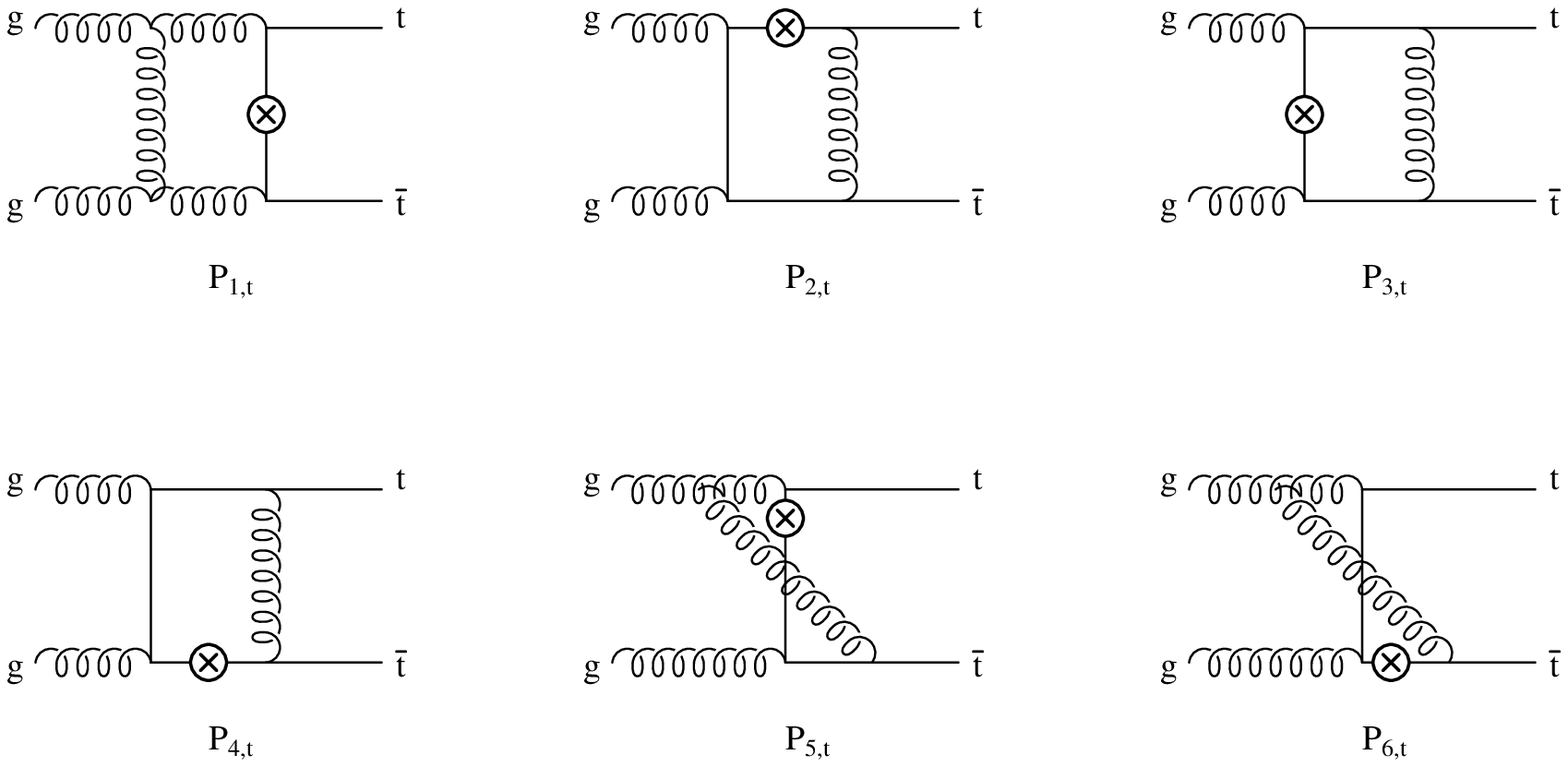} 
\caption[]{${\cal O}(\alpha_s)$ virtual corrections to $gg\to
  t\bar{t}h$: pentagon diagrams. The circled crosses denote all
  possible insertions of the final Higgs boson leg, each insertion
  corresponding to a different diagram. All $t$-channel diagrams
  (labeled as $P_{i,t}$) have corresponding $u$-channel diagrams.}
\label{fg:pentagons_gg}
\end{center}
\end{figure}

The amplitude of each virtual diagram (${\cal A}_{D_{i,j}}$) is
calculated as a linear combination of fundamental Dirac structures
with coefficients that depend on both tensor and scalar one-loop
Feynman integrals with up to five denominators.  The tensor integrals
are further reduced in terms of scalar one-loop integrals using
standard techniques \cite{'tHooft:1979xw,Passarino:1979jh}. The ${\cal
  O}(\alpha_s)$ virtual corrections to $gg\to t\bar{t}h$ involve
pentagon tensor integrals of rank higher than one, i.e. Feynman
integrals with five denominators and more than one Lorentz tensor
index.  These pentagon tensor integrals are not present in the
corresponding corrections for $q\bar{q}\to t\bar{t}h$. This introduces
a new difficulty in the calculation, due to the numerical
instabilities that may arise as a consequence of the proportionality
of the tensor integral coefficients to higher powers of the inverse
Gram determinant (GD) of the full $gg\to t\bar{t}h$ phase space.
Indeed, the standard techniques introduced in
Refs.~\cite{'tHooft:1979xw,Passarino:1979jh} allow us to rewrite a
tensor integral as a linear combination of the linearly independent
tensor structures that can be built, for a given tensor rank, out of
the independent external momenta and the metric tensor.  The
coefficients of the linearly independent tensor structures can be
found by solving a system of linear equations, one for each
independent tensor structure. As a result, they are proportional to
inverse powers of the so called Gram determinant (GD), of the form
${\rm GD}\!=\!\mbox{det}(p_i\!\cdot\!p_j)$ with $p_i$ and $p_j$
generic independent external momenta (for $i,j\!=\!1,\dots,4$, since
only four out of the five external momenta are independent). The
higher the rank of the original tensor integral, the higher the
inverse power of GD that appears in the coefficients of its tensor
decomposition.

To briefly illustrate the problem, we parameterize the Gram
determinant in terms of the $t\bar{t}h$ phase space variables as
\begin{eqnarray}
\label{eq:gram_det}
{\rm GD}&=&-\frac{[s-(2 m_t+M_h)^2]}{64}
[M_h^4+(s-\bar s_{t\bar t})^2-2M_h^2(s+\bar s_{t\bar t})]\, s \, 
\bar s_{t\bar t}
\sin^2\theta_{t \bar t} \sin^2\phi_{t \bar t} \sin^2\theta\,\,\,, 
\nonumber \\
\end{eqnarray}
where $s\!=\!x_1x_2s_{\sss H}$ is the partonic center-of-mass energy
squared, and the $t\bar th$ phase space has been expressed in terms of
a time-like invariant $\bar s_{t\bar t}\!=\!(p_t+p_t^\prime)^2$, polar
angles ($\theta, \theta_{t\bar t}$) and azimuthal angles ($\phi,
\phi_{t\bar t}$) in the center-of-mass frames of the incoming gluons
and of the $t\bar t$ pair, respectively.  As can be seen in
Eq.~(\ref{eq:gram_det}), the Gram determinant vanishes when two
momenta become degenerate, i.e. at the boundaries of phase space.
Near the boundary of phase space it can become arbitrary small, giving
rise to spurious divergences which cause serious numerical
difficulties, since they appear in various parts of the calculation
that are normally numerically, not analytically, combined.  In the
case of a $2\to 3$ process, this problem arises for pentagon tensor
integrals, when all the independent external momenta are involved, and
it becomes more serious for higher rank tensor integrals.  The
probability that the Monte Carlo integration hits a point close to the
boundary of phase space is not negligible and these points cannot just
be discarded.

We use two methods to overcome this problem and find agreement within
the statistical uncertainty of the Monte Carlo phase space
integration. In the first method, we impose kinematic cuts to avoid
the phase space regions where the Gram determinant vanishes, and then
extrapolate from the numerically safe to the numerically unsafe region
using different algorithms. We have used extrapolations based on
polynomial or trigonometric functions. We have also reproduced the
analytic dependence of each pentagon diagram on the Gram determinant,
tested it in the safe region of phase space, and used it to
extrapolate to the unsafe region. A phase space point is kept only if
the true and the extrapolated results come very close to each other,
after repeated iterations.  Each extrapolation has been repeated
imposing cuts on different kinematic variables, until a stable answer,
independent of the kinematic cuts, can be found. The details of the
extrapolation procedure are very technical and we do not think they
can be of interest to this discussion.  In the second method, after
having interfered the pentagon amplitudes with the Born matrix
element, we eliminate all pentagon tensor integrals by simplifying
scalar products of the loop momentum in the numerator against the
propagators in the denominator wherever possible. The resulting
expressions are very large, but numerically very stable, and we have
used them to confirm the results obtained using the extrapolation
methods explained above.

After the tensor integral reduction is performed, the fundamental
building blocks are one-loop scalar integrals with up to five
denominators. They may be finite or contain both ultraviolet (UV) and
infrared (IR) divergences. The finite scalar integrals are evaluated
using the method described in Ref.~\cite{Denner:1993kt} and cross
checked with the numerical package FF~\cite{vanOldenborgh:1990wn}. The
singular scalar integrals are calculated analytically by using
dimensional regularization in $d\!=\!4-2\epsilon$ dimensions. The most
difficult integrals arise from IR divergent pentagon diagrams with
several external and internal massive particles. We calculate them as
linear combination of box integrals using the method of
Ref.~\cite{Bern:1993em,Bern:1994kr} and of Ref.~\cite{Denner:1993kt}.
Details of the box and pentagon scalar integrals used in this
calculation are given in Appendix~\ref{sec:app_ir_integrals}. All
other scalar integrals, with two or three denominators, are commonly
found in the literature.

Inserting all diagram contributions into Eq.~(\ref{eq:amp2_virt_gen}),
we obtain the complete ${\cal O}(\alpha_s^3)$ contribution to the
virtual amplitude squared, and integrating over the final state phase
space we calculate $\hat\sigma^{gg}_{virt}$ in
Eq.~(\ref{eq:delta_sigmahat}). The UV singularities of the virtual
cross section are regularized in $d\!=\!4-2 \epsilon_{\sss UV}$
dimensions and renormalized by introducing a suitable set of
counterterms, while the residual renormalization scale dependence is
checked from first principles using renormalization group arguments.
The detailed renormalization procedure adopted in this calculation is
explained in Section~\ref{subsec:virtual_uv}. The IR singularities of
the virtual cross section are extracted in $d\!=\!4-2 \epsilon_{\sss
  IR}$ dimensions and are cancelled by analogous singularities in the
${\cal O}(\alpha_s^3)$ real cross section.  The structure of the IR
singular part of the virtual cross section is presented in
Section~\ref{subsec:virtual_ir}, while the IR singularities of the
real cross section are discussed in Section~\ref{sec:real}. The
explicit cancellation of IR singularities in the total inclusive NLO
cross section for $gg\to t\bar{t}h$ is outlined in
Sections~\ref{sec:real} and \ref{sec:total}.

Finally, we note that the tree level amplitude ${\cal A}_0$ in
Eq.~(\ref{eq:amp2_virt_gen}) has generically to be considered as the
$d$-dimensional tree level amplitude. This matters when the ${\cal
  A}_{D_{i,j}}$ amplitudes in Eq.~(\ref{eq:amp2_virt_gen}) are UV or
IR divergent. Actually, as will be shown in the following, both UV and
IR divergences are always proportional to the tree level amplitude or
parts of it and they can be formally cancelled without having to
explicitly specify the dimensionality of the tree level amplitude(s).
After UV and IR singularities have been cancelled, everything is
calculated in $d\!=\!4$ dimensions.
\subsection{Virtual corrections: UV singularities and counterterms}
\label{subsec:virtual_uv}

Self-energy and vertex one loop corrections to the tree level $gg\to
t\bar th$ process give rise to UV divergences. These singularities are
cancelled by a set of counterterms fixed by well defined
renormalization conditions. As required by renormalization group
arguments, the renormalization of the fundamental propagators and
interaction vertices of the theory reduces to introducing counterterms
for the external field wave functions of top quarks and gluons
($\delta Z_2^{(t)}$, $\delta Z_3$), for the top mass ($\delta m_t$),
and for the strong coupling constant ($\delta Z_{\alpha_s}$).  The
counterterm for the top quark Yukawa coupling,
$g_{t\bar{t}h}\!=\!m_t/v$, coincides with the counterterm for the top
mass, since the SM Higgs vacuum expectation value $v$ is not
renormalized at one loop in QCD.

By carefully grouping subsets of self-energy and vertex diagrams, we
can factor out the UV singularities of the ${\cal O}(\alpha_s^3)$
virtual amplitude and write them in terms of the tree level partial
amplitudes ${\cal A}_{0,s}$ , ${\cal A}_{0,t}$, and ${\cal A}_{0,u}$
introduced in Eq.~(\ref{eq:a0_ab_nab}) and defined in
Appendix~\ref{sec:app_tree_level}. According to the notation
introduced in Figs.~\ref{fg:self_gg}-\ref{fg:pentagons_gg}, we denote
by $D_{i,j}$ (with $D\!=\!S,V$, $i\!=\!1,2,\ldots$, and $j\!=\!s,t,u$)
a class of diagrams with a given self-energy or vertex correction
insertion, summed over all possible insertions of the external Higgs
field, one for each different diagram.  We now define
$\Delta_{UV}({\cal A}_{D_{i,j}})$ to be the UV pole part of the
corresponding amplitude.  Using this notation, we find
\begin{eqnarray}
\label{eq:virtual_uv}
&&\Delta_{UV}({\cal A}_{S_{1,s}})=\frac{\alpha_s}{4\pi}
\left[{\cal N}_s\,\left(\frac{5}{3}N-\frac{2}{3}n_{lf}\right)-
{\cal N}_t\,\frac{2}{3}\right]\left(\frac{1}{\epsilon_{\sss UV}}\right)
[T^A,T^B]{\cal A}_{0,s}\,\,\,,
\nonumber\\
&&\Delta_{UV}({\cal A}_{V_{1,s}})=\frac{\alpha_s}{4\pi}\left[
{\cal N}_s\,
\left(-\frac{2}{3}N+\frac{2}{3}n_{lf}\right)+{\cal N}_t\,\frac{2}{3}\right]
\left(\frac{1}{\epsilon_{\sss UV}}\right)
[T^A,T^B]{\cal A}_{0,s}\,\,\,,
\nonumber\\
&&\Delta_{UV}({\cal A}_{V_{2,s}}+{\cal A}_{V_{7,t}}+{\cal A}_{V_{7,u}})=
\frac{\alpha_s}{4\pi}{\cal N}_t\,
\left(\frac{3}{2}N-\frac{1}{2N}\right)
\left(\frac{1}{\epsilon_{\sss UV}}\right){\cal A}_0\,\,\,,
\nonumber\\
&&\Delta_{UV}({\cal A}_{V_{8,t}}+{\cal A}_{V_{8,u}})=
\frac{\alpha_s}{4\pi}{\cal N}_t\,
\left(\frac{3}{2}N-\frac{1}{2N}\right)
\left(\frac{1}{\epsilon_{\sss UV}}\right)\times\nonumber\\
&&\quad\quad\quad\quad\quad
\left(\frac{1}{2}({\cal A}_{0,t}-{\cal A}_{0,u})[T^A,T^B]+
\frac{1}{2}({\cal A}_{0,t}+{\cal A}_{0,u})\{T^A,T^B\}\right)\,\,\,,
\nonumber\\
&&\Delta_{UV}({\cal A}_{V_{3,s}}+{\cal A}_{V_{9,t}}+{\cal A}_{V_{9,u}})=
\frac{\alpha_s}{4\pi}{\cal N}_t\,
\left(\frac{N}{2}-\frac{1}{2N}\right)
\left(\frac{4}{\epsilon_{\sss UV}}\right){\cal A}_0\,\,\,,
\nonumber\\
&&\Delta_{UV}({\cal A}_{S_{2,s}}+
{\cal A}_{S_{3,t}}+{\cal A}_{S_{3,u}}+
{\cal A}_{S_{4,t}}+{\cal A}_{S_{4,u}})=
\frac{\alpha_s}{4\pi}{\cal N}_t\,
\left(\frac{N}{2}-\frac{1}{2N}\right)
\left(-\frac{1}{\epsilon_{\sss UV}}\right)\times\nonumber\\
&&\quad\quad\quad\quad\quad
\left({\cal A}_0+\frac{1}{2}({\cal A}_{0,t}-{\cal A}_{0,u})[T^A,T^B]+
\frac{1}{2}({\cal A}_{0,t}+{\cal A}_{0,u})\{T^A,T^B\}\right)\,\,\,,
\nonumber\\
\end{eqnarray}
where $n_{lf}\!=\!5$ corresponds to the number of light quark flavors,
$N\!=\!3$ is the number of colors, ${\cal N}_s$ and ${\cal N}_t$ are
defined as:
\begin{equation}
\label{eq:nsnt}
{\cal N}_s=\left(\frac{4\pi\mu^2}{s}\right)^\epsilon
\Gamma(1+\epsilon)\,\,\,,\,\,\,
{\cal N}_t=\left(\frac{4\pi\mu^2}{m_t^2}\right)^\epsilon
\Gamma(1+\epsilon)\,\,\,,
\end{equation}
and we have already included in the top quark self-energy diagrams 
the top mass counterterm.

We notice that some of the UV divergent virtual corrections
($V_{1,s}$, $V_{7,(t,u)}$, and $V_{8,(t,u)}$), as well as $\delta
Z_2^{(t)}$ and $\delta Z_3$ in Eqs.~(\ref{eq:z2_t}) and (\ref{eq:z3})
below, have also IR singularities. In this section we limit the
discussion to the UV singularities only, while the IR structure of
these terms will be considered in Section~\ref{subsec:virtual_ir}. To
this purpose we have explicitly denoted by $\epsilon_{\sss UV}$ the
pole parameter.

The corresponding counterterms are defined as follows.  For the
external fields, we fix the wave-function renormalization constants of
the external top quark fields using the on-shell subtraction scheme:
\begin{equation}
\label{eq:z2_t}
\big(\delta Z_2^{(t)}\big)_{UV}=-\frac{\alpha_s}{4\pi}{\cal N}_t
\left(\frac{N}{2}-\frac{1}{2N}\right)
 \left(\frac{1}{\epsilon_{\sss UV}}+4\right)\,\,\,,
\end{equation}
while we renormalize the wave-function of external gluons in the
$\overline{MS}$ subtraction scheme:
\begin{equation}
\label{eq:z3}
\left(\delta Z_3\right)_{UV}=\frac{\alpha_s}{4\pi}
(4\pi)^\epsilon\Gamma(1+\epsilon)
\left\{
\left(\frac{5}{3}N-\frac{2}{3}n_{lf}\right)\frac{1}{\epsilon_{\sss UV}}-
\frac{2}{3}\left[\frac{1}{\epsilon_{\sss UV}}+
\ln\left(\frac{\mu^2}{m_t^2}\right)\right]
\right\}\,\,\,,
\end{equation}
according to which we also need to consider the insertion of a finite
self-energy correction on the external gluon legs. This amounts to an
extra contribution
\begin{equation}
\delta_{\sss UV}=\frac{\alpha_s}{4\pi}
(4\pi)^\epsilon\Gamma(1+\epsilon)
\left(\frac{5}{3}N-\frac{2}{3}n_{lf}\right)
\ln\left(\frac{\mu^2}{m_t^2}\right)\,\,\,,
\end{equation}
which is important in order to obtain the correct scale dependence of
the NLO cross section.

We define the subtraction condition for the top-quark mass $m_t$ in
such a way that $m_t$ is the pole mass, in which case the top-mass
counterterm is given by:
\begin{equation}
\label{eq:mt_ct}
\frac{\delta m_t}{m_t}=-\frac{\alpha_s}{4\pi}{\cal N}_t
\left(\frac{N}{2}-\frac{1}{2N}\right)
\left(\frac{3}{\epsilon_{\sss UV}}+4\right)\,\,\,.
\end{equation}
This counterterm has to be used twice: to renormalize the top-quark
mass, in all diagrams that contain a top quark self-energy insertion,
and to renormalize the top quark Yukawa coupling. As previously noted,
the expressions in Eq.~(\ref{eq:virtual_uv}) already include the
top-mass counterterm.

Finally, for the renormalization of $\alpha_s$ we use the
$\overline{MS}$ scheme, modified to decouple the top quark
\cite{Collins:1978wz,Nason:1989zy}.  The first $n_{lf}$ light flavors
are subtracted using the $\overline{MS}$ scheme, while the divergences
associated with the top-quark loop are subtracted at zero momentum:
\begin{equation}
\label{eq:alphas_ct}
\delta Z_{\alpha_s}=\frac{\alpha_s}{4\pi}(4\pi)^\epsilon
\Gamma(1+\epsilon)
\left\{
\left(\frac{2}{3}n_{lf}-\frac{11}{3}N\right)
\frac{1}{\epsilon_{\sss UV}}+
\frac{2}{3}\left[\frac{1}{\epsilon_{\sss UV}}+
\ln\left(\frac{\mu^2}{m_t^2}\right)\right]
\right\}\,\,\,,
\end{equation}
such that, in this scheme, the renormalized strong coupling constant
$\alpha_s(\mu)$ evolves with $n_{lf}=5$ light flavors.

Using the results in Eqs.~(\ref{eq:virtual_uv})-(\ref{eq:alphas_ct}) it
is easy to verify that the UV pole part of $\hat\sigma_{virt}^{gg}$:
\begin{eqnarray}
\label{eq:sigma_virt_uv_pole}
(\hat\sigma_{virt}^{gg})_{UV-pole}&=&
\int d(PS_3)\,\sum_{D_{i,j}} \overline{\sum}2\,{\cal R}e\left(
{\cal A}_0\,\Delta_{\sss UV}({\cal A}_{D_{i,j}}^*)\right)+\nonumber\\
&&2 \hat\sigma^{gg}_{\sss LO}\left[
\big(\delta Z_2^{(t)}\big)_{\sss UV}+
\left(\delta Z_3\right)_{\sss UV}+\delta_{\sss UV}+
\frac{\delta m_t}{m_t}+\delta Z_{\alpha_s}\right]
\end{eqnarray}
is free of UV singularities and has a residual renormalization scale
dependence of the form:
\begin{equation}
\label{eq:mudep_res_uv}
\hat\sigma_{\sss LO}^{gg} \frac{\alpha_s(\mu)}{2\pi}
\left(-\frac{2}{3}n_{lf}+\frac{11}{3} N\right)
\ln\left(\frac{\mu^2}{s}\right)\,\,\,,
\end{equation}
as expected by renormalization group arguments (see the first term of
Eq.~(\ref{eq:mudep_coeff})). We note that the presence of $s$ in the
argument of the logarithm of Eq.~(\ref{eq:mudep_res_uv}) has no
particular relevance. Choosing a different argument would amount to
reabsorbing some $\mu$-independent logarithms in $f_1^{ij}$ of
Eq.~(\ref{eq:f_ij_nlo}).
\subsection{Virtual corrections: IR singularities}
\label{subsec:virtual_ir}

The structure of the IR singularities originating from the ${\cal
  O}(\alpha_s)$ virtual corrections to the tree level amplitude for
$gg\to t\bar{t}h$ is more involved than for the UV singularities.
However it simplifies considerably when given at the level of the
amplitude squared, and this is what we present in this section.

The IR divergent part of the ${\cal O}(\alpha_s^3)$ virtual amplitude
squared of Eq.~(\ref{eq:amp2_virt_gen}) can be written in the
following compact form:
\begin{equation}
\sum_{D_{i,j}} \overline{\sum}2{\cal R}e\left(
{\cal A}_0\,\Delta_{\sss IR}({\cal A}_{D_{i,j}}^*)\right)=
\frac{\alpha_s}{2\pi}{\cal N}_t  \overline{\sum}
\left(C_1 {\cal M}_{V,\epsilon}^{(1)}+C_2 {\cal M}_{V,\epsilon}^{(2)}+
C_3 {\cal M}_{V,\epsilon}^{(3)}\right)\,\,\,,
\end{equation}
where ${\cal N}_t$ is defined in Eq.~(\ref{eq:nsnt}) and we denote by
$\Delta_{\sss IR}({\cal A}_{D_{i,j}})$ the IR pole part of the
amplitude of a given $D_{i,j}$ class of diagrams. The result is
organized in terms of leading and sub-leading color factors:
\begin{eqnarray}
\label{eq:color_factors}
C_1&=&\frac{N^2}{4}(N^2-1)\,\,\,,\nonumber\\
C_2&=&-\frac{1}{4}(N^2-1)\,\,\,,\nonumber\\
C_3&=&\left(1+\frac{1}{N^2}\right)(N^2-1)\,\,\,,
\end{eqnarray}
and the corresponding matrix elements squared $M_{V,\epsilon}^{(1)}$, 
$M_{V,\epsilon}^{(2)}$, and $M_{V,\epsilon}^{(3)}$ are given by:
\begin{eqnarray}
\label{eq:m2_virt_ir_poles}
{\cal M}_{V,\epsilon}^{(1)}&=&
\left[-\frac{4}{\epsilon_{\sss IR}^2}+\frac{2}{\epsilon_{\sss IR}}
(-2+\Lambda_\sigma)\right]
\left(|{\cal A}_0^{nab}|^2+|{\cal A}_0^{ab}|^2\right)\nonumber\\
&+&\frac{1}{\epsilon_{\sss IR}}\left[
\left(\Lambda_{\tau_1}+\Lambda_{\tau_2}\right)
|{\cal A}_{0,s}+{\cal A}_{0,t}|^2+
\left(\Lambda_{\tau_3}+\Lambda_{\tau_4}\right)
|{\cal A}_{0,u}-{\cal A}_{0,s}|^2\right]\,\,\,,
\nonumber\\
{\cal M}_{V,\epsilon}^{(2)}&=&\left[-\frac{8}{\epsilon_{\sss IR}^2}+
\frac{4}{\epsilon_{\sss IR}}\left(-2+\Lambda_{\tau_1}+\Lambda_{\tau_2}+
\Lambda_{\tau_3}+\Lambda_{\tau_4}\right)\right]|{\cal A}_0^{ab}|^2\nonumber\\
&+&\frac{2}{\epsilon_{\sss IR}}\frac{\bar s_{t\bar{t}}-2m_t^2}
{\bar s_{t\bar{t}}\beta_{t\bar{t}}} \Lambda_{t\bar{t}}
\left(|{\cal A}_0^{nab}|^2+|{\cal A}_0^{ab}|^2\right)\,\,\,,\nonumber\\
{\cal M}_{V,\epsilon}^{(3)}&=&\frac{1}{\epsilon_{\sss IR}}
\frac{\bar{s}_{t\bar{t}}-2m_t^2}{\bar{s}_{t\bar{t}}\beta_{t\bar{t}}}
\Lambda_{t\bar{t}}
|{\cal A}_0^{ab}|^2 \,\,\,,
\end{eqnarray}
where the IR nature of the pole terms has been made explicit. ${\cal
  A}_0^{ab}$ and ${\cal A}_0^{nab}$ are defined in
Eq.~(\ref{eq:a0_ab_nab}), while ${\cal A}_{0,s}$, ${\cal A}_{0,t}$,
and ${\cal A}_{0,u}$ are given explicitly in
Appendix~\ref{sec:app_tree_level}. Moreover, we have defined:
\begin{equation}
\label{eq:stt_betatt}
\bar s_{t\bar{t}}=(p_t+p_t^\prime)^2\,\,\,\,,\,\,\,\,
\beta_{t\bar{t}}=\sqrt{1-\frac{4m_t^2}{\bar{s}_{t\bar{t}}}}\,\,\,,
\, \, \, \Lambda_{t\bar{t}}=\ln\left(\frac{1+\beta_{t\bar{t}}}
{1-\beta_{t\bar{t}}}\right) \, ,
\end{equation}
and we have introduced the notation:
$\Lambda_{\sigma}=\ln(\sigma/m_t^2)$ and
$\Lambda_{\tau_i}=\ln(\tau_i/m_t^2)$ where
\begin{eqnarray}
\label{eq:kinematic_invariants}
\sigma&=&(q_1+q_2)^2\,\,\,,\nonumber\\
\tau_1 &=& m_t^2-(q_1-p_t)^2=2\,q_1\cdot p_t\,\,\,, \nonumber\\
\tau_2 &=& m_t^2-(q_2-p_t^\prime)^2=2\,q_2\cdot p_t^\prime \,\,\,,
\nonumber\\
\tau_3 &=& m_t^2-(q_2-p_t)^2=2\,q_2\cdot p_t \,\,\,,\nonumber\\
\tau_4 &=& m_t^2-(q_1-p_t^\prime)^2=2\,q_1\cdot p_t^\prime \,\,\,.
\end{eqnarray}
When we add the IR singularities coming from the counterterms that we
have introduced in Section~\ref{subsec:virtual_uv}, we can write the
complete pole part of the IR singular ${\cal O}(\alpha_s^3)$ virtual
cross section as:
\begin{eqnarray}
\label{eq:sigma_virt_ir_poles}
(\hat\sigma_{virt}^{gg})_{IR-pole}&=& \int d(PS_3) 
\sum_{D_{i,j}} \overline{\sum}2{\cal R}e\left(
{\cal A}_0\,\Delta_{\sss IR}({\cal A}_{D_{i,j}}^*)\right)+
2\hat \sigma^{gg}_{\sss LO} \left(
\big(\delta Z_2^{(t)}\big)_{\sss IR}+
\left(\delta Z_3\right)_{\sss IR}\right)\nonumber\\
&=& \int d(PS_3)\frac{\alpha_s}{2\pi}{\cal N}_t  \overline{\sum}
\left(C_1 {\cal M}_{V,\epsilon}^{(1)}+C_2 {\cal M}_{V,\epsilon}^{(2)}+
C_3 {\cal M}_{V,\epsilon}^{(3)}\right)\nonumber\\
&+&\frac{\alpha_s}{2\pi}{\cal N}_t
\left(\frac{2}{3}n_{lf}-\frac{8}{3}N+\frac{1}{N}\right)
\frac{1}{\epsilon_{\sss IR}} \hat\sigma^{gg}_{\sss LO}\,\,\,.
\end{eqnarray}
As will be demonstrated in Section~\ref{sec:real}, the IR singularities
of $\hat \sigma_{virt}^{gg}$ are cancelled by the corresponding IR
singularities of $\hat \sigma_{real}^{gg}$.
\boldmath
\section{NLO real QCD corrections to $gg\to t\bar{t}h$: the 
  $\hat \sigma_{real}^{gg}$ and $\hat \sigma_{real}^{qg}$ cross
  sections} \unboldmath
\label{sec:real}

The NLO real cross section $\hat\sigma_{real}^{gg}$ in
Eq.~(\ref{eq:delta_sigmahat}) corresponds to the ${\cal O}(\alpha_s)$
corrections to $gg\to t\bar{t}h$ due to the emission of a real gluon,
i.e. to the process $gg\to t\bar{t}h+g$, examples of which are
illustrated in Fig.~\ref{fg:real_gg}. It contains IR singularities
which cancel the analogous singularities present in the ${\cal
  O}(\alpha_s)$ virtual corrections (see
Section~\ref{subsec:virtual_ir}) and in the NLO parton distribution
functions.  These singularities can be either \emph{soft}, when the
energy of the emitted gluon becomes very small, or \emph{collinear},
when the final state gluon is emitted collinear to one of the initial
gluons. There is no collinear radiation from the final $t$ and
$\bar{t}$ quarks because they are massive. At the same order in
$\alpha_s$, the $\hat\sigma_{real}^{qg}$ cross section corresponds to
the tree level processes $(q,\bar{q})g\to t\bar{t}h+(q,\bar{q})$, an
example of which is also illustrated in Fig.~\ref{fg:real_gg}. This
part of the NLO cross section develops IR singularities entirely due
to the collinear emission of a final state quark or antiquark from one
of the initial state massless partons.  The IR singularities can be
conveniently isolated by \emph{slicing} the $gg\to t\bar{t}h+g$ and
$(q,\bar{q})g\to t\bar{t}h+(q,\bar{q})$ phase spaces into different
regions defined by suitable cutoffs, a method which goes under the
general name of \emph{Phase Space Slicing} (PSS).  The dependence on
the arbitrary cutoff(s) introduced in \emph{slicing} the phase space
of the final state particles is not physical, and cancels at the level
of the total real hadronic cross section, i.e. in $\sigma_{real}$, as
well as at the level of the real cross section for each separate
channel, i.e. in $\sigma_{real}^{gg}$, $\sigma_{real}^{qg}$, and
$\sigma_{real}^{q\bar{q}}$.  This cancellation constitutes an
important check of the calculation and will be discussed in detail in
Section~\ref{sec:total}.
\begin{figure}[t]
\begin{center}
\includegraphics[scale=0.85]{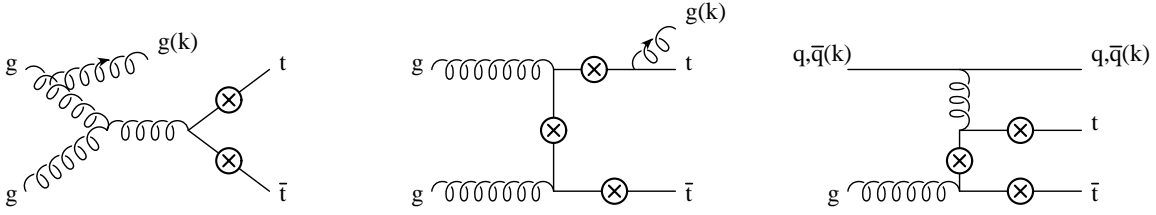} 
\caption[]{Examples of ${\cal O}(\alpha_s)$ real corrections to $gg\to
  t\bar{t}h$ (first two diagrams) and of the tree level
  $(q,\bar{q})g\to t\bar{t}h(q,\bar{q})$ processes (third diagram).
  The circled crosses denote all possible insertions of an external
  Higgs boson leg, each insertion corresponding to a different
  diagram.}
\label{fg:real_gg}
\end{center}
\end{figure}

We have calculated the cross section for the processes
\[
g(q_1)+g(q_2)\to t(p_t)+\bar{t}(p_t^\prime)+h(p_h)+g(k)\,\,\,,
\]
and
\[
(q,\bar{q})(q_1)+g(q_2)\to
t(p_t)+\bar{t}(p_t^\prime)+h(p_h)+(q,\bar{q})(k)\,\,\,, 
\]
with $q_1+q_2=p_t+p_t^\prime+p_h+k$, using two different
implementations of the PSS method which we call the \emph{two-cutoff}
and \emph{one-cutoff} methods respectively, depending on the number of
cutoffs introduced.  The \emph{two-cutoff} implementation of the PSS
method was originally developed to study QCD corrections to dihadron
production \cite{Bergmann:1989th} and has since then been applied to a
variety of processes (for a review see, e.g. \cite{Harris:2001sx}).
The \emph{one-cutoff} PSS method was developed for massless quarks in
Ref.~\cite{Giele:1992vf,Giele:1993dj} and extended to the case of
massive quarks in Ref.~\cite{Keller:1998tf}.

In the next two sections we discuss the application of the PSS method
to our case, using the \emph{two-cutoff} implementation in
Section~\ref{subsec:two_cutoff} and the \emph{one-cutoff}
implementation in Section~\ref{subsec:one_cutoff}. The results for
$\sigma_{real}$ obtained using PSS with one or two cutoffs agree
within the statistical errors of the Monte Carlo integration.  In
spite of the fact that both methods are realizations of the general
idea of phase space slicing, they have very different characteristics
and finding agreement between the two represents an important check of
our calculation.

\subsection{Phase Space Slicing method with two cutoffs}
\label{subsec:two_cutoff}

The general implementation of the PSS method using two cutoffs
proceeds in two steps. First, by introducing an arbitrary small
\emph{soft} cutoff $\delta_s$, we separate the overall integration of
the $gg\to t\bar{t}h+g$ phase space into two regions according to
whether the energy of the final state gluon ($k^0\!=\!E_g$) is
\emph{soft}, i.e.  $E_g\le\delta_s\sqrt{s}/2$, or \emph{hard}, i.e.
$E_g>\delta_s\sqrt{s}/2$. The partonic real cross section of
Eq.~(\ref{eq:delta_sigmahat}) can then be written as:
\begin{equation}
\label{eq:sigma_real_two_cutoff}
\hat{\sigma}^{gg}_{real} =
\hat{\sigma}^{gg}_{soft}+\hat{\sigma}^{gg}_{hard}\,\,\,,
\end{equation}
where $\hat\sigma^{gg}_{soft}$ is obtained by integrating over the
\emph{soft} region of the gluon phase space, and contains all the IR
soft divergences of $\hat\sigma_{real}^{gg}$.  To isolate the
remaining collinear divergences from $\hat\sigma^{gg}_{hard}$, we
further split the integration over the gluon phase space according to
whether the final state gluon is ($\hat\sigma^{gg}_{hard/coll}$) or is
not ($\hat\sigma^{gg}_{hard/non-coll}$) emitted within an angle
$\theta$ from the initial state gluons such that
$(1-\cos\theta)<\delta_c$, for an arbitrary small \emph{collinear}
cutoff $\delta_c$:
\begin{equation}
\label{eq:sigma_gg_hard}
\hat{\sigma}^{gg}_{hard}=\hat{\sigma}^{gg}_{hard/coll}+
\hat{\sigma}^{gg}_{hard/non-coll}\,\,\,.
\end{equation}
In the same way, we isolate the collinear divergences in the cross
section for the $(q,\bar{q})g$ initiated processes and write the
corresponding cross section as:
\begin{equation}
\label{eq:sigma_qg_real}
\hat{\sigma}^{qg}_{real}=\hat{\sigma}^{qg}_{coll}+
\hat{\sigma}^{qg}_{non-coll}\,\,\,.
\end{equation}
The hard non-collinear part of the real $gg$-initiated cross section,
$\hat{\sigma}^{gg}_{hard/non-coll}$, and the non-collinear part of the
$(q,\bar{q})g$-initiated cross section,
$\hat{\sigma}^{qg}_{non-coll}$, are finite and can be computed
numerically.

On the other hand, in the soft and collinear regions the integration
over the phase space of the emitted gluon or quark can be performed
analytically, thus allowing us to isolate and extract the IR
divergences of $\hat{\sigma}^{gg}_{real}$ and
$\hat{\sigma}^{qg}_{real}$.  More details on the calculation of
$\hat{\sigma}^{gg}_{soft}$ and $\hat\sigma^{gg}_{hard}$ are given in
Section~\ref{subsubsec:two_cutoff_soft} and
Section~\ref{subsubsec:two_cutoff_hard}, respectively.  The
calculation of $\hat\sigma^{qg}_{real}$ is described in
Section~\ref{subsubsec:two_cutoff_qg}.
\subsubsection{Real gluon emission, $gg\to t\bar{t}h+g$: soft region}
\label{subsubsec:two_cutoff_soft}

The soft region of the phase space for the gluon emission process
\begin{equation}
\label{eq:gg_ttbg}
g^A(q_1)+g^B(q_2)\to t(p_t)+\bar{t}(p_t^\prime)+h(p_h)+g^C(k)
\end{equation}
is defined by demanding that the energy of the emitted gluon
($k^0\!=\!E_g$) satisfies the condition
\begin{equation}
\label{eq:eg_cut}
E_g\le\delta_s\frac{\sqrt{s}}{2}
\end{equation}
for an arbitrary small value of the \emph{soft} cutoff $\delta_s$.  In
the \emph{soft limit} ($E_g\to 0$), the amplitude for this process can
be written as:
\begin{eqnarray}
&&{\cal A}_{soft}(gg\to t\bar{t}h+g)=\nonumber\\
&&T^CT^AT^B
\left(\frac{p_t\!\cdot\!\epsilon^*}{p_t\!\cdot\!k}-
\frac{q_1\!\cdot\!\epsilon^*}{q_1\!\cdot\!k}\right)
\left({\cal A}_{0,t}+{\cal A}_{0,s}\right)  
+
T^CT^BT^A
\left(\frac{p_t\!\cdot\!\epsilon^*}{p_t\!\cdot\!k}-
\frac{q_2\!\cdot\!\epsilon^*}{q_2\!\cdot\!k}\right)
\left({\cal A}_{0,u}-{\cal A}_{0,s}\right)\nonumber\\
&-&
T^AT^BT^C
\left(\frac{p_t^{\prime}\!\cdot\!\epsilon^*}{p_t^\prime\!\cdot\!k}-
\frac{q_2\!\cdot\!\epsilon^*}{q_2\!\cdot\!k}\right)
\left({\cal A}_{0,t}+{\cal A}_{0,s}\right) 
-
T^BT^AT^C
\left(\frac{p_t^{\prime}\!\cdot\!\epsilon^*}{p_t^\prime\!\cdot\!k}-
\frac{q_1\!\cdot\!\epsilon^*}{q_1\!\cdot\!k}\right)
\left({\cal A}_{0,u}-{\cal A}_{0,s}\right)\nonumber\\
&+&
T^AT^CT^B
\left(\frac{q_1\!\cdot\!\epsilon^*}{q_1\!\cdot\!k}-
\frac{q_2\!\cdot\!\epsilon^*}{q_2\!\cdot\!k}\right)
\left({\cal A}_{0,t}+{\cal A}_{0,s}\right)
+
T^BT^CT^A
\left(\frac{q_2\!\cdot\!\epsilon^*}{q_2\!\cdot\!k}-
\frac{q_1\!\cdot\!\epsilon^*}{q_1\!\cdot\!k}\right)
\left({\cal A}_{0,u}-{\cal A}_{0,s}\right)\;, 
\nonumber\\
\end{eqnarray}
where $A,B$, and $C$ are the color indices of the external gluons,
while $\epsilon^{\mu}(k,\lambda)$ (for $\lambda\!=\!1,2$) is the
polarization vector of the emitted soft gluon. Moreover, in the soft
region the $gg\to t\bar{t}h+g$ phase space factorizes as:
\begin{eqnarray}
\label{eq:ps_soft_lim}
d(PS_4)(gg\to t\bar{t}h+g)
& \stackrel{soft}{\longrightarrow} & d(PS_3)(gg\to t\bar{t}h)
d(PS_g)_{soft}\nonumber\\
&=& d(PS_3)(gg\to t\bar{t}h)
\frac{d^{(d-1)}k}{(2\pi)^{(d-1)}2E_g} 
\theta\left(\delta_s\frac{\sqrt{s}}{2}-E_g\right)
\nonumber\,\,\,,\\
\end{eqnarray}
where $d(PS_4)$ and $d(PS_3)$ have been defined in
Section~\ref{sec:framework}, while $d(PS_g)_{soft}$ denotes the
integration over the phase space of the soft gluon.  Since the
contribution of the soft gluon is now completely factorized, we can
perform the integration over $d(PS_g)_{soft}$ analytically, using
dimensional regularization in $d\!=\!4-2\epsilon$ to extract the soft
poles that will have to cancel the corresponding singularities in
Eqs.~(\ref{eq:sigma_virt_ir_poles}) and (\ref{eq:m2_virt_ir_poles}).
The integrals that we have used to perform the integration over the
phase space of the soft gluon are collected in
Appendix~\ref{sec:app_soft_integrals}.

After squaring the soft amplitude ${\cal A}_{soft}$, summing over the
polarization of the radiated soft gluon, and integrating over the soft
gluon momentum, the pole part of the parton level soft cross section
reads
\begin{eqnarray}
\label{eq:soft_ir_poles}
(\hat\sigma^{gg}_{soft})_{pole}&=&
\int d(PS_3)
\left(\int d(PS_g)_{soft}\overline{\sum}
|{\cal A}_{soft}(gg\to t\bar{t}h+g)|^2\right)_{pole}
\nonumber\\ 
&=&\int d(PS_3) \frac{\alpha_s}{2\pi}{\cal N}_t \overline{\sum} 
\left(C_1 {\cal M}_{S,\epsilon}^{(1)}+C_2 {\cal M}_{S,\epsilon}^{(2)}+
C_3 {\cal M}_{S,\epsilon}^{(3)}\right)\,\,\,,
\end{eqnarray}
where $C_1$, $C_2$, and $C_3$ are defined in
Eq.~(\ref{eq:color_factors}), while ${\cal M}_{S,\epsilon}^{(1)}$,
${\cal M}_{S,\epsilon}^{(2)}$, and ${\cal M}_{S,\epsilon}^{(3)}$
represent the IR pole parts of the corresponding matrix elements
squared, and can be written as:
\begin{eqnarray}
\label{eq:soft_a2_poles}
{\cal M}_{S,\epsilon}^{(1)}&=&-{\cal M}_{V,\epsilon}^{(1)}-
\frac{2}{\epsilon}(1+4 \ln(\delta_s))
\left(|{\cal A}_0^{nab}|^2+|{\cal A}_0^{ab}|^2\right)\,\,\,,\nonumber\\
{\cal M}_{S,\epsilon}^{(2)}&=&-{\cal M}_{V,\epsilon}^{(2)}-
\frac{16}{\epsilon} \ln(\delta_s) |{\cal A}_0^{ab}|^2
+\frac{2}{\epsilon} \left(|{\cal A}_0^{nab}|^2+|{\cal A}_0^{ab}|^2\right) 
\,\,\,,\nonumber\\
{\cal M}_{S,\epsilon}^{(3)}&=&-{\cal M}_{V,\epsilon}^{(3)}+
\frac{1}{\epsilon}|{\cal A}_0^{ab}|^2~.
\end{eqnarray}
Note that in this section we do not explicitly denote the IR poles as
poles in $\epsilon_{\sss IR}$, since all singularities present in
$\sigma^{gg,qg}_{real}$ are of IR origin.

After adding the IR divergent part of the parton level virtual cross
section of Eq.~(\ref{eq:sigma_virt_ir_poles}) we obtain:
\begin{eqnarray}\label{eq:irsv}
\hat\sigma^{gg}_{s+v}\equiv(\hat\sigma^{gg}_{soft})_{pole}+
(\hat\sigma^{gg}_{virt})_{IR-pole}&=&
\frac{\alpha_s}{2\pi}{\cal N}_t 
\left[-4N\ln(\delta_s)-\frac{1}{3}(11N-2n_{lf})\right] 
\frac{1}{\epsilon} \hat\sigma^{gg}_{\sss LO}\,\,\,,\nonumber\\
\end{eqnarray}
where we can see that the IR poles of the parton level virtual cross
section are exactly canceled by the corresponding singularities in the
parton level soft gluon emission cross section. The residual
divergences will be canceled by the soft+virtual part of the PDF
counterterm when convoluting with the gluon PDFs as will be
demonstrated in Section~\ref{sec:total}.  The finite contribution to
the parton level soft cross section is finally given by
\begin{eqnarray}
\label{eq:soft_ir_finite}
(\hat \sigma^{gg}_{soft})_{finite}&=&
\int d(PS_3) \left(\int d(PS_g)_{soft}\overline{\sum}
|{\cal A}_{soft}(gg\to t\bar{t}h+g)|^2\right)_{finite}
\nonumber\\ 
&=&\int d(PS_3) \frac{\alpha_s}{2\pi}{\cal N}_t\overline{\sum}
\left(C_1{\cal M}_S^{(1)}+C_2{\cal M}_S^{(2)}+
C_3{\cal M}_S^{(3)}\right)\,\,\,,
\end{eqnarray}
where the finite parts of the ${\cal M}_S^{(1)}$, ${\cal M}_S^{(2)}$,
and ${\cal M}_S^{(3)}$ matrix element squared are explicitly given
by:
\begin{eqnarray}
\label{eq:soft_a2_finite}
{\cal M}_S^{(1)}&=&\left[
-\frac{4}{3}\pi^2
+4\Lambda_\sigma\ln(\delta_s)+8\ln^2(\delta_s)
-2\Lambda_\sigma-4\ln(\delta_s)+\frac{2}{\beta_{t\bar{t}}} 
\Lambda_{t\bar{t}}\right]\times\nonumber\\
&&\left(|{\cal A}_0^{nab}|^2+|{\cal A}_0^{ab}|^2\right)
\nonumber\\
&+&\left[
\left(\Lambda_\sigma+2\ln(\delta_s)\right)
\left(\Lambda_{\tau_1}+\Lambda_{\tau_2}\right)
+\frac{1}{2}F(q_1,p_t)+\frac{1}{2}F(q_2,p_t^\prime)\right]
|{\cal A}_0^{nab}+{\cal A}_0^{ab}|^2
\nonumber\\
&+&\left[
\left(\Lambda_\sigma+2\ln(\delta_s)\right)
\left(\Lambda_{\tau_3}+\Lambda_{\tau_4}\right)
+\frac{1}{2}F(q_2,p_t)+\frac{1}{2}F(q_1,p_t^\prime)\right]
|{\cal A}_0^{nab}-{\cal A}_0^{ab}|^2\,\,\,,
\nonumber\\
{\cal M}_S^{(2)}&=&\left\{
\frac{\bar s_{t\bar t}-2m_t^2}{\bar s_{t\bar{t}}}\left[
\left(2\Lambda_\sigma+4\ln(\delta_s)\right)\frac{1}{\beta_{t\bar{t}}} 
\Lambda_{t\bar{t}}
-\frac{1}{\beta_{t\bar{t}}}\Lambda_{t\bar{t}}^2-\frac{4}{\beta_{t\bar{t}}}
\mbox{Li}_2\left(\frac{2\beta_{t\bar{t}}}{1+\beta_{t\bar{t}}}\right)\right]
\right.\nonumber\\
&&\left.\phantom{\frac{1}{2}}-2\Lambda_\sigma-4\ln(\delta_s)+
\frac{2}{\beta_{t\bar{t}}}\Lambda_{t\bar{t}}\right\}
\left(|{\cal A}_0^{nab}|^2+|{\cal A}_0^{ab}|^2\right)\nonumber\\
&+&2\left[-\frac{4}{3}\pi^2-2\Lambda_\sigma^2+8\ln^2(\delta_s)
+2\left(\Lambda_\sigma+2\ln(\delta_s)\right)
\left(\Lambda_{\tau_1}+\Lambda_{\tau_2}+\Lambda_{\tau_3}+\Lambda_{\tau_4}\right)
\right.\nonumber\\
&&\quad\,
+F(q_1,p_t)+F(q_2,p^\prime_t)+F(q_2,p_t)+F(q_1,p^\prime_t)\nonumber\\
&&\left.\phantom{\frac{1}{2}}-4\Lambda_\sigma-8\ln(\delta_s)+
\frac{4}{\beta_{t\bar{t}}}\Lambda_{t\bar{t}}\right]|{\cal A}_0^{ab}|^2\,\,\,,
\nonumber\\
{\cal M}_S^{(3)}&=&\frac{1}{2}\left\{
\frac{\bar s_{t\bar t}-2m_t^2}{\bar s_{t\bar t}}\left[
\left(2\Lambda_\sigma+4\ln(\delta_s)\right)\frac{1}{\beta_{t\bar{t}}} 
\Lambda_{t\bar{t}}-
\frac{1}{\beta_{t\bar{t}}} \Lambda_{t\bar{t}}^2-\frac{4}{\beta_{t\bar{t}}}
\mbox{Li}_2\left(\frac{2\beta_{t\bar{t}}}{1+\beta_{t\bar{t}}}\right)
\right]
\right.\nonumber\\
&&\left.\phantom{\frac{1}{2}}
-2\Lambda_\sigma-4\ln(\delta_s)+\frac{2}{\beta_{t\bar{t}}}\Lambda_{t\bar{t}}
\right\}|{\cal A}_0^{ab}|^2\,\,\,.
\end{eqnarray}
We note that $\bar{s}_{t\bar{t}}$, $\beta_{t\bar{t}}$, and
$\Lambda_{t\bar{t}}$ are defined in Eq.~(\ref{eq:stt_betatt}), while
the function $F(p_i,p_f)$ can be found in
Appendix~\ref{sec:app_soft_integrals} (Eq.~(\ref{eq:f_if})).

\subsubsection{Real gluon emission, $gg\to t\bar{t}h+g$: hard region}
\label{subsubsec:two_cutoff_hard}

The hard region of the final state gluon phase space is defined by
requiring that the energy of the emitted gluon is above a given
threshold.  As we discussed earlier, this is expressed by the
condition that
\begin{equation}
E_g >\delta_s \frac{\sqrt{s}}{2}\,\,\,,
\end{equation}
for an arbitrary small \emph{soft} cutoff $\delta_s$, which
automatically assures that $\hat{\sigma}^{gg}_{hard}$ does not contain
soft singularities.  However, a hard gluon can still give rise to
singularities when it is emitted at a small angle, i.e.
\emph{collinear}, to a massless incoming or outgoing parton. In order
to isolate these divergences and compute them analytically, we further
divide the hard region of the $gg\to t\bar{t}h+g$ phase space into a
\emph{hard/collinear} and a \emph{hard/non-collinear} region, by
introducing a second small \emph{collinear} cutoff $\delta_c$.  The
\emph{hard/non-collinear} region is defined by the condition that both
\begin{equation}
\label{eq:deltac_cuts}
\frac{2 q_1\!\cdot\! k}{E_g \sqrt{s}}> \delta_c\,\,\,\,\,\,\,
\mbox{and}\,\,\,\,\,\,\,
\frac{2 q_2\!\cdot\! k}{E_g\sqrt{s}}>\delta_c
\end{equation}
are true. The contribution from the \emph{hard/non-collinear} region,
$\hat{\sigma}^{gg}_{hard/non-coll}$, is finite and we compute it
numerically by using standard Monte Carlo integration techniques.

In the $\emph{hard/collinear}$ region, one of the conditions in
Eq.~(\ref{eq:deltac_cuts}) is not satisfied and the hard gluon is
emitted collinear to one of the incoming partons.  In this region, the
initial state parton $i$ with momentum $q_i$ is considered to split
into a hard parton $i^\prime$ and a collinear gluon $g$, $i\to
i^\prime g$, with $q_{i^\prime}\!=\!z q_i$ and $k\!=\!(1-z)q_i$.  The
matrix element squared for $ij\to t\bar{t}h+g$ factorizes into the
Born matrix element squared and the unregulated Altarelli-Parisi
splitting function $P_{ii'}= P_{ii'}^4+\epsilon P_{ii'}^\prime$ for
$i\to i^\prime g$, i.e.:
\begin{equation}
\label{eq:m2_coll_lim}
\overline{\sum}|{\cal A}_{real}(ij\to t\bar{t}h+g)|^2 
\stackrel{collinear}{\longrightarrow}
(4\pi\alpha_s)\overline{\sum}|{\cal A}_0(i^\prime j\to t\bar{t}h)|^2
\frac{2P_{ii^\prime}(z)}{z \, s_{ig}} \,\,\,,
\end{equation}
where $P_{ii^\prime}^4$ and $P_{ii^\prime}^\prime$ denote the
coefficients of the ${\cal O}(1)$ and ${\cal O}(\epsilon)$ parts of
$P_{ii^\prime}$, while $s_{ig}\!=\!2q_i\!\cdot\!k$. In the case of
$gg\to t\bar{t}h+g$ the initial partons are gluons and the unregulated
splitting function in $d$ dimensions is ($P_{gg}^\prime\!=\!0$):
\begin{equation}
\label{eq:apgg}
P_{ii^\prime}(z)=P_{gg}(z)= 
2 N\left(\frac{z}{1-z} +\frac{1-z}{z}+z(1-z)\right)\,\,\,.
\end{equation}
Moreover, in the collinear limit, the $ij\rightarrow t\bar th+g$ phase
space also factorizes as:
\begin{eqnarray}
\label{eq:ps_coll_lim}
d(PS_4)(ij\to t\bar{t}h+g)
&\stackrel{collinear}{\longrightarrow}& 
d(PS_3)(i^\prime j\to t\bar{t}h)
\frac{z\,d^{(d-1)}k}{(2\pi)^{(d-1)}2E_g}
\theta\left(E_g-\delta_s\frac{\sqrt{s}}{2}\right)\times\nonumber \\
&&\theta\left(\cos\theta_{ig}-(1-\delta_c)\right)\nonumber\\
&&\!\!\!\!\!\!\!\!\!\!\!\!\!\!\!\!\!\!\!\!\!\! \stackrel{d=4-2 \epsilon}{=}
d(PS_3)(i^\prime j\to t\bar{t}h)
\frac{1}{\Gamma(1-\epsilon)}
\frac{\left(4\pi\right)^\epsilon}{16 \pi^2}\,z\,dz\,ds_{ig} 
\left[(1-z) s_{ig}\right]^{-\epsilon}\times\nonumber\\
&&
\theta\left(\frac{(1-z)}{z}s^\prime\frac{\delta_c}{2}-s_{ig}\right)
\nonumber\,\,\,,\\
\end{eqnarray}
where $d(PS_4)$ and $d(PS_3)$ have been defined in
Section~\ref{sec:framework}, while the integration range for $s_{ig}$
in the collinear region is given in terms of the collinear cutoff, and
we have defined $s^\prime\!=\!2q_{i^\prime}\!\cdot\!q_j$.  The
integral over the collinear gluon degrees of freedom can then be
performed analytically, and this allows us to explicitly extract the
collinear singularities of
$\hat{\sigma}^{gg}_{hard}$ \cite{Harris:2001sx,Baur:1998kt} as:
\begin{eqnarray}
\label{eq:coll_pole}
\hat{\sigma}^{gg}_{hard/coll}&=&
\left[\frac{\alpha_s}{2\pi}\frac{1}{\Gamma(1-\epsilon)}
\left(\frac{4\pi\mu^2}{m_t^2}\right)^\epsilon\right]
\left(-\frac{1}{\epsilon}\right)\delta_c^{-\epsilon}\times\nonumber\\
&&\left\{
\int_{0}^{1-\delta_s} dz
\left[\frac{(1-z)^2}{2 z} \frac{s^\prime}{m_t^2}\right]^{-\epsilon} 
P_{ii^\prime}(z) \,
\hat{\sigma}_{\sss LO}^{gg}(i^\prime j\rightarrow t\bar t h)
+ (i\leftrightarrow j)\right\}\,\,\,,
\end{eqnarray}
where $i,i^\prime$, and $j$ are all gluons.  As usual, these initial
state collinear divergences are absorbed into the gluon distribution
functions as will be described in detail in Section~\ref{sec:total}.

\subsubsection{The tree level processes $(q,\bar{q})g\to t\bar{t}h+(q,\bar{q})$}
\label{subsubsec:two_cutoff_qg}

The extraction of the collinear singularities of $\hat
\sigma^{qg}_{real}$ is done in the same way as described in
Section~\ref{subsubsec:two_cutoff_hard} for the $gg$ initial state.
In the collinear region, $\cos\theta_{iq}>1-\delta_c$, the initial
state parton $i$ with momentum $q_i$ is considered to split into a
hard parton $i^\prime$ and a collinear quark $q$, $i\rightarrow
i^\prime q$, with $q_{i^\prime}\!=\!z q_i$ and $k\!=\!(1-z)q_i$. The
matrix element squared for $ij\to t\bar{t}h+q$ factorizes into the
unregulated Altarelli-Parisi splitting functions in $d$ dimensions:
$P_{ii'}= P_{ii'}^4+\epsilon P_{ii'}^\prime$ and the corresponding
Born matrix elements squared.  The $ij\to t\bar{t}h+q$ phase space
factorizes into the $i'j\to t\bar{t}h$ phase space and the phase space
of the collinear quark. As a result, after integrating over the phase
space of the collinear quark, the collinear singularity of $\hat
\sigma^{qg}_{real}$ can be extracted as:
\begin{eqnarray}
\label{eq:qgcoll_pole}
\hat{\sigma}^{qg}_{coll}&=&
\left[\frac{\alpha_s}{2\pi}\frac{1}{\Gamma(1-\epsilon)}
\left(\frac{4\pi\mu^2}{m_t^2}\right)^\epsilon\right]
\left(-\frac{1}{\epsilon}\right)\delta_c^{-\epsilon}
\int_{0}^{1} dz
\left[\frac{(1-z)^2}{2 z} \frac{s^\prime}{m_t^2}\right]^{-\epsilon} 
\times\nonumber\\
&&
\left[ P_{qg}(z) \,
\hat{\sigma}_{\sss LO}^{gg}(g(q_{1^\prime})g(q_2)\to t\bar{t}h)
+ P_{gq}(z) \,
\hat{\sigma}_{\sss LO}^{q\bar{q}}(q(q_1)\bar{q}(q_{2^\prime})\to t\bar{t}h)
\right]\,\,\,. 
\end{eqnarray}
The collinear radiation of an antiquark in $\bar{q}g\to
t\bar{t}h+\bar{q}$ is treated analogously. In the case of
$(q,\bar{q})g\to t\bar{t}h+(q,\bar{q})$ we have two possible
splittings: $(q,\bar{q})\rightarrow g(q,\bar{q})$ and $g\to q\bar{q}$.
The $O(1)$ and $O(\epsilon)$ parts of the corresponding splitting
functions are:
\begin{eqnarray}
\label{eq:apqg}
P_{gq}^4(z) &=& \frac{1}{2}\left(z^2+(1-z)^2\right)\,\,\,, 
\nonumber\\
P^\prime_{gq}(z) &=& -z(1-z)\,\,\,,\nonumber\\
P_{qg}^4(z) &=&
\frac{N^2-1}{2N}\left(\frac{1+(1-z)^2}{z}\right)\,\,\,, 
\nonumber\\
P^\prime_{qg}(z) &=& -\frac{N^2-1}{2N} z\,\,\,.
\end{eqnarray}
Again, these initial state collinear divergences are absorbed into the
parton distribution functions as will be described in detail in
Section~\ref{sec:total}.
\subsection{Phase Space Slicing method with one cutoff}
\label{subsec:one_cutoff}

An alternative way of isolating both soft and collinear singularities
is to divide the phase space for the radiated parton into only two
regions, according to whether all partons can be resolved (the
\emph{hard} region) or not (the \emph{infrared} or \emph{ir} region).
In the case of $gg\to t\bar{t}h+g$, the \emph{hard} and \emph{ir}
regions are defined according to whether the final state gluon can be
resolved. The emitted gluon is considered as not resolved, and
therefore part of the \emph{ir} cross section, when
\begin{equation}
\label{eq:smin_conditiong}
{s}_{ig}=2p_i\cdot k<s_{min}\,\,\,\,\,\,\,(p_i\!=\!q_1,q_2,p_t,p_t^\prime)
\,\,\,,
\end{equation} 
for an arbitrary small cutoff $s_{min}$, with $k$ the final state
gluon momentum which becomes soft or collinear.  In the case of
$(q,\bar{q})g\to t\bar{t}h+(q,\bar{q})$, the emitted (anti)quark is
considered as not resolved, and therefore part of the \emph{ir} cross
section, when
\begin{equation}
\label{eq:smin_conditionq}
{s}_{iq}=2p_i\cdot k<s_{min}\,\,\,\,\,\,\,(p_i\!=\!q_1,q_2,p_t,p_t^\prime)
\,\,\,,
\end{equation} 
with $k$ the final state (anti)quark momentum which becomes collinear.
The partonic real cross sections can then be written as the sum of two
terms ($ij=gg,qg$):
\begin{equation}
\label{eq:sigma_real_one_cutoff}
\hat\sigma_{real}^{ij}=\hat\sigma_{ir}^{ij}+\hat\sigma_{hard}^{ij}\,\,\,,
\end{equation}
where $\hat\sigma_{ir}^{ij}$ includes the IR singularities, both soft
and collinear, while $\hat\sigma_{hard}^{ij}$ is finite. Following the
general idea of PSS, we calculate $\hat\sigma_{ir}^{ij}$ analytically,
while we evaluate $\hat\sigma_{hard}^{ij}$ numerically, using standard
Monte Carlo integration techniques. Both $\hat\sigma_{ir}^{ij}$ and
$\hat\sigma_{hard}^{ij}$ depend on the cutoff $s_{min}$, but the
hadronic real cross section, $\sigma_{real}$, is cutoff independent
after mass factorization, as will be shown in Section~\ref{sec:total}.

In order to calculate $\hat\sigma_{ir}^{gg}$ we apply the formalism
developed in
Refs.~\cite{Giele:1992vf,Giele:1993dj,Keller:1998tf,Reina:2001bc} as
follows.
\begin{itemize}
\item[(a)] We consider the crossed process 
\begin{equation} 
\label{eq:h_ggttbg}
h(p_h)\rightarrow t(p_t)+\bar{t}(p_t^\prime)+g^A(q_1)+g^B(q_2)+g^C(k)\;,
\end{equation}
obtained from $gg\to t\bar{t}h+g$ by crossing all the initial state
colored partons to the final state, while crossing the Higgs boson to
the initial state. All colored partons are hence considered as final
state partons. For a systematic extraction of the IR singularities
within the one-cutoff method, we organize the amplitude for $h\to
ggt\bar{t}+g$, ${\cal A}^{h\to ggt\bar{t}+g}_{real}$, in terms of six
colored ordered amplitudes, ${\cal A}_{ijk}$, which are the
coefficients of all possible permutations of the color matrices
$T^A,T^B,T^C$, i.e.:
\begin{equation}
\label{eq:color_ordered_amp}
{\cal A}^{h\to ggt\bar{t}+g}_{real} = \sum_{i,j,k=A,B,C \atop i\ne j\ne k} 
{\cal A}_{ijk}\,T^i\,T^j\,T^k\; .
\end{equation}
The explicit color ordered amplitudes have very lengthy expressions
and we do not give them in this paper. Since they are tree level
amplitudes, they can be easily obtained. In the following we will
however discuss in detail their properties in both the soft and
collinear regions of the phase space of the extra emitted gluon. In
this respect, we note that decomposing ${\cal A}^{h\to
  ggt\bar{t}+g}_{real}$ in terms of color ordered amplitudes ${\cal
  A}_{ijk}$ allows us to write the partonic real cross section as:
\begin{equation}
\label{eq:sigmareal_crossed}
\hat{\sigma}^{h\to ggt\bar{t}+g}_{real}=
\int d(PS_5)\overline{\sum}|{\cal A}^{h\to ggt\bar{t}+g}_{real}|^2\,\,\,,
\end{equation}
where
\begin{eqnarray}
\label{eq:a2_real_1c}
&&\overline{\sum}|{\cal A}^{h\to ggt\bar{t}+g}_{real}|^2=
\frac{1}{2}\left[\,\,C_1\sum_{i,j,k=A,B,C \atop i\ne j\ne k} 
|{\cal A}_{ijk}|^2\right.\nonumber\\
&&+C_2\bigg(
 |{\cal A}_{\sss ABC}+{\cal A}_{\sss ACB}+{\cal A}_{\sss CAB}|^2
+|{\cal A}_{\sss CBA}+{\cal A}_{\sss BAC}+{\cal A}_{\sss BCA}|^2
+|{\cal A}_{\sss CAB}+{\cal A}_{\sss CBA}+{\cal A}_{\sss BCA}|^2+
\nonumber\\
&&\quad\quad\,\,\,\,\,
 |{\cal A}_{\sss ABC}+{\cal A}_{\sss BAC}+{\cal A}_{\sss ACB}|^2
+|{\cal A}_{\sss CAB}+{\cal A}_{\sss CBA}+{\cal A}_{\sss ACB}|^2
+|{\cal A}_{\sss ABC}+{\cal A}_{\sss BAC}+{\cal A}_{\sss BCA}|^2
\bigg)\nonumber\\
&&+\left.
C_3\,\frac{1}{4}\,\bigg|\sum_{i,j,k=A,B,C \atop i\ne j\ne k} 
{\cal A}_{ijk}\bigg|^2 
\right]\,\,\,
\end{eqnarray}
is the full real amplitude squared, including both leading and
subleading color factors (see Eq.~(\ref{eq:color_factors}) for a
definition of $C_1$, $C_2$, and $C_3$). Each sub-amplitude squared in
Eq.~(\ref{eq:a2_real_1c}) has very definite factorization properties
in the soft or collinear regions of the phase space of the extra
emitted gluon.
\item[(b)] Using the one-cutoff PSS method and the factorization
  properties of soft and collinear divergences of the various
  amplitudes squared in Eq.~(\ref{eq:a2_real_1c}), we extract the IR
  singularities from $\hat\sigma^{h\to ggt\bar{t}+g}_{real}$ in
  $d\!=\!4-2\epsilon$ dimensions. In the soft and collinear limits we
  obtain:
\begin{eqnarray}
\label{eq:sigsoft_crossed}
\hat \sigma^{h\to ggt\bar t+g}_{real}& \stackrel{soft}{\longrightarrow}& 
\hat \sigma_{soft}^{h\to ggt\bar{t}+g}
=\int d(PS_4) d(PS_g)_{soft} \overline{\sum}
|{\cal A}_{soft}^{h\to ggt\bar{t}+g}|^2
\,\,\,,\\
\label{eq:sigcoll_crossed}
\hat \sigma^{h\to ggt\bar{t}+g}_{real}
&\stackrel{collinear}{\longrightarrow}& 
\hat \sigma_{coll}^{h\to ggt\bar{t}+g}
=\int d(PS_4) d(PS_g)_{coll} \overline{\sum}
|{\cal A}_{coll}^{h\to ggt\bar{t}+g}|^2\,\,\,,
\end{eqnarray}
where we denote by $d(PS_g)_{soft}$ ($d(PS_g)_{coll}$) the phase space
of the gluon $g^C$ in the soft (collinear) limit.  In both the soft
and the collinear limits, the cross section for $h\to ggt\bar{t}+g$
integrated over the phase space of the IR singular gluon has the form:
\begin{eqnarray} 
\label{eq:sigma_soft_coll_colamp}
\hat \sigma^{h\to ggt\bar{t}+g}_{soft,\,coll}
&=& \int d(PS_4) \frac{1}{N}\, \overline{\sum} \nonumber \\
&&\left\{ C_1 \left[K_{S,C}(t;1,2;\bar{t})\,
|{\cal A}_{0,s}^{(c)}+{\cal A}_{0,t}^{(c)}|^2+
K_{S,C}(t;2,1;\bar{t})\,
|{\cal A}_{0,u}^{(c)}-{\cal A}_{0,s}^{(c)}|^2\right]\right.\nonumber\\ 
&&+C_2 \left[\,\,2\,K_{S,C}(t;\bar{t})
\left(|{\cal A}_{0}^{ab,(c)}|^2+|{\cal A}_{0}^{nab,(c)}|^2\right)+
\right.\nonumber\\
&&\quad\quad\quad\left.
4\,\left(K_{S,C}(t;1;\bar{t})+K_{S,C}(t;2;\bar{t})\right)
|{\cal A}_{0}^{ab,(c)}|^2\right]\nonumber\\
&&+ \left. C_3\,K_{S,C}(t;\bar{t})\,|{\cal A}_{0}^{ab,(c)}|^2 \right\}\; ,
\end{eqnarray}
where the tree level amplitudes for the crossed process $h\to
ggt\bar{t}$, denoted by ${\cal A}_{0,s}^{(c)}$, ${\cal
  A}_{0,t}^{(c)}$, and ${\cal A}_{0,u}^{(c)}$, as well as their linear
combinations ${\cal A}_{0}^{ab,(c)}$ and ${\cal A}_{0}^{nab,(c)}$, can
be obtained from the corresponding amplitudes for $gg\to t\bar{t}h$
given explicitly in Appendix~\ref{sec:app_tree_level} by flipping
momenta and helicities of the crossed particles.  The functions $K$
are either evaluated in the soft ($K_S$) or in the collinear limit
($K_C$), and will be explicitly given in Eqs.~(\ref{eq:k_soft}) and
(\ref{eq:k_coll}). Moreover, we notice that the arguments of the
$K_{S,C}$ functions are indices $i\!=1,2,t,\bar{t}$ denoting the
external partons $g^A(q_1)$, $g^B(q_2)$, $t(p_t)$, and
$\bar{t}(p_t^\prime)$ respectively. The explicit forms of both the
pole and finite parts of $\hat{\sigma}^{h\to ggt\bar{t}+g}_{soft}$ and
$\hat{\sigma}^{h\to ggt\bar{t}+g}_{coll}$ are given in
Sections~\ref{subsubsec:one_cutoff_soft} and
\ref{subsubsec:one_cutoff_collinear} respectively.
\item[(c)]
Finally, the IR singular contribution $\hat\sigma_{ir}^{gg}$ of
Eq.~(\ref{eq:sigma_real_one_cutoff}) consists of two terms:
\begin{equation}\label{eq:sigma_ir_gg}
\hat\sigma_{ir}^{gg}=\hat{\bar{\sigma}}_{ir}^{gg}+
\hat\sigma_{crossing}^{gg}\,\,\,.
\end{equation}
As described in Section~\ref{subsubsec:one_cutoff_ir} ,
$\hat{\bar{\sigma}}^{gg}_{ir}$ is obtained by crossing $g^A, g^B$ to
the initial state and $h$ to the final state in the sum of $\hat
\sigma_{soft}^{h\to ggt\bar{t}+g}$ and $\hat \sigma_{coll}^{h\to
  ggt\bar{t}+g}$, while $\hat\sigma_{crossing}^{gg}$ corrects for the
difference between the collinear gluon radiation from initial and
final state partons \cite{Giele:1993dj,Reina:2001bc}.  The IR
singularities of $\hat \sigma_{virt}^{gg}$ of
Section~\ref{subsec:virtual_ir} are exactly canceled by the
corresponding singularities in $\hat{\bar{\sigma}}_{ir}^{gg}$. On the
other hand, $\hat\sigma_{crossing}^{gg}$ still contains collinear
divergences that will be canceled by the PDF counterterms when the
parton cross section is convoluted with the gluon PDFs, as we will
show in Section~\ref{sec:total}.
\end{itemize}
When calculating the cross section for $qg\to t\bar{t}h+q$ in the
collinear limit using the procedure described above, the resulting IR
singular cross section $\hat\sigma_{ir}^{qg}$ is simply given by the
initial state $qg$ splitting functions of Eq.~(\ref{eq:apqg})
convoluted with the corresponding Born cross sections (see, e.g.
Ref.~\cite{Giele:1993dj})
\begin{eqnarray}  
\label{eq:sigma_ir_qg}
\hat\sigma_{ir}^{qg}&=& 
\left[\frac{\alpha_s}{2\pi}\frac{1}{\Gamma(1-\epsilon)}
\left(\frac{4\pi\mu^2}{s_{min}}\right)^\epsilon\right]
\left(-\frac{1}{\epsilon}\right) \int_0^1 dz(1-z)^{-\epsilon} 
\times\nonumber\\
&& 
\left[ P_{qg}(z) \,
\hat{\sigma}_{\sss LO}^{gg}(g(q_{1^\prime})g(q_2)\to t\bar{t}h)
+ P_{gq}(z) \,
\hat{\sigma}_{\sss LO}^{q\bar{q}}(q(q_1)\bar{q}(q_{2^\prime})\to t\bar{t}h)
\right]\,\,\,, 
\end{eqnarray}
where the prime identifies the parton that originates from the
splitting of a similar or different parent parton.  The cross section
for $\bar{q}g\to t\bar{t}h+\bar{q}$ in the collinear limit is obtained
in complete analogy with Eq.~(\ref{eq:sigma_ir_qg}).

Finally, the hard part of the parton level cross section,
$\hat\sigma^{ij}_{hard}$ ($ij\!=\!gg,qg,\bar{q}g$), is finite and can
be calculated numerically. In this respect we note that, in the one
cutoff method, the soft and collinear limits of the real cross
section, and consequently $\hat\sigma^{ij}_{hard}$, are more sensitive
to the smallness of the cutoff. A cut on the full invariant masses
$s_{ig}$ is more drastic than two separate cuts on either the energy
or the angle of emission of the extra gluon ($q$ or $\bar{q}$), and
can be felt even by terms in the amplitude squared that do not contain
singularities.  These effects are very small, but large enough to
affect the results at the level of precision of our calculation. It is
therefore crucial, in particular for $\hat\sigma^{gg}_{hard}$, to
model the Monte Carlo integration for each term in
Eqs.~(\ref{eq:sigmareal_crossed})-(\ref{eq:a2_real_1c}) separately,
and to enforce term by term only the cuts on the $s_{ig}$ invariants
that are actually present in each term.
\subsubsection{Real gluon emission $h\to ggt\bar{t}+g$: soft region}
\label{subsubsec:one_cutoff_soft}

We first consider the case of soft singularities, when, in the limit
of $E_g\!\to\!0$ (soft limit), one or more ${s}_{ig}\!<\!s_{min}$.  In
the soft limit, the $h\to ggt\bar{t}+g$ phase space, as well as the
full parton level real amplitude squared, factorize the dependence on
the degrees of freedom of the soft emitted gluon, as illustrated in
Eq.~(\ref{eq:sigsoft_crossed}). The soft part of the parton level
cross section can be calculated analytically according to
Eq.~(\ref{eq:sigma_soft_coll_colamp}). The soft limit of the $K$
functions, $K_S$, is explicitly given by:
\begin{eqnarray}
\label{eq:k_soft}
K_S(t;i,j;\bar{t})&=& \frac{N g_s^2}{2}\int d(PS_g)_{soft}
  [f_{ti}(g)+f_{ij}(g)+f_{j\bar{t}}(g)]=S_{ti}+S_{ij}+S_{j\bar{t}}
\,\,\,,\nonumber \\
K_S(t;i;\bar{t})&=& \frac{N g_s^2}{2}\int d(PS_g)_{soft}  
[f_{ti}(g)+f_{i\bar{t}}(g)]= S_{ti}+S_{i\bar{t}}\,\,\,,\nonumber \\
K_S(t;\bar{t})&=& \frac{N g_s^2}{2}\int d(PS_g)_{soft}
  f_{t\bar{t}}(g)= S_{t\bar{t}}\,\,\,,
\end{eqnarray} 
where $i,j\!=\!1,2$ denote the two external hard gluons with momenta
$q_1$ and $q_2$.  For any pair of partons $(a,b)$ excluding the soft
gluon, the soft $f_{ab}(g)$ functions introduced in
Eq.~(\ref{eq:k_soft}) are given by:
\begin{equation}
\label{eq:fab}
f_{ab}(g)\equiv\frac{4s_{ab}}{s_{ag}s_{bg}} 
-\frac{4m_a^2}{s_{ag}^2}-\frac{4m_b^2}{s_{bg}^2}\,\,\,,
\end{equation}
with ${s}_{ij}\equiv 2 p_i\cdot p_j$ both for massless and massive
partons, and the corresponding integrated soft functions $S_{ab}$ are
consequently defined to be:
\begin{equation}
\label{eq:sab}
S_{ab}= \frac{g_s^2N}{2}\int d(PS_g)_{soft}(a,b,g) f_{ab}(g) \,\,\,.
\end{equation}
In the one-cutoff PSS method, the explicit form of the soft gluon phase
space is given by \cite{Keller:1998tf}:
\begin{eqnarray}
\label{eq:ps_soft}
d(PS_g)_{soft}(a,b,g)&=&\frac{(4\pi)^\epsilon}{16\pi^2}
\frac{\lambda^{\left(\epsilon-\frac{1}{2}\right)}}{\Gamma(1-\epsilon)}
\left[{s}_{ag}{s}_{bg}{s}_{ab}
-m_b^2{s}_{ag}^2-m_a^2{s}_{bg}^2\right]^{-\epsilon}
d{s}_{ag}d{s}_{bg}\times\nonumber\\
&&\quad\quad 
\theta(s_{min}-{s}_{ag})\theta(s_{min}-{s}_{bg})\,\,\,,
\end{eqnarray}
with $\lambda={s}_{ab}^2-4m_a^2m_b^2$, while the integration
boundaries for ${s}_{ag}$ and ${s}_{bg}$ vary according to whether $a$
and $b$ are massive or massless quarks.

The explicit form of the integrated soft functions $S_{ab}$ is
obtained by carrying out the integration in Eq.~(\ref{eq:sab}) and
analytic expressions for the $S_{ab}$ are collected in
Appendix~\ref{sec:app_sab}. Finally, using
Eq.~(\ref{eq:sigma_soft_coll_colamp}),
Eqs.~(\ref{eq:k_soft})-(\ref{eq:ps_soft}), and the results in
Appendix~\ref{sec:app_sab}, the pole part of the parton level soft
cross section can be written as:
\begin{eqnarray}
\label{eq:sigma_soft_1c_pole}
&&(\hat\sigma_{soft}^{h\to ggt\bar t+g})_{pole}=
\int d(PS_4) \frac{\alpha_s}{2\pi}{\cal N}_t \times 
\nonumber \\
&\overline{\sum}& \left\{
C_1\left[\frac{4}{\epsilon^2}+\frac{2}{\epsilon}+
\frac{2}{\epsilon}\Lambda_\sigma-\frac{8}{\epsilon}\Lambda_{min}\right]
\left(|{\cal A}_0^{ab,(c)}|^2+|{\cal A}_0^{nab,(c)}|^2\right)\right.
\nonumber\\
&& \left.+2C_2\left[\frac{1}{\epsilon} \left(1
-\frac{\bar s_{t\bar{t}}-2m_t^2}{\bar s_{t\bar{t}}\beta_{t\bar{t}}}
\Lambda_{t\bar t}\right)
\left(|{\cal A}_0^{ab,(c)}|^2+|{\cal A}_0^{nab,(c)}|^2\right)\right.\right. 
\nonumber \\
&& \left.\left.\quad\quad\; +4\left(
\frac{1}{\epsilon^2}+\frac{1}{\epsilon}-\frac{2}{\epsilon}\Lambda_{min}
\right)|{\cal A}_0^{ab,(c)}|^2 \right]+C_3 \frac{1}{\epsilon}\left[1
-\frac{\bar s_{t\bar t}-2m_t^2}{\bar s_{t\bar t}\beta_{t\bar{t}}}
\Lambda_{t\bar t}\right]
|{\cal A}_0^{ab,(c)}|^2\right\}\; ,\nonumber\\
\end{eqnarray}
while the corresponding finite contribution is:
\begin{eqnarray}
\label{eq:sigma_soft_1c_finite}
&&(\hat\sigma_{soft}^{h\to ggt\bar t+g})_{finite}=
\int d(PS_4)\frac{\alpha_s}{2\pi}{\cal N}_t\times 
\nonumber \\
&\overline{\sum}& \left\{ C_1\left[ \left(
\Lambda_\sigma^2-4\Lambda_\sigma\Lambda_{min}-4\Lambda_{min}+
8\Lambda_{min}^2-\pi^2\right) 
\left(|{\cal A}_0^{ab,(c)}|^2+|{\cal A}_0^{nab,(c)}|^2\right)\right.\right.
\nonumber\\
&&\quad+\left. \left.\left(
-\frac{1}{2}\Lambda_{\tau_1}^2-\frac{1}{2}\Lambda_{\tau_2}^2
+\Lambda_{\tau_1}+\Lambda_{\tau_2}
+\frac{m_t^2}{s_{t1}}+\frac{m_t^2}{s_{\bar{t}2}}\right)
|{\cal A}_{0,s}^{(c)}+{\cal A}_{0,t}^{(c)}|^2\right.\right. 
\nonumber\\
&&\quad\left.\left.+\left(
-\frac{1}{2}\Lambda_{\tau_3}^2-\frac{1}{2}\Lambda_{\tau_4}^2
+\Lambda_{\tau_3}+\Lambda_{\tau_4}
+\frac{m_t^2}{s_{t2}}+\frac{m_t^2}{s_{\bar{t}1}}\right)
|{\cal A}_{0,u}^{(c)}-{\cal A}_{0,s}^{(c)}|^2\right]\right. 
\nonumber\\
&+& \left. C_2\left[\left(
-2\Lambda_{min}\left(1-\frac{\bar s_{t\bar{t}}-2m_t^2}
{\bar s_{t\bar{t}}\beta_{t\bar{t}}}\Lambda_{t\bar{t}}\right)
+\frac{2m_t^2}{\sqrt{\lambda_{t\bar{t}}}}\left(J_a+J_b\right)\right)
\left(|{\cal A}_0^{ab,(c)}|^2+|{\cal A}_0^{nab,(c)}|^2\right)\right.\right. 
\nonumber\\
&&\quad \left. \left.+4\left(
-4\Lambda_{min}+4\Lambda_{min}^2-\frac{2}{3}\pi^2
-\frac{1}{2}\Lambda_{\tau_1}^2-\frac{1}{2}\Lambda_{\tau_2}^2
-\frac{1}{2}\Lambda_{\tau_3}^2-\frac{1}{2}\Lambda_{\tau_4}^2 
\right.\right.\right.\nonumber\\
&&\quad\left.\left.\left.+\Lambda_{\tau_1}+\Lambda_{\tau_2}+
\Lambda_{\tau_3}+\Lambda_{\tau_4}
+\frac{m_t^2}{s_{t1}}+\frac{m_t^2}{s_{t2}}
+\frac{m_t^2}{s_{\bar{t}1}}+\frac{m_t^2}{s_{\bar{t}2}}\right)
|{\cal A}_{0}^{ab,(c)}|^2\right]
\right.\nonumber\\
&+&C_3\left.\left[
-\Lambda_{min}\left(1-\frac{\bar{s}_{t\bar{t}}-2m_t^2}
{\bar{s}_{t\bar{t}}\beta_{t\bar{t}}}\Lambda_{t\bar{t}}\right)
+\frac{m_t^2}{\sqrt{\lambda_{t\bar{t}}}}\left(J_a+J_b\right)\right]
|{\cal A}_0^{ab,(c)}|^2\right\} \; , 
\end{eqnarray}
where $\sigma$, $\tau_i$, $\Lambda_\sigma$, and $\Lambda_{\tau_i}$ are
defined in Eq.~(\ref{eq:kinematic_invariants}) and right before it, 
$\bar{s}_{t\bar{t}}, \beta_{t\bar{t}}$, and $\Lambda_{t\bar{t}}$ are
defined in Eq.~(\ref{eq:stt_betatt}), $\lambda_{t\bar{t}}$ is given in
Eq.~(\ref{eq:lambdatt_taupm}), $\Lambda_{min}$ is:
\begin{equation}
\label{eq:lambda_min}
\Lambda_{min}=\ln\left(\frac{s_{min}}{m_t^2}\right)\;,
\end{equation}
and the functions $J_a$ and $J_b$ are given in
Eq.~(\ref{eq:ttbar_jsjajb}).
\subsubsection{Real gluon emission $h\to ggt\bar{t}+g$: collinear region}
\label{subsubsec:one_cutoff_collinear}

We now turn to the case of collinear singularities, which arise when
one of the two final state gluons $i$ ($i\!=\!g^A,g^B$) and the hard
extra gluon $g$ ($g\!=\!g^C$) become collinear and cluster to form a
new parton $i^\prime$ (also a gluon), $i+g\to i^\prime$, with the
collinear kinematics: $q_i\!=\!zq_{i^\prime}$ and
$k\!=\!(1-z)q_{i^\prime}$.  In the collinear limit, the $h\to
ggt\bar{t}+g$ phase space as well as the full parton level real
amplitude squared factorize the dependence on the degrees of freedom
of the collinear emitted gluon, as illustrated in
Eq.~(\ref{eq:sigcoll_crossed}). The collinear part of the parton level
cross section can be calculated analytically according to
Eq.~(\ref{eq:sigma_soft_coll_colamp}). The collinear limit of the $K$
functions, $K_C$, is explicitly given by:
\begin{eqnarray}
\label{eq:k_coll}
K_C(t;i,j;\bar{t})&=& \int d(PS_g)_{coll}
 \frac{N g_s^2}{2} [f^{gg\to i}_{tj}+
f^{gg\to j}_{i\bar{t}}+2 n_{lf} f^{q\bar{q}\to g}] \nonumber \\
&=& -\frac{\alpha_s N}{2\pi} \frac{1}{\Gamma(1-\epsilon)}
\left(\frac{4\pi\mu^2}{s_{min}}\right)^{\epsilon} \frac{1}{\epsilon}
\biggl[
I_{gg\to g}\left(\frac{s_{min}}{s_{ti}},\frac{s_{min}}{s_{ij}}\right)+
I_{gg\to g}\left(\frac{s_{min}}{s_{ij}},\frac{s_{min}}{s_{j\bar{t}}}\right)
\nonumber\\
&& +2 n_{lf} I_{q\bar{q}\to g}(0,0)\biggr]\,\,\,,\nonumber\\
K_C(t;i;\bar{t})&=& \int d(PS_g)_{coll} \frac{N g_s^2}{2} 
[f^{gg\to i}_{t\bar{t}}+n_{lf} f^{q\bar q\to g}]\nonumber\\
&=&
-\frac{\alpha_s N}{2\pi}\frac{1}{\Gamma(1-\epsilon)}
\left(\frac{4\pi\mu^2}{s_{min}}\right)^{\epsilon}\frac{1}{\epsilon}
\left[I_{gg\to g}(\frac{s_{min}}{s_{ti}},\frac{s_{min}}{s_{i\bar t}})
+n_{lf} I_{q\bar q\to g}(0,0)\right]\,\,\,,
\nonumber \\
K_C(t;\bar{t})&=& 0\,\,\,,
\end{eqnarray} 
where $i,j\!=\!1,2$ denote the two external hard gluons with momenta
$q_1$ and $q_2$.  

In the one-cutoff PSS method, the collinear gluon phase space is
\begin{equation}
d(PS_g)_{coll}(i,j,z)=\frac{(4\pi)^\epsilon}{16\pi^2}
\frac{1}{\Gamma(1-\epsilon)} {s}_{ig}^{-\epsilon} 
d {s}_{ig}[z(1-z)]^{-\epsilon} dz \theta(s_{min}
-{s}_{ig}) \; .
\end{equation}
The collinear functions $f^{ig\to i^\prime}_{ab}$ are proportional to
the Altarelli-Parisi splitting function for $ig\to i^\prime$ and
explicitly factorize the corresponding collinear pole, i.e.
\cite{Giele:1992vf,Giele:1993dj}:
\begin{equation}
\label{eq:f_ig_ip}
f^{ig\to i^\prime}_{ab}=
\frac{2}{N}\frac{P_{ig\to i^\prime}(z)}{s_{ig}}\,\,\,,
\end{equation}
where both $i$ and $i^\prime$ are gluons and therefore $P_{ig\to
  i^\prime}$ corresponds to the $P_{gg}(z)$ splitting function given
in Eq.~(\ref{eq:apgg}). The lower indices $a$ and $b$ have been used
to specify the integration boundaries on $z$. In order to avoid double
counting between soft and collinear regions of the phase space of
$g^C$, it is crucial to impose that only one $s_{ig}$ at a time
becomes singular, i.e. satisfies the condition $s_{ig}\!<\!s_{min}$.
The advantage of having reorganized the amplitude in terms of color
ordered amplitudes, as in Eq.~(\ref{eq:color_ordered_amp}), is that
the $f^{ig\to i^\prime}_{ab}$ collinear functions have a very definite
structure: they are all proportional to $(s_{ai}s_{ig}s_{gb})^{-1}$,
for $a,b\!=g^A,g^B,t,\bar{t}$, and the integration boundaries are then
found by imposing that:
\begin{eqnarray}
\label{eq:coll_dz_boundaries}
s_{ai}&=&zs_{ai^\prime}>s_{min} \,\,\,\longrightarrow\,\,\, 
z>z_1=\frac{s_{min}}{s_{ai^\prime}}\,\,\,,\nonumber\\
s_{gb}&=&(1-z)s_{i^\prime b}>s_{min}\,\,\,\longrightarrow\,\,\, 
z< 1-z_2=1-\frac{s_{min}}{s_{i^\prime b}}\,\,\,.
\end{eqnarray}
The terms proportional to $f^{q\bar{q}\to g}$ come from the fact that
a pair of collinear final state massless quarks ($n_{lf}\!=\!5$ is the
number of massless flavors) can also mimic a gluon. The corresponding
collinear function is:
\begin{equation}
\label{eq:f_qq_g}
f^{q\bar{q}\to g}=
\frac{2}{N}\frac{P_{q\bar{q}\to g}(z)}{s_{q\bar{q}}}\,\,\,,
\end{equation}
where both the ${\cal O}(1)$ and ${\cal O}(\epsilon)$ parts of the
splitting function $P_{q\bar{q}\to g}$ are defined in
Eq.~(\ref{eq:apqg}). Note that we do not attach any lower index to
$f^{q\bar{q}\to g}$ because the integration over $z$ has no
singularities and can be performed over the entire range from
$z\!=\!0$ to $z\!=\!1$.

The analytic form of the integrated collinear functions $I_{gg\to
  g}(z_1,z_2)$ and $I_{q\bar{q}\to g}(0,0)$ is obtained by carrying
out the integration in Eq.~(\ref{eq:k_coll}), and is explicitly
given by \cite{Giele:1993dj}:
\begin{eqnarray}
\label{eq:coll_functions_Iij}
I_{gg\to g}(z_1,z_2)&=&
\frac{1}{\epsilon}\left(z_1^{-\epsilon}+z_2^{-\epsilon}-2\right)
-\frac{11}{6}+\left(\frac{\pi^2}{3}-\frac{67}{18}\right)\epsilon+
{\cal O}(\epsilon^2)\,\,\,,\nonumber\\
I_{q\bar{q}\to g}(0,0)&=&
\frac{1}{N}\left(\frac{1}{3}+\frac{5}{9}\epsilon\right)+
{\cal O}(\epsilon^2)\,\,\,.
\end{eqnarray}

Finally, using Eq.~(\ref{eq:sigma_soft_coll_colamp}) and
Eqs.~(\ref{eq:k_coll})-(\ref{eq:coll_functions_Iij}), the pole part of
the parton level collinear cross section can be written as:
\begin{eqnarray}
\label{eq:sigma_coll_1c_pole}
&&(\hat\sigma_{coll}^{h\to ggt\bar{t}+g})_{pole}= 
\int d(PS_4) \frac{\alpha_s}{2\pi}{\cal N}_t\frac{1}{\epsilon}\times
\nonumber\\
&\overline{\sum}&\left\{C_1
\left[\left(-4\Lambda_\sigma+8\Lambda_{min} 
+2\left(\frac{11}{3}-\frac{2}{3}\frac{n_{lf}}{N}\right)\right) 
\left(|{\cal A}_0^{ab,(c)}|^2+|{\cal A}_0^{nab,(c)}|^2\right) 
\right. \right. \nonumber \\
&&\quad\quad-\left(\Lambda_{\tau_1}+\Lambda_{\tau_2}\right)  
|{\cal A}_{0,s}^{(c)}+{\cal A}_{0,t}^{(c)}|^2 
-\left(\Lambda_{\tau_3}+\Lambda_{\tau_4}\right)
|{\cal A}_{0,u}^{(c)}-{\cal A}_{0,s}^{(c)}|^2
\biggr]\nonumber \\  
&&+ \left. C_2 \left[
- 4 \left(\Lambda_{\tau_1}+\Lambda_{\tau_2}
+\Lambda_{\tau_3}+\Lambda_{\tau_4}\right)
+16\Lambda_{min}+4\left(\frac{11}{3}-
\frac{2}{3}\frac{n_{lf}}{N}\right)\right]
|{\cal A}_0^{ab,(c)}|^2 \,\right\}\,\,\,,\nonumber\\
\end{eqnarray}  
while the corresponding finite contribution is:
\begin{eqnarray}
\label{eq:sigma_coll_1c_finite}
&&(\hat\sigma_{coll}^{h\to ggt\bar{t}+g})_{finite}=
\int d(PS_4) \frac{\alpha_s}{2\pi}{\cal N}_t\times
\nonumber\\
&\overline{\sum}&\left\{C_1\left[\left( 
-2\Lambda_\sigma^2-12\Lambda_{min}^2
+8\Lambda_\sigma\Lambda_{min}-2\Lambda_{min}
\biggl(\frac{11}{3}-\frac{2}{3}\frac{n_{lf}}{N}\biggr)
\right.\right.\right.\nonumber\\ 
&&\left.\quad\quad
-2\left(\frac{2}{3}\pi^2-\frac{67}{9}+\frac{10}{9}\frac{n_{lf}}{N}\right)
\right) \left(|{\cal A}_0^{ab,(c)}|^2+|{\cal A}_0^{nab,(c)}|^2\right)
\nonumber\\
&&\quad\quad +\left(
-\frac{1}{2}\Lambda_{\tau_1}^2-\frac{1}{2}\Lambda_{\tau_2}^2
+2\Lambda_{min}\left(\Lambda_{\tau_1}+\Lambda_{\tau_2}\right)\right)
|{\cal A}_{0,s}^{(c)}+{\cal A}_{0,t}^{(c)}|^2\nonumber\\
&&\quad\quad \left. +\left(
-\frac{1}{2}\Lambda_{\tau_3}^2-\frac{1}{2}\Lambda_{\tau_4}^2
+2\Lambda_{min}\left(\Lambda_{\tau_3}+\Lambda_{\tau_4}\right)\right)
|{\cal A}_{0,u}^{(c)}-{\cal A}_{0,s}^{(c)}|^2\right]\nonumber\\
&&+C_2\biggl[
8\Lambda_{min}\left(\Lambda_{\tau_1}+\Lambda_{\tau_2}
+\Lambda_{\tau_3}+\Lambda_{\tau_4}\right)-24\Lambda_{min}^2
-2\left(\Lambda_{\tau_1}^2 +\Lambda_{\tau_2}^2
+\Lambda_{\tau_3}^2+\Lambda_{\tau_4}^2\right)
\nonumber\\
&&\quad\quad\left.\left.
-4\Lambda_{min}\left(\frac{11}{3}-\frac{2}{3}\frac{n_{lf}}{N}\right)
-4\left(\frac{2}{3}\pi^2-\frac{67}{9}+\frac{10}{9}\frac{n_{lf}}{N}\right)
\right]|{\cal A}_0^{ab,(c)}|^2 \right\} \; ,\nonumber\\
\end{eqnarray}
where $\sigma$, $\tau_i$, $\Lambda_\sigma$, and $\Lambda_{\tau_i}$ are
defined in Eq.~(\ref{eq:kinematic_invariants}) and right before it, 
while $\Lambda_{min}$ is defined in Eq.~(\ref{eq:lambda_min}).
\subsubsection{IR Singular Gluon Emission: Complete Result for 
${\hat\sigma}_{ir}^{gg}$}
\label{subsubsec:one_cutoff_ir}

Summing both soft and collinear contributions to the $h\to
ggt\bar{t}+g$ cross section of
Sections~\ref{subsubsec:one_cutoff_soft},
\ref{subsubsec:one_cutoff_collinear}, and crossing the final state
gluons $g^A, g^B$ to the initial state and the Higgs boson to the
final state (which flips both helicities and momenta of these
particles), yields $\hat{\bar{\sigma}}_{ir}^{gg}$ of
Eq.~(\ref{eq:sigma_ir_gg}) as
\begin{equation}
\label{eq:sigma_ir_bar}
\hat{\bar{\sigma}}_{ir}^{gg}=(\hat\sigma_{soft}^{h\to ggt\bar{t}+g}+
\hat\sigma_{coll}^{h\to ggt\bar{t}+g})_{crossed}\,\,\,.
\end{equation}
The IR pole part of $\hat{\bar{\sigma}}_{ir}^{gg}$ is given by
\begin{eqnarray}
\label{eq:sigma_ir_bar_pole}
(\hat{\bar{\sigma}}_{ir}^{gg})_{pole}&=&
((\hat \sigma_{soft}^{h\to ggt\bar{t}+g})_{pole}+
(\hat \sigma_{coll}^{h\to ggt\bar{t}+g})_{pole})_{crossed} 
\nonumber\\
&=& \int d(PS_3)\frac{\alpha_s}{2\pi}{\cal N}_t \overline{\sum}
\left(C_1 {\cal M}_{ir,\epsilon}^{(1)}+C_2 {\cal M}_{ir,\epsilon}^{(2)}+
C_3 {\cal M}_{ir,\epsilon}^{(3)}\right) \nonumber\\
&&+\frac{\alpha_s}{2\pi}{\cal N}_t
\left(-\frac{2}{3}n_{lf}+\frac{8}{3}N-\frac{1}{N}\right)
\frac{1}{\epsilon} \hat\sigma^{gg}_{\sss LO}\,\,\,,
\end{eqnarray}
where ${\cal M}_{ir,\epsilon}^{(i)}=-{\cal M}_{V,\epsilon}^{(i)}$ (see
Eq.~(\ref{eq:m2_virt_ir_poles})) and therefore
$(\hat{\bar{\sigma}}_{ir}^{gg})_{pole}$ completely cancels the IR
singularities of the virtual cross section
$(\hat\sigma_{virt}^{gg})_{IR-pole}$ in Eq.~(\ref{eq:sigma_virt_ir_poles}).
The IR finite part of $\hat{\bar{\sigma}}_{ir}^{gg}$ is given by
\begin{eqnarray}
(\hat{\bar{\sigma}}_{ir}^{gg})_{finite}&=&
((\hat \sigma_{soft}^{h\to ggt\bar t+g})_{finite}+
(\hat\sigma_{coll}^{h\to ggt\bar{t}+g})_{finite})_{crossed}\,\,\,,
\end{eqnarray}
with $(\hat\sigma_{soft,coll}^{h\to ggt\bar{t}+g})_{finite}$ given in
Eqs.~(\ref{eq:sigma_soft_1c_finite}) and
(\ref{eq:sigma_coll_1c_finite}).
 
Finally, as described in Section~\ref{subsec:one_cutoff}, the partonic
cross section for the IR singular real gluon radiation for the process
$gg\to t\bar{t}h+g$ using the one-cutoff PSS method,
$\hat\sigma_{ir}^{gg}$, is obtained from
$\hat{\bar{\sigma}}_{ir}^{gg}$ by adding $\hat\sigma_{crossing}^{gg}$
(see Eq.~(\ref{eq:sigma_ir_gg})).  The cross section
$\hat\sigma_{crossing}^{gg}$ accounts for the difference between
initial and final state collinear gluon radiation and contributes to
the hadronic cross section as
\begin{equation}
\label{eq:sigma_crossing}
\sigma_{crossing}^{gg}=\alpha_s(\mu)
\int dx_1 dx_2 f_g(x_1)
\int_{x_2}^1 \frac{dz}{z}f_g(\frac{x_2}{z}) 
X_{g\to g}(z){\hat\sigma}_{\sss LO}^{gg}
+(x_1\leftrightarrow x_2)\,\,\,,
\end{equation}
with $X_{g\to g}$ given by \cite{Giele:1993dj}:
\begin{eqnarray}
X_{g\to g}(z) &=&-\frac{N}{2\pi}
\left(\frac{4\pi \mu^2}{s_{min}}\right)^{\epsilon} 
\frac{1}{\Gamma(1-\epsilon)}\left(\frac{1}{\epsilon}\right)\times
\nonumber\\
&& \left\{\left[\frac{11}{6}-\frac{1}{3}\frac{n_{lf}}{N}-
\epsilon\left(\frac{\pi^2}{3}-\frac{67}{18}
+\frac{5}{9}\frac{n_{lf}}{N}\right)\right]\delta(1-z)\right.
\nonumber \\  &&
+2\left.
\left[\frac{z}{[(1-z)^{1+\epsilon}]_{+}}+
\frac{(1-z)^{1-\epsilon}}{z}+z(1-z)^{1-\epsilon}
\right]\right\} \,\,\,,
\end{eqnarray}
in terms of regularized \emph{plus} functions (see
Ref.~\cite{Giele:1993dj} for the exact definition). As will be
demonstrated in Section~\ref{sec:total}, these remaining IR
singularities will be absorbed into the gluon PDFs when including the
effects of mass factorization.
\boldmath
\section{The total cross section for $pp\to t\bar{t}h$ at NLO QCD}
\label{sec:total}
\unboldmath
The total inclusive hadronic cross section for $pp\to t\bar{t}h$ is the
sum of the contribution from the $gg$ initial state, the $q\bar{q}$
initial state and the $(q,\bar{q})g$ initial states
\begin{equation}
\label{eq:sigma_nlo_ij}
\sigma_{\sss NLO}(pp\to t\bar{t}h)=
\sigma_{\sss NLO}^{gg}(pp\to t\bar{t}h)+
\sigma_{\sss NLO}^{q\bar q}(pp\to t\bar{t}h)+
\sigma_{\sss NLO}^{qg}(pp\to t\bar{t}h)\,\,\,.
\end{equation}
As described in Section~\ref{sec:framework}, $\sigma_{\sss
  NLO}^{ij}(pp\to t\bar{t}h)$ is obtained by convoluting the parton
level NLO cross section $\hat\sigma_{\sss NLO}^{ij}(pp\to t\bar{t}h)$
with the NLO PDFs ${\cal F}_i^{p}(x,\mu)$ ($i=q,g$), thereby absorbing
the remaining initial state singularities of $\delta\hat\sigma_{\sss
  NLO}^{ij}$ into the renormalized PDFs. In the following we
demonstrate in detail how this cancellation works in the case of the
$gg$ and $(q,\bar{q})g$ initiated processes. The case of the $q\bar{q}$
initiated process is discussed in Section~V of
Ref.~\cite{Reina:2001bc}, where we presented in detail the
contribution of the $q\bar{q}$ initial state to $\sigma_{\sss
  NLO}(p\bar{p}\to t\bar{t}h)$. $\sigma_{\sss NLO}^{q\bar{q}}(pp\to
t\bar{t}h)$ can be obtained from there with obvious modifications, and
will not be repeated here.

First the parton level cross section is convoluted with the {\em bare}
parton distribution functions ${\cal F}_i^{p}(x)$ and subsequently the
${\cal F}_i^{p}(x)$ are replaced by the renormalized parton
distribution functions, ${\cal F}_i^{p}(x,\mu_f)$, defined in some
subtraction scheme at a given factorization scale $\mu_f$. Using the
${\overline{MS}}$ scheme, the scale-dependent NLO parton distribution
functions are given in terms of the bare ${\cal F}_i^{p}(x)$ and the
QCD NLO parton distribution function counterterms
\cite{Harris:2001sx,Giele:1993dj} as follows:
\begin{itemize}
\item[(a)] For the case where an initial state gluon, quark or antiquark
  ($b=g,(q,\bar{q})$) split respectively into a $q\bar{q}$ or
  $(q,\bar{q})g$ pair ($b'=(q,\bar{q}),g$):
\begin{eqnarray}
\label{eq:pdfqg_mu}
{\cal F}_{b'}^{p}(x,\mu_f)&=& {\cal F}_{b'}^{p}(x) 
+\left[\frac{\alpha_s}{2\pi}
\left(\frac{4\pi\mu_r^2}{\mu_f^2}\right)^\epsilon
\frac{1}{\Gamma(1-\epsilon)}\right]
\int_{x}^{1} \frac{dz}{z}
\left(-\frac{1}{\epsilon}\right) P_{bb'}^4(z) 
{\cal F}_{b}^p\left(\frac{x}{z}\right)\,\,\,,\nonumber\\
\end{eqnarray}
for both the one-cutoff and two-cutoff PSS methods, where $P^4_{ij}$
is defined in Eq.~(\ref{eq:apqg}).
\item[(b)] For the case of $g\to gg$ splitting:
\begin{itemize}
\item[b.1)]\underline{two-cutoff PSS method}:
\begin{eqnarray}
\label{eq:pdfgg_mu2}
{\cal F}_g^{p}(x,\mu_f)&=&
{\cal F}_g^{p}(x) \left[1- 
\frac{\alpha_s}{2\pi}
\left(\frac{4\pi\mu_r^2}{\mu_f^2}\right)^\epsilon
\frac{1}{\Gamma(1-\epsilon)}
\left(\frac{1}{\epsilon}\right) N
\left(2 \ln(\delta_s)+\frac{11}{6}-\frac{1}{3}\frac{n_{lf}}{N}\right)
\right]\nonumber\\
&&+\left[\frac{\alpha_s}{2\pi}
\left(\frac{4\pi\mu_r^2}{\mu_f^2}\right)^\epsilon
\frac{1}{\Gamma(1-\epsilon)}\right]
\int_{x}^{1-\delta_s} \frac{dz}{z}
\left(-\frac{1}{\epsilon}\right) P_{gg}(z)
{\cal F}_g^{p}\left(\frac{x}{z}\right)\,\,\,,\nonumber\\
\end{eqnarray}
where $P_{gg}$ is Altarelli-Parisi splitting function given in
Eq.~(\ref{eq:apgg}).
\item[b.2)]\underline{one-cutoff PSS method}:
\begin{eqnarray}
\label{eq:pdfgg_mu1}
&&{\cal F}_g^{p}(x,\mu)=
{\cal F}_g^{p}(x) \left[1- 
\frac{\alpha_s}{2\pi}
\left(\frac{4\pi\mu_r^2}{\mu_f^2}\right)^\epsilon
\frac{1}{\Gamma(1-\epsilon)}
\left(\frac{1}{\epsilon}\right) 
N\left(\frac{11}{6}-\frac{1}{3}\frac{n_{lf}}{N}\right)
\right]\nonumber\\
&+&\left[\frac{\alpha_s}{2\pi}
\left(\frac{4\pi\mu_r^2}{\mu_f^2}\right)^\epsilon
\frac{1}{\Gamma(1-\epsilon)}\right]
\int_{x}^{1} \frac{dz}{z}
\left(-\frac{1}{\epsilon}\right)P_{gg}^{(+)}(z) 
{\cal F}_g^{p}\left(\frac{x}{z}\right)\,\,\,,\nonumber\\
\end{eqnarray}
where $P_{gg}^{(+)}$ is the regulated Altarelli-Parisi splitting
function given by:
\begin{equation}
\label{eq:pgg_plus}
P_{gg}^{(+)}(z)=2N\left(\frac{z}{(1-z)_{+}}+\frac{1-z}{z}+z (1-z)\right)
\,\,\,.
\end{equation}
\end{itemize}
\end{itemize}
The ${\cal O}(\alpha_s)$ terms in the previous equations are
calculated from the ${\cal O}(\alpha_s)$ corrections to the
$b\rightarrow b' j$ splittings, in the PSS formalism.  Moreover, note
that in Eqs.~(\ref{eq:pdfqg_mu})-(\ref{eq:pdfgg_mu1}) we have
carefully separated the dependence on the factorization ($\mu_f$) and
renormalization scale $(\mu_r$). It is understood that
$\alpha_s\!=\!\alpha_s(\mu_r)$. The definition of the subtracted PDFs
is indeed the only place where both scales play a role, and the only
place where we have a dependence on $\mu_f$. In the rest of this paper
we have always set $\mu_r\!=\!\mu_f\!=\!\mu$ and we will also give the
master formulas for the total NLO cross section,
Eqs.~(\ref{eq:sigmatot_gg2})-(\ref{eq:sigmatot_qg1}), using
$\mu_r\!=\!\mu_f\!=\!\mu$. We have checked that this simplifying
assumption has a negligible effect on our results and we will comment
more about this in Section~\ref{sec:results}.

In the two-cutoff PSS method, when convoluting the parton $gg$ cross
section with the renormalized gluon distribution function of
Eq.~(\ref{eq:pdfgg_mu2}), the IR singular counterterm of
Eq.~(\ref{eq:pdfgg_mu2}) exactly cancels the remaining IR poles of
$\hat{\sigma}_{virt}^{gg}+\hat\sigma_{soft}^{gg}$ and
$\hat{\sigma}_{hard/coll}^{gg}$. In the one-cutoff PSS method, the IR
singular counterterm of Eq.~(\ref{eq:pdfgg_mu1}) exactly cancels the
IR poles of $\hat\sigma_{crossing}^{gg}$.  Finally, the complete
${\cal O}(\alpha_s^3)$ inclusive total cross section for $pp\to
t\bar{t}h$ in the ${\overline{MS}}$ factorization scheme when only the
$gg$ initial state is included, i.e.  $\sigma_{\sss NLO}^{gg}(pp\to
t\bar{t}h)$ of Eq.~(\ref{eq:sigma_nlo_ij}), can be written as follows:\\
\begin{itemize}
\item[1)]\underline{two-cutoff PSS method}
\begin{eqnarray}
\label{eq:sigmatot_gg2}
\sigma_{\sss NLO}^{gg} &=&\frac{1}{2}\int dx_1 dx_2 
\left\{ {\cal F}_g^p(x_1,\mu)
{\cal F}_{g}^{p}(x_2,\mu) \left[
\hat{\sigma}^{gg}_{\sss LO}(x_1,x_2,\mu)+
(\hat{\sigma}_{virt}^{gg})_{finite}(x_1,x_2,\mu) \right. \right.\nonumber \\ 
&+&\left. \left. (\hat \sigma_{soft}^{gg})_{finite}(x_1,x_2,\mu) +
\hat \sigma_{s+v+ct}^{gg}(x_1,x_2,\mu) + (1\leftrightarrow 2) \right] \right\} 
\nonumber\\
&+&\frac{1}{2}\int dx_1 dx_2 \left\{
\int_{x_1}^{1-\delta_s}\frac{dz}{z}
\left[{\cal F}_g^p(\frac{x_1}{z},\mu) {\cal F}_{g}^{p}(x_2,\mu)+
{\cal F}_g^{p}(x_2,\mu) {\cal F}_{g}^p(\frac{x_1}{z},\mu)\right]
\right. \nonumber\\
&&\times \left. \hat{\sigma}_{\sss LO}^{gg}(x_1, x_2,\mu)
\frac{\alpha_s}{2\pi} \ln\left(\frac{s}{\mu^2}\frac{(1-z)^2}{z}
\frac{\delta_c}{2}\right) P_{gg}(z)
+(1\leftrightarrow 2) \right\}\nonumber\\
&+& \frac{1}{2} \int dx_1 dx_2 \left\{ {\cal F}_g^p(x_1,\mu)
{\cal F}_{g}^{p}(x_2,\mu) \, 
\hat{\sigma}_{hard/non-coll}^{gg}(x_1,x_2,\mu)+(1 \leftrightarrow 2) 
\right\}\,\,\,,\nonumber\\
\end{eqnarray}
where $\hat\sigma^{gg}_{s+v+ct}$ is obtained from the sum of
$(\hat\sigma^{gg}_{virt})_{UV-pole}$ of Eq.~(\ref{eq:sigma_virt_uv_pole}),
$\hat\sigma_{s+v}^{gg}$ of Eq.~(\ref{eq:irsv}), and the PDF counterterm in
Eq.~(\ref{eq:pdfgg_mu2}) as follows
\begin{equation}
\hat\sigma^{gg}_{s+v+ct}=
\frac{\alpha_s}{2\pi} \left[4 N \ln(\delta_s) \ln\left(\frac{s}{\mu^2}\right) 
+\left(\frac{11}{3} N-\frac{2 n_{lf}}{3}+4N \ln(\delta_s)\right) 
\ln\left(\frac{m_t^2}{s}\right)\right]\hat\sigma^{gg}_{\sss LO}\; .
\end{equation}
\item[2)]\underline{one-cutoff PSS method}
\begin{eqnarray}
\label{eq:sigmatot_gg1}
\sigma_{\sss NLO}^{gg} &=&
\frac{1}{2} \int dx_1 dx_2 \left\{ {\cal F}_g^p(x_1,\mu)
{\cal F}_{g}^{p}(x_2,\mu) \left[
\hat{\sigma}^{gg}_{\sss LO}(x_1,x_2,\mu)+
(\hat{\sigma}_{virt}^{gg})_{finite}(x_1,x_2,\mu) \right. \right.\nonumber \\ 
&+&\left.\left.(\hat{\bar{\sigma}}_{ir}^{gg})_{finite}(x_1,x_2,\mu) +
\hat \sigma_{v+ir+ct}^{gg}(x_1,x_2,\mu)+(1\leftrightarrow 2) 
\right]\right\}\nonumber\\
&+&\frac{1}{2}\int dx_1 dx_2 \left\{
\int_{x_1}^{1}\frac{dz}{z}
\left[{\cal F}_g^p(\frac{x_1}{z},\mu) {\cal F}_{g}^{p}(x_2,\mu)+
{\cal F}_g^{p}(x_2,\mu) {\cal F}_{g}^p(\frac{x_1}{z},\mu)\right]
\right. \nonumber \\
&&\times \hat{\sigma}_{\sss LO}^{gg}(x_1, x_2,\mu)
\frac{\alpha_s}{2\pi} 2 N \ln\left(\frac{s}{\mu^2} \frac{s_{min}}{s}\right)
\left(\frac{z}{(1-z)_{+}} + \frac{1-z}{z} +z (1-z)\right)\nonumber\\
&&+(1\leftrightarrow 2) \biggr\}\nonumber\\
&+&\frac{1}{2}\int dx_1 dx_2 \left\{
\int_{x_1}^{1}\frac{dz}{z}
\left[{\cal F}_g^p(\frac{x_1}{z},\mu) {\cal F}_{g}^{p}(x_2,\mu)+
{\cal F}_g^{p}(x_2,\mu) {\cal F}_{g}^p(\frac{x_1}{z},\mu)\right]
\right. \nonumber\\
&&\times \hat{\sigma}_{\sss LO}^{gg}(x_1, x_2,\mu)
\frac{\alpha_s}{2\pi} 2 N 
\left[\left(\frac{1-z}{z} +z (1-z)\right)\ln(1-z)+z \left(\frac{\ln(1-z)}{1-z}\right)_{+} \right]\nonumber\\
&&+(1\leftrightarrow 2) \biggr\}\nonumber\\
&+& \frac{1}{2} \int dx_1 dx_2 \left\{ {\cal F}_g^p(x_1,\mu)
{\cal F}_{g}^{p}(x_2,\mu) \, 
\hat{\sigma}_{hard}^{gg}(x_1,x_2,\mu)+(1 \leftrightarrow 2) \right\}
\,\,\,,\nonumber\\
\end{eqnarray}
where $\hat\sigma^{gg}_{v+ir+ct}$ is obtained from the sum of
$(\hat\sigma^{gg}_{virt})_{UV-pole}$ of
Eq.~(\ref{eq:sigma_virt_uv_pole}),
$(\hat\sigma^{gg}_{virt})_{IR-pole}$ of
Eq.~(\ref{eq:sigma_virt_ir_poles}),
$(\hat{\bar{\sigma}}_{ir}^{gg})_{pole}$ of
Eq.~(\ref{eq:sigma_ir_bar_pole}), the part proportional to
$\delta(1-z)$ of $\hat\sigma_{crossing}^{gg}$ of
Eq.~(\ref{eq:sigma_crossing}), and the PDF counterterm in
Eq.~(\ref{eq:pdfgg_mu1}), and can be written as:
\begin{equation}
\hat\sigma^{gg}_{v+ir+ct}=
\frac{\alpha_s}{2\pi} \left[
\left(\frac{11}{3} N-\frac{2 n_{lf}}{3}\right) \ln\left(\frac{s_{min}}{s}\right)
+2 N \left(\frac{\pi^2}{3}-\frac{67}{18}+\frac{5 n_{lf}}{9 N} \right)\right]  
\hat \sigma^{gg}_{\sss LO}\; .
\end{equation}
\end{itemize}

We note that $\sigma_{\sss NLO}^{gg}$ is finite, since, after mass
factorization, both soft and collinear singularities have been
canceled between $\hat{\sigma}_{virt}^{gg}+\hat\sigma_{soft}^{gg}$ and
$\hat{\sigma}_{hard/coll}^{gg}$ in the two-cutoff PSS method, and
between $\hat{\sigma}_{virt}^{gg}$ and $\hat{\bar{\sigma}}_{ir}^{gg}$
in the one-cutoff PSS method.  The last terms respectively describe
the finite real gluon emission of Eq.~(\ref{eq:sigma_gg_hard}) and
(\ref{eq:sigma_real_one_cutoff}).  Note that when collecting all the
terms in Eqs.~(\ref{eq:sigmatot_gg2}) and (\ref{eq:sigmatot_gg1}) that
are proportional to $\ln(\mu^2/s)$, one obtains exactly the last two
terms in Eq.~(\ref{eq:mudep_coeff}), as predicted by renormalization
group arguments.

For the $(q,\bar{q})g$ initiated processes we find
\begin{itemize}
\item[1)]\underline{two-cutoff PSS method}
\begin{eqnarray}
\label{eq:sigmatot_qg2}
\sigma^{qg}_{\sss NLO} &=&
\frac{\alpha_s}{2\pi} \sum_{i=q,\bar{q}}\int dx_1dx_2 
\left\{ 
\int_{x_1}^{1}\frac{dz}{z} 
{\cal F}_i^p(\frac{x_1}{z},\mu) {\cal F}_{g}^{p}(x_2,\mu) 
\times \right. \nonumber \\
&&\left. \hat{\sigma}_{\sss LO}^{gg}(x_1, x_2,\mu) 
\left[P^4_{ig}(z) 
\ln\left(\frac{s}{\mu^2}\frac{(1-z)^2}{z}\frac{\delta_c}{2}\right)-
P^{\prime}_{ig}(z) \right]\right.\nonumber\\
&+& \left. \int_{x_1}^{1}\frac{dz}{z} 
{\cal F}_g^p(\frac{x_1}{z},\mu) {\cal F}_{i}^{p}(x_2,\mu) 
\times \right. \nonumber\\
&&\left. \hat{\sigma}_{\sss LO}^{q\bar{q}}(x_1, x_2,\mu) 
\left[
{P}^4_{gi}(z) \ln\left(\frac{s}{\mu^2}\frac{(1-z)^2}{z}
\frac{\delta_c}{2}\right)-
{P}^{\prime}_{gi}(z)\right] +(1\leftrightarrow 2)\right\}\nonumber\\
&+&\sum_{i=q,\bar{q}} \int dx_1 dx_2 
\left\{ {\cal F}_i^p(x_1,\mu) {\cal F}_{g}^{p}(x_2,\mu) \, 
\hat{\sigma}_{non-coll}^{qg}(x_1,x_2,\mu)+(1 \leftrightarrow 2)
\right\}\,\,\,,\nonumber\\
\end{eqnarray}
\item[2)]\underline{one-cutoff PSS method}
\begin{eqnarray}
\label{eq:sigmatot_qg1}
\sigma^{qg}_{\sss NLO} &=&
\frac{\alpha_s}{2\pi} \sum_{i=q,\bar{q}}\int dx_1dx_2
\left\{ 
\int_{x_1}^{1}\frac{dz}{z} 
{\cal F}_g^p(\frac{x_1}{z},\mu) {\cal F}_{i}^{p}(x_2,\mu) 
\times \right. \nonumber \\
&&\left. \hat{\sigma}_{\sss LO}^{gg}(x_1, x_2,\mu) 
\left[{P}^4_{ig}(z) \ln\left(\frac{s_{min} (1-z)}{\mu^2}\right)-
{P}^{\prime}_{ig}(z)\right]\right.\nonumber\\
&+& \left. \int_{x_1}^{1}\frac{dz}{z} 
{\cal F}_i^p(\frac{x_1}{z},\mu) {\cal F}_{g}^{p}(x_2,\mu) 
\times \right. \nonumber\\
&&\left. \hat{\sigma}_{\sss LO}^{q\bar{q}}(x_1, x_2,\mu) 
\left[ {P}^4_{gi}(z)
\ln\left(\frac{s_{min} (1-z)}{\mu^2}\right)-
{P}^{\prime}_{gi}(z)
\right]+(1\leftrightarrow 2)\right\}\nonumber\\
&+&\sum_{i=q,\bar{q}}\int dx_1dx_2
\left\{ {\cal F}_i^p(x_1,\mu) {\cal F}_{g}^{\bar p}(x_2,\mu) \, 
\hat{\sigma}_{hard}^{qg}(x_1,x_2,\mu)+(1 \leftrightarrow 2) 
\right\}\,\,\,,\nonumber\\
\end{eqnarray}
where ${P}^4_{ij}$ and ${P}^\prime_{ij}$ are the ${\cal O}(1)$ and
${\cal O}(\epsilon)$ contributions to the splitting functions as given
in Eq.~(\ref{eq:apqg}). The last terms respectively describe the
finite gluon/quark emission of Eqs.~(\ref{eq:sigma_qg_real}) and
(\ref{eq:sigma_real_one_cutoff}).

\end{itemize}

We would like to conclude this section by showing explicitly that the
total NLO cross section, $\sigma_{\sss NLO}$, does not depend on the
arbitrary cutoffs introduced by the PSS method, i.e. on $s_{min}$ for
the one-cutoff method and on $\delta_s$ and $\delta_c$ for the
two-cutoff method. The cancellation of the PSS cutoff dependence is
realized in $\sigma_{real}$ by matching contributions that are
calculated either analytically, in the IR-unsafe region below the
cutoff(s), or numerically, in the IR-safe region above the cutoff(s).
While the analytical calculation in the IR-unsafe region reproduces
the form of the cross section in the soft or collinear limits and is
therefore only accurate for small values of the cutoff(s), the
numerical integration in the IR-safe region becomes unstable for very
small values of the cutoff(s). Therefore, obtaining a convincing
cutoff independence involves a delicate balance between the previous
antagonistic requirements and ultimately dictates the choice of
neither too large nor too small values for the cutoff(s). The Monte
Carlo phase space integration has been performed using the adaptive
multi-dimensional integration program VEGAS \cite{Lepage:1978sw} as
well as multichannel integration techniques
\cite{Berends:1985gf,Hilgart:1993xu,Denner:1999gp}.

In Figs.~\ref{fg:ds_dependence} and \ref{fg:dc_dependence} we consider
the two-cutoff PSS method and study the independence of $\sigma_{\sss
  NLO}(pp\to t\bar{t}h)$ on $\delta_s$ and $\delta_c$ separately, by
varying only one of the two cutoffs while the other is kept fixed. In
Fig.~\ref{fg:ds_dependence}, $\delta_s$ is varied between $10^{-5}$
and $10^{-3}$ with $\delta_c\!=\!10^{-5}$, while in
Fig.~\ref{fg:dc_dependence}, $\delta_c$ is varied between $10^{-6}$
and $10^{-4}$ with $\delta_s\!=\!10^{-4}$. In both plots, we show in
the upper window the overall cutoff dependence cancellation between
$\sigma_{soft}^{gg}+\sigma_{hard/coll}^{gg}$ and
$\sigma_{hard/non-coll}^{gg}$ in $\sigma_{real}^{gg}$. We do not
include the corresponding contributions from the Born and the virtual
cross sections since they are, of course, cutoff independent.  Similar
plots could be obtained for the other two sub-channels, $q\bar{q}$ and
$qg+\bar{q}g$.  We illustrate the point using just the $gg$ channel,
since the $q\bar{q}$ channel has already been thoroughly studied in
Ref.~\cite{Reina:2001bc}, while the cutoff dependence of the
$qg+\bar{q}g$ channel is trivial.  In the lower window of the same
plots we complement this information by reproducing the full
$\sigma_{\sss NLO}$, including all channels, on a larger scale that
magnifies the details of the cutoff dependence cancellation. The
statistical errors from the Monte Carlo phase space integration are
also shown.  Both Figs.~\ref{fg:ds_dependence} and
\ref{fg:dc_dependence} show a clear plateau over a wide range of
$\delta_s$ and $\delta_c$ and the NLO cross section is proven to be
cutoff independent. The results presented in Section~\ref{sec:results}
have been obtained by using the two-cutoff PSS method with
$\delta_s\!=\!10^{-4}$ and $\delta_c\!=\!10^{-5}$.
\begin{figure}[t]
\begin{center}
\includegraphics[scale=0.60]{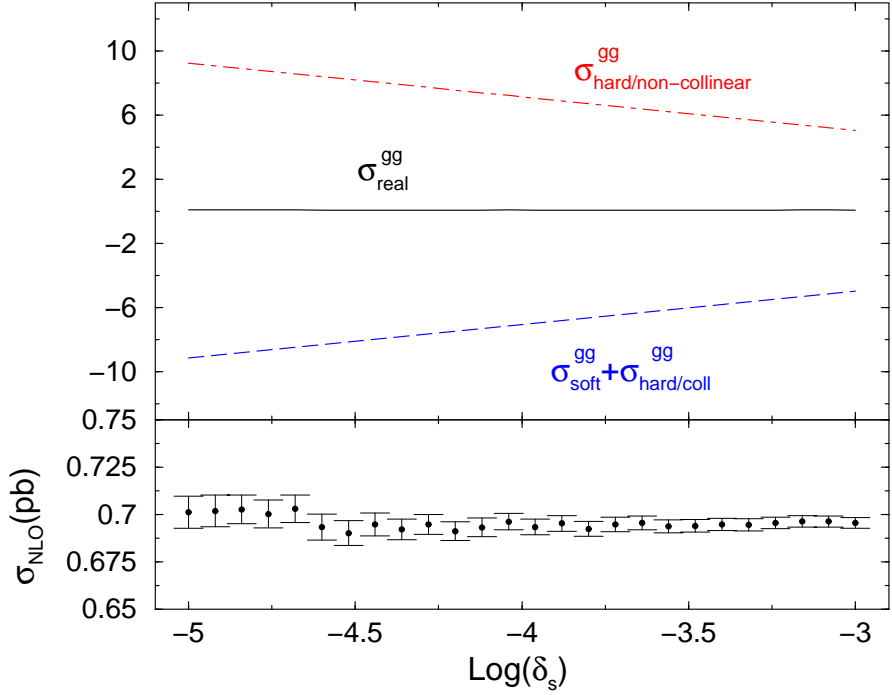}
\caption[]{Dependence of $\sigma_{\sss NLO}(pp\to t\bar{t}h)$ on the
  soft cutoff $\delta_s$ of the two-cutoff PSS method, at
  $\sqrt{s_{\sss H}}\!=\!14$~TeV, for $M_h\!=\!120$~GeV,
  $\mu\!=\!m_t+M_h/2$, and $\delta_c\!=\!10^{-5}$. The upper plot
  shows the cancellation of the $\delta_s$-dependence between
  $\sigma_{soft}^{gg}+\sigma_{hard/coll}^{gg}$ and
  $\sigma_{hard/non-coll}^{gg}$. The lower plot shows, on an enlarged
  scale, the dependence of the full $\sigma_{\sss NLO}=\sigma_{\sss
    NLO}^{gg}+\sigma_{\sss NLO}^{q\bar{q}}+\sigma_{\sss NLO}^{qg}$ on
  $\delta_s$ with the corresponding statistical errors.}
\label{fg:ds_dependence}
\end{center}
\end{figure}
\begin{figure}[t]
\begin{center}
\includegraphics[scale=0.60]{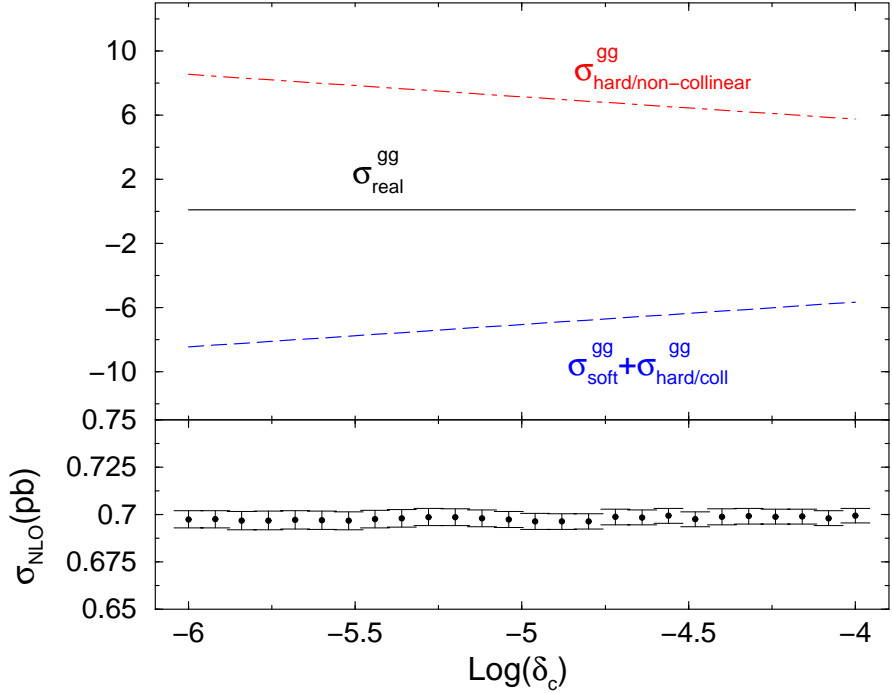}
\caption[]{Dependence of $\sigma_{\sss NLO}(pp\to t\bar{t}h)$ on the
  collinear cutoff $\delta_c$ of the two-cutoff PSS method, at
  $\sqrt{s_{\sss H}}\!=\!14$~TeV, for $M_h\!=\!120$~GeV,
  $\mu\!=\!m_t+M_h/2$, and $\delta_s\!=\!10^{-4}$. The upper plot
  shows the cancellation of the $\delta_s$-dependence between
  $\sigma_{soft}^{gg}+\sigma_{hard/coll}^{gg}$, and
  $\sigma_{hard/non-coll}^{gg}$. The lower plot shows, on an enlarged
  scale, the dependence of the full $\sigma_{\sss NLO}=\sigma_{\sss
    NLO}^{gg}+\sigma_{\sss NLO}^{q\bar{q}}+\sigma_{\sss NLO}^{qg}$ on
  $\delta_c$ with the corresponding statistical errors.}
\label{fg:dc_dependence}
\end{center}
\end{figure}

We now turn to the one-cutoff PSS method and, following the same
criterion adopted for the case of the two-cutoff PSS method, we
summarize in the upper window of Fig.~\ref{fg:smin_dependence} the
behavior of the different cutoff dependent contributions to the real
$gg\to t\bar{t}h$ cross section, i.e.  $\sigma_{ir}^{gg}$ and
$\sigma_{hard}^{gg}$, as well as the resulting cutoff independence of
$\sigma_{real}^{gg}$. The lower window of
Fig.~\ref{fg:smin_dependence} shows the full $\sigma_{\sss NLO}$,
where all $t\bar{t}h$ subprocesses are included, on an enlarged scale.
The statistical deviations due to the Monte Carlo integration are also
shown, and therefore the stability of the integration procedure can be
appreciated. In Fig.~\ref{fg:smin_dependence} $s_{min}$ is varied over
several orders of magnitude and the presence of a clear plateau over
most of the $s_{min}$ range is evident. The results presented in
Section~\ref{sec:results} have been cross-checked using the one-cutoff
PSS method with $s_{min}\!=\!10$~GeV$^2$.
\begin{figure}[t]
\begin{center}
\includegraphics[scale=0.60]{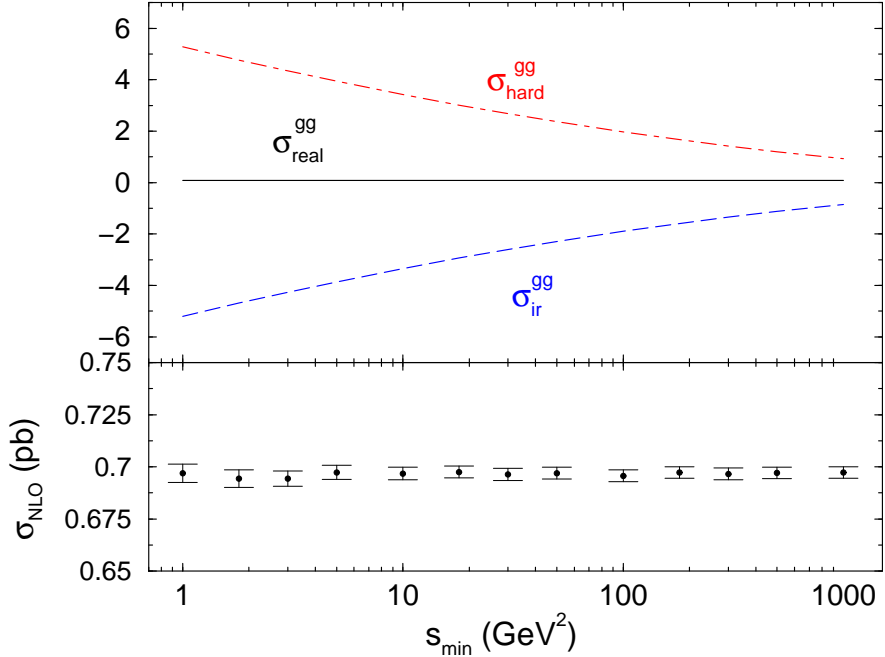}
\caption[]{Dependence of $\sigma_{\sss NLO}(pp\to t\bar{t}h)$ on the
  $s_{min}$ cutoff of the one-cutoff PSS method, at $\sqrt{s_{\sss
      H}}\!=\!14$~TeV, for $M_h\!=\!120$~GeV, and $\mu\!=\!m_t+M_h/2$.
  The upper plot shows the cancellation of the $s_{min}$-dependence
  between $\sigma_{ir}^{gg}$ and $\sigma_{hard}^{gg}$. The lower plot
  shows, on an enlarged scale, the dependence of the full
  $\sigma_{\sss NLO}=\sigma_{\sss NLO}^{gg}+\sigma_{\sss
    NLO}^{q\bar{q}}+\sigma_{\sss NLO}^{qg}$ on $s_{min}$ with the
  corresponding statistical errors.}
\label{fg:smin_dependence}
\end{center}
\end{figure}
\section{Numerical results}
\label{sec:results}

In this section we summarize the most important numerical results for
$\sigma_{\sss NLO}(pp\to t\bar{t}h)$ and illustrate the impact of NLO
QCD corrections on the tree level cross section. In particular, we
discuss the renormalization/factorization scale dependence of
$\sigma_{\sss NLO}$ with respect to $\sigma_{\sss LO}$, and illustrate
the dependence of both LO and NLO cross sections on the Higgs boson
mass. Results for $\sigma_{\sss LO}$ are obtained using the 1-loop
evolution of $\alpha_s(\mu)$ and CTEQ5L parton distribution functions
\cite{Lai:1999wy}, while results for $\sigma_{\sss NLO}$ are obtained
using the 2-loop evolution of $\alpha_s(\mu)$ and CTEQ5M parton
distribution functions, with $\alpha_s^{\sss NLO}(M_Z)\!=\!0.118$.
According to the renormalization prescription adopted in this paper
and explained in Section~\ref{subsec:virtual_uv}, throughout our
calculation we use for the input parameter $m_t$ the top quark pole
mass. Results are presented for $m_t\!=\!174$~GeV, while the
uncertainty introduced by varying $m_t$ within its experimental
uncertainty is discussed later in this section.  We define the top
quark Yukawa coupling to be $g_{t\bar{t}h}\!=\!m_t/v$ where
$v\!=(G_F\sqrt{2})^{-1/2}\!=\!246$~GeV is the vacuum expectation value
of the SM Higgs field, given in terms of the Fermi constant $G_F$.

In Fig.~\ref{fg:mu_dependence}, we illustrate the dependence of both
$\sigma_{\sss NLO}$ and $\sigma_{\sss LO}$ on the renormalization and
factorization scales when the two scales are identical, i.e. when
$\mu_r\!=\!\mu_f\!=\!\mu$.  We have also studied the behavior of
$\sigma_{\sss NLO}$ when the renormalization and factorization scales
are varied independently and noticed no appreciable difference with
respect to the case in which the two scales are identical. This
justifies our decision to present results only for
$\mu_r\!=\!\mu_f\!=\!\mu$.  We also illustrate in
Fig.~\ref{fg:mu_dependence_qq_qg_gg} the $\mu$ dependence of the NLO
cross section for each parton level channel independently. We use
$M_h\!=\!120$~GeV for the purpose of these plots. As expected,
Fig.~\ref{fg:mu_dependence} shows that the NLO cross section has a
much weaker scale dependence and represents a much more stable
theoretical prediction.
\begin{figure}[t]
\begin{center}
\includegraphics[scale=0.60]{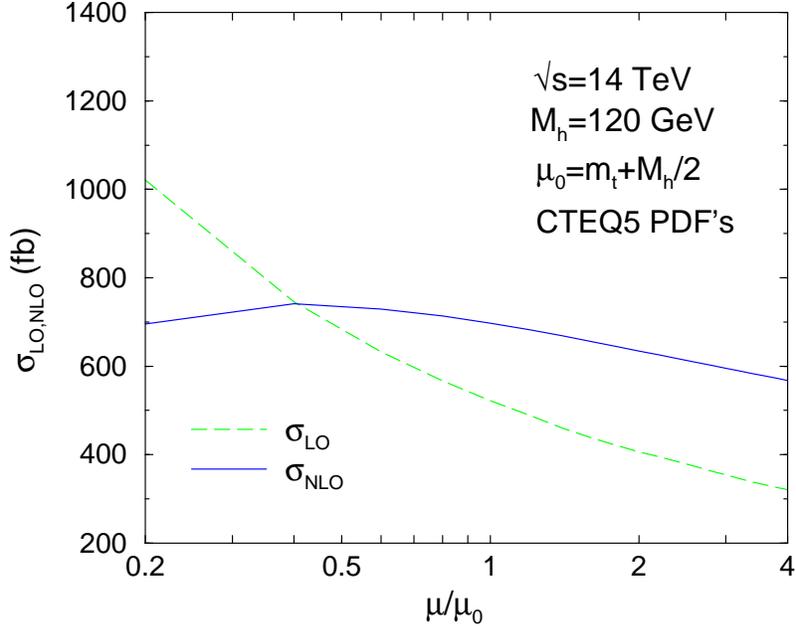}
\caption[]{Dependence of $\sigma_{\sss LO,NLO}(pp\to t\bar{t}h)$ on 
  the renormalization/factorization scale $\mu$, at $\sqrt{s_{\sss
      H}}\!=\!14$~TeV, for $M_h\!=\!120$ GeV.}
\label{fg:mu_dependence}
\end{center}
\end{figure}
\begin{figure}[t]
\begin{center}
\includegraphics[scale=0.60]{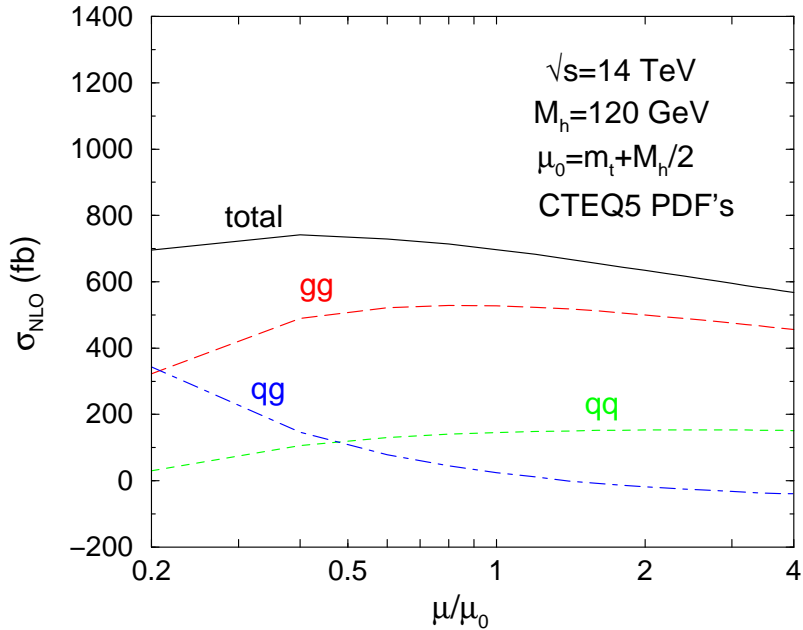}
\caption[]{Dependence of $\sigma_{\sss NLO}(gg,q\bar{q},qg+\bar{q}g 
  \to t\bar{t}h)$ on the renormalization/factorization scale $\mu$, at
  $\sqrt{s_{\sss H}}\!=\!14$~TeV, for $M_h\!=\!120$ GeV.}
\label{fg:mu_dependence_qq_qg_gg}
\end{center}
\end{figure}
\begin{figure}[t]
\begin{center}
\includegraphics[scale=0.60]{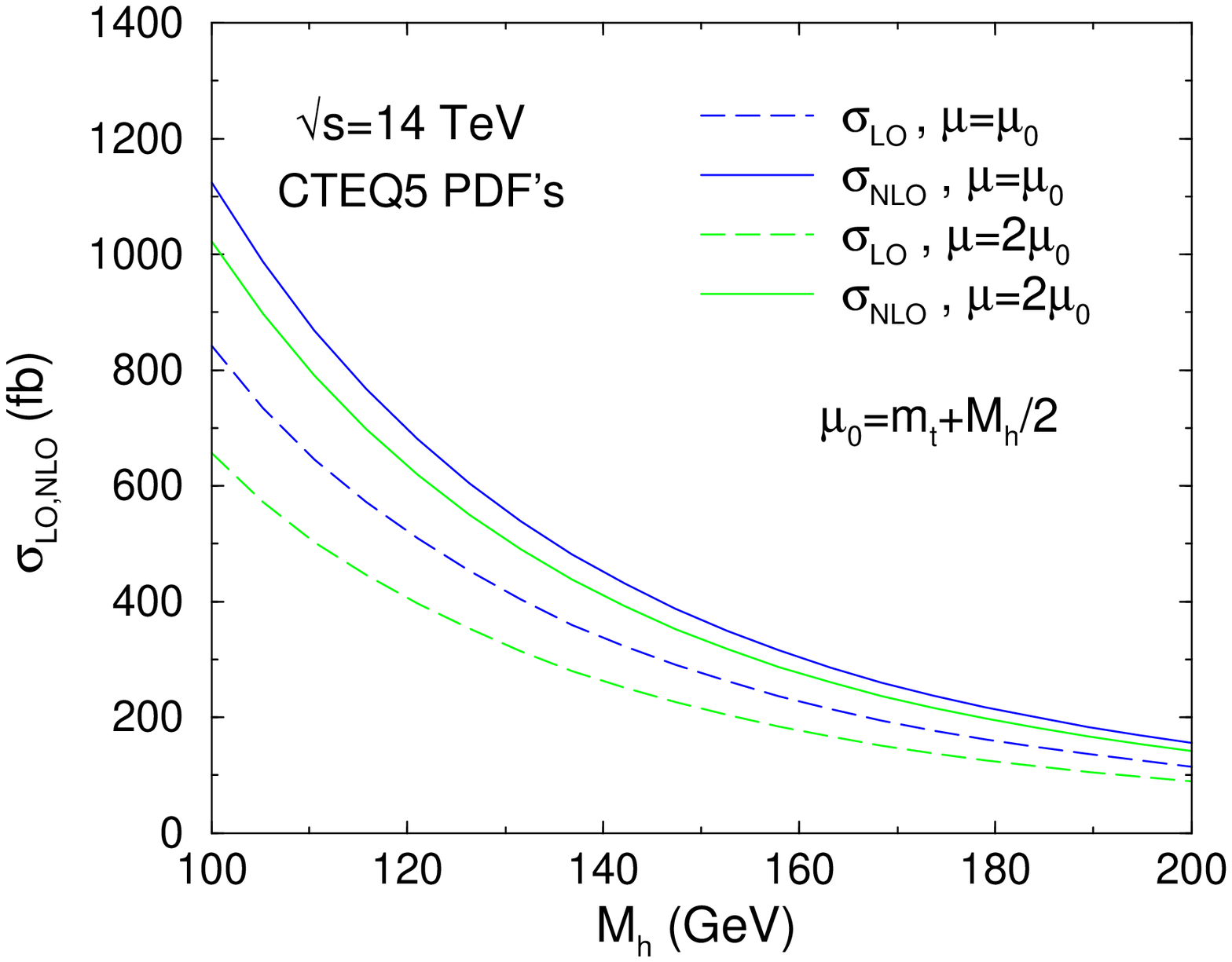}
\caption[]{$\sigma_{\sss NLO}(pp\to t\bar{t}h)$ and $\sigma_{\sss
    LO}(pp\to t\bar{t}h)$ as functions of $M_h$, at $\sqrt{s_{\sss
      H}}\!=\!14$~TeV, for $\mu\!=m_t+M_h/2$ and $\mu\!=\!2m_t+M_h$.}
\label{fg:mh_dependence}
\end{center}
\end{figure}
In Fig.~\ref{fg:mh_dependence}, we plot $\sigma_{\sss LO}(pp\to
t\bar{t}h)$ and $\sigma_{\sss NLO}(pp\to t\bar{t}h)$ as functions of
the Higgs boson mass, for $\sqrt{s_{\sss H}}\!=\!14$~TeV and two
values of the common renormalization/factorization scale,
$\mu\!=\!m_t+M_h/2$ and $\mu\!=\!2m_t+M_h$.  We consider
$100\,\mbox{GeV}\!\le\!M_h\!\le\!200\,\mbox{GeV}$ since the production
of a Higgs boson in association with a pair of top quarks at the LHC
will play a crucial role only for relatively light Higgs bosons. The
information gathered from this plot nicely complements what has
already been shown in Fig.~\ref{fg:mu_dependence}. We summarize a
sample of results from both Figs.~\ref{fg:mu_dependence} and
\ref{fg:mh_dependence} in Table~\ref{tab:results_mu_mh}, where we also
provide the LO cross section, $\overline\sigma_{\sss LO}$, calculated
using the 2-loop evolution of $\alpha_s(\mu)$ and CTEQ5M PDFs. This
can be useful to separately evaluate the impact of the full set of NLO
QCD corrections as opposed to the subset of them that mainly
correspond to the NLO running of $\alpha_s(\mu)$.  The error we quote
on our values is the statistical error of the numerical integration
involved in evaluating the total cross section.  We estimate the
remaining theoretical uncertainty on the NLO results to be of the
order of $15-20\%$. This is mainly due to: the left over
$\mu$-dependence (about $15\%$), the dependence on the PDFs (about
$6\%$), and the error on $m_t$ (about $7\%$) which particularly plays
a role in the top quark Yukawa coupling.
\begin{table*}
\begin{center}
\begin{tabular}{|c|c|c|c|c|c|}
\hline\hline
$M_h$ (GeV) & $\mu$ & $\sigma_{\sss LO}$ (fb) & 
$\overline\sigma_{\sss LO}$ (fb) & $\sigma_{\sss NLO}$ (fb) \\
\hline\hline
    & $m_t$       & 582.92 $\pm$ 0.06 & 616.81 $\pm$ 0.07 & 718.64 $\pm$ 3.71 \\
120 & $m_t+M_h/2$ & 520.47 $\pm$ 0.06 & 553.25 $\pm$ 0.06 & 697.27 $\pm$ 3.20 \\
    & $2m_t$      & 450.09 $\pm$ 0.05 & 480.80 $\pm$ 0.05 & 662.66 $\pm$ 2.77 \\
    & $2m_t+M_h$  & 405.54 $\pm$ 0.04 & 434.59 $\pm$ 0.05 & 634.36 $\pm$ 2.34 \\
\hline
    & $m_t$       & 316.27 $\pm$ 0.03 & 336.41 $\pm$ 0.04 & 380.95 $\pm$ 1.81 \\
150 & $m_t+M_h/2$ & 275.44 $\pm$ 0.03 & 294.35 $\pm$ 0.03 & 367.38 $\pm$ 1.52 \\
    & $2m_t$      & 243.47 $\pm$ 0.03 & 261.03 $\pm$ 0.03 & 352.71 $\pm$ 1.35 \\ 
    & $2m_t+M_h$  & 214.43 $\pm$ 0.02 & 230.60 $\pm$ 0.02 & 334.48 $\pm$ 1.18 \\
\hline
    & $m_t$       & 187.44 $\pm$ 0.02 & 200.46 $\pm$ 0.02 & 221.63 $\pm$ 1.01 \\
180 & $m_t+M_h/2$ & 159.32 $\pm$ 0.02 & 171.15 $\pm$ 0.02 & 214.01 $\pm$ 0.85 \\
    & $2m_t$      & 143.77 $\pm$ 0.02 & 154.74 $\pm$ 0.02 & 206.59 $\pm$ 0.77 \\ 
    & $2m_t+M_h$  & 123.85 $\pm$ 0.01 & 133.65 $\pm$ 0.02 & 194.42 $\pm$ 0.70 \\
 \hline\hline
\end{tabular}
\caption{\label{tab:results_mu_mh}Values of both 
  $\sigma_{\sss LO}(pp\to t\bar{t}h)$, $\overline\sigma_{\sss
    LO}(pp\to t\bar{t}h)$, and $\sigma_{\sss NLO}(pp\to t\bar{t}h)$,
  at $\sqrt{s_{\sss H}}\!=\!14$~TeV, for a sample of different values
  of $M_h$ and of the renormalization/factorization scales
  $\mu\!=\!\mu_r\!=\!\mu_f$.}
\end{center}
\end{table*}

It can also be useful to quote the impact of NLO corrections in terms
of a so called $K$-factor, defined as the ratio between the NLO and LO
cross sections:
\begin{equation}
\label{eq:k_factor}
K=\frac{\sigma_{\sss NLO}(pp\to t\bar{t}h)}{\sigma_{\sss LO}(pp\to t\bar{t}h)}
\,\,\,.
\end{equation}
We can see in Fig.~\ref{fg:mu_dependence} that, for a SM Higgs boson
of mass $M_h\!=\!120$~GeV, the $K$-factor for $pp\to t\bar{t}h$ is
larger than unity when $\mu\!\ge\!0.4\mu_0$ for $\mu_0\!=\!m_t+M_h/2$.
Therefore, over a broad range of the commonly used
renormalization/factorization scales, NLO QCD corrections increase the
LO cross section. Using the results of Table~\ref{tab:results_mu_mh},
the K-factors for a sample of Higgs boson masses and
renormalization/factorization scales can easily be calculated, both
using $\sigma_{\sss LO}$ and $\overline{\sigma}_{\sss LO}$.  We
notice, however, that the $K$-factor defined in
Eq.~(\ref{eq:k_factor}) is affected by a very strong scale dependence,
the same as $\sigma_{\sss LO}$. Therefore, when the $K$-factor is used
to obtain $\sigma_{\sss NLO}$ from $\sigma_{\sss LO}$, care must be
used in matching $\sigma_{\sss LO}$ and $K$ corresponding to the same
$\mu$-scale.

\section{Conclusions}
\label{sec:conclusions}

The associated production of a Higgs boson with a pair of top quarks
will play a very important role in the discovery of a low mass Higgs
boson at the LHC with a center of mass energy of $\sqrt{s_{\sss
    H}}\!=\!14$~TeV.  With the statistics expected at the LHC, $pp\to
t\bar{t}h$, with $h\to b\bar{b},\tau^+\tau^-,W^+W^-,\gamma\gamma$ will
also play a crucial role in determining the couplings of the
discovered Higgs boson, and will give the only handle on a direct
measurement of the top quark Yukawa coupling.

In this paper the inclusive cross section for $pp\to t\bar{t}h$
production has been calculated, in the Standard Model, including full
NLO QCD corrections. The NLO cross section shows a drastically reduced
renormalization and factorization scale dependence, and leads to
increased confidence in predictions based on these results. The
overall uncertainty on the theoretical prediction, including the
errors coming from parton distribution functions and the top quark
mass, is reduced to only 15-20\%, as opposed to the 100-200\%
uncertainty of the LO cross section.  Including NLO QCD corrections
increases the LO cross section for a broad range of commonly used
renormalization and factorization scales, and over the entire Higgs
boson mass range considered in this paper. This is summarized by
saying that the $K$-factor for renormalization and factorization
scales in the range $m_t\!<\!\mu\!\le\!2m_t+M_h$ and Higgs boson
masses in the range $100\,\mbox{GeV}\!\le M_h\!\le\! 200\,\mbox{GeV}$
is between 1.2 and 1.6.

The calculation of the NLO QCD cross section for $pp\to t\bar{t}h$
contains several technical difficulties that have been thoroughly
explained in this paper (see also Ref.~\cite{Reina:2001bc}). The NLO
virtual corrections involve pentagon diagrams and consequently require
the evaluation of both scalar and tensor pentagon integrals with
several external and internal massive particles. Detailed information
about the method used as well as explicit results for all the IR
singular integrals appearing in the calculation are presented in a
series of Appendices.  Tensor pentagon integrals are affected by
numerical instabilities and we discuss in this paper how we have
calculated them in a numerically stable form. The NLO real corrections
are complicated by the presence of IR divergences. We have calculated
them in two different variations of the phase space slicing method,
involving one or two arbitrary cutoffs respectively. The
correspondence between the two phase space slicing methods is made
explicit, and the agreement between them constitutes a powerful check
of the technicalities used in their implementations. The techniques
developed in this paper and in Ref.~\cite{Reina:2001bc} can now be
applied to similar higher order calculations, in particular to the
case of the associated $b\bar{b}h$ production at both the Tevatron and
the LHC.
\begin{acknowledgments}
  We thank U.~Baur, Z.~Bern, and F.~Paige for valuable discussions and
  encouragement. We are grateful to the authors of
  Ref.~\cite{Beenakker:2001rj} for a detailed comparison of the
  results.  The work of S.D. (C.J., L.H.O., L.R.) is supported in part
  by the U.S.  Department of Energy under grant DE-AC02-76CH00016
  (DE-FG02-97ER41022, DE-FG-02-91ER40685, DE-FG02-97ER41022).
\end{acknowledgments}
\appendix
\boldmath
\section{Tree level amplitude for $gg\to t\bar{t}h$}
\label{sec:app_tree_level}
\unboldmath

The amplitudes ${\cal A}_{0,s}$, ${\cal A}_{0,t}$, and ${\cal A}_{0,u}$
introduced in Section~\ref{eq:sigma_nlo} can be written as:
\begin{eqnarray}
\label{eq:a0_stu}
{\cal A}_{0,s}&=& ig_s^2\,g_{t\bar{t}h}\,\epsilon_\mu(q_1)\,\epsilon_\nu(q_2)\, 
\bar{u}_t {\cal A}_{0,s}^{\mu\nu}v_{\bar{t}}\,\,\,,\nonumber\\
{\cal A}_{0,t}&=& ig_s^2\,g_{t\bar{t}h}\,\epsilon_\mu(q_1)\,\epsilon_\nu(q_2)\, 
\bar{u}_t {\cal A}_{0,t}^{\mu\nu} v_{\bar{t}}\nonumber\,\,\,,\\
{\cal A}_{0,u}&=& ig_s^2\,g_{t\bar{t}h}\,\epsilon_\mu(q_1)\,\epsilon_\nu(q_2)\, 
\bar{u}_t {\cal A}_{0,u}^{\mu\nu} v_{\bar{t}}\,\,\,, 
\end{eqnarray}
where $g_{t\bar{t}h}\!=\!m_t/v$ is the top quark Yukawa coupling, with
$v\!=\!246$~GeV the SM Higgs boson vacuum expectation value, while
${\cal A}_{0,s}^{\mu\nu}$, ${\cal A}_{0,t}^{\mu\nu}$, and ${\cal
  A}_{0,u}^{\mu\nu}$ represent the total $s-$channel, $t-$channel, and
$u-$channel amplitudes, corresponding to the diagrams in
Fig.~\ref{fg:tree_gg}. More explicitly:
\begin{eqnarray}
\label{eq:A0_stu_munu}
{\cal A}_{0,s}^{\mu\nu}&=&{\cal A}_{0,s}^{(1)\mu\nu}+
{\cal A}_{0,s}^{(2)\mu\nu}\,\,\,,\nonumber\\
{\cal A}_{0,t}^{\mu\nu}&=&{\cal A}_{0,t}^{(1)\mu\nu}+
{\cal A}_{0,t}^{(2)\mu\nu}+{\cal A}_{0,t}^{(3)\mu\nu}\,\,\,,\nonumber\\
{\cal A}_{0,u}^{\mu\nu}&=&{\cal A}_{0,u}^{(1)\mu\nu}+
{\cal A}_{0,u}^{(2)\mu\nu}+{\cal A}_{0,u}^{(3)\mu\nu}\,\,\,,
\end{eqnarray}
where
\begin{eqnarray}
\label{eq:A0_stu_munu_123}
{\cal A}_{0,s}^{(1),\mu\nu}&=& 
\frac{1}{s}\frac{\not p_t+\not p_h+m_t}{[(p_t+p_h)^2-m_t^2]} 
\gamma_{\alpha} V^{\mu\nu\alpha}\,\,\,,\nonumber\\
{\cal A}_{0,s}^{(2),\mu\nu}&=& 
\frac{1}{s}\gamma_{\alpha}\frac{-\not p_t^\prime-\not p_h+m_t} 
{[(p_t^\prime+p_h)^2-m_t^2]} V^{\mu\nu\alpha}\,\,\,,\nonumber\\
{\cal A}_{0,t}^{(1),\mu\nu}&=& 
\frac{\not p_t+\not p_h+m_t}{[(p_t+p_h)^2-m_t^2]} \gamma^\mu 
\frac{\not q_2-\not p_t^\prime+m_t}{[(q_2-p_t^\prime)^2-m_t^2]} 
\gamma^\nu\,\,\,,\nonumber\\
{\cal A}_{0,t}^{(2),\mu\nu}&=& 
\gamma_\mu \frac{\not p_t-\not q_1+m_t}{[(p_t-q_1)^2-m_t^2]} 
\frac{\not q_2-\not p_t^\prime+m_t}{[(q_2-p_t^\prime)^2-m_t^2]} 
\gamma_\nu\,\,\,, \nonumber\\
{\cal A}_{0,t}^{(3),\mu\nu}&=& 
\gamma_\mu\frac{\not p_t-\not q_1+m_t}{[(p_t-q_1)^2-m_t^2]} 
\gamma^\nu\frac{-\not p_t^\prime-\not p_h+m_t}
{[(p_t^\prime+p_h)^2-m_t^2]}\,\,\,,\nonumber\\
{\cal A}_{0,u}^{(1),\mu\nu}&=&{\cal A}_{0,t}^{(1),\mu\nu} 
(\mu\leftrightarrow\nu,q_1\leftrightarrow q_2)\,\,\,,\nonumber\\ 
{\cal A}_{0,u}^{(2),\mu\nu}&=&{\cal A}_{0,t}^{(2),\mu\nu} 
(\mu\leftrightarrow\nu,q_1\leftrightarrow q_2)\,\,\,,\nonumber\\ 
{\cal A}_{0,u}^{(3),\mu\nu}&=&{\cal A}_{0,t}^{(3),\mu\nu} 
(\mu\leftrightarrow\nu,q_1\leftrightarrow q_2)\,\,\,, 
\end{eqnarray}
with
\[
V^{\mu\nu\alpha}=(q_1-q_2)^\alpha g^{\mu\nu}+(q_1+2 q_2)^\mu
g^{\nu\alpha}-(2 q_1+q_2)^\nu g^{\mu\alpha}\,\,\,,
\]
are the individual amplitudes for the $s-$channel, $t-$channel, and
$u-$channel diagrams in Fig.~\ref{fg:tree_gg}.
\section{Box and Pentagon integrals}
\label{sec:app_ir_integrals}

We label the various one-loop box and pentagon scalar and tensor
integrals appearing in the calculation of the ${\cal O}(\alpha_s)$
virtual corrections to
\[
g(q_1)+g(q_2)\rightarrow t(p_t)+\bar{t}(p_t^\prime)+h(p_h)
\]
according to the diagram where they are encountered. Moreover we
denote by $D0$, $D1^\mu$, $D2^{\mu\nu}$, and $D3^{\mu\nu\rho}$ the
scalar and tensor box integrals with one, two, and three tensor
indices, and by $E0$, $E1^\mu$, $E2^{\mu\nu}$, and $E3^{\mu\nu\rho}$
the analogous scalar and tensor pentagon integrals. With this
convention $D0_{D_{i,j}^{(k)}}$, for instance, is the scalar box
integral appearing in box diagram $D_{i,j}^{(k)}$, as labeled in
Fig.~\ref{fg:boxes_gg}.  The external momenta are labeled as shown
above, where $q_1,q_2$ are incoming and $p_t,p_t^\prime,p_h$ are
outgoing momenta with $q_1+q_2=p_t+p_t^\prime+p_h$.  It is convenient
to express our results in terms of the kinematic invariants of
Eq.~(\ref{eq:kinematic_invariants}) and:
\begin{eqnarray}
\label{eq:kinematic_invariants_all}
\omega_1 &=& (p_t+p_h)^2-m_t^2 \,\,\,,\nonumber\\
\omega_2 &=& (p_t^\prime+p_h)^2-m_t^2\,\,\,.
\end{eqnarray}
These kinematic invariants do not form a linearly independent set, but
are related by:
\begin{equation}
\tau_3=\sigma-\tau_1-\omega_2\,\,\,\,\,\,\,\,\,\,\mbox{and}
\,\,\,\,\,\,\,\,\,\,
\tau_4=\sigma-\tau_2-\omega_1\,\,\,.
\end{equation}
We also make frequent use of the shorthand notation
$\Lambda_a\!\equiv\!\ln(a/m_t^2)$ with $a=\sigma,\tau_i,\omega_i$.

In the following, we explicitly give only the box and pentagon
integrals that contain IR divergences. The IR divergences are
extracted using dimensional regularization with $d\!=\!4-2\epsilon$.
We only give results for integrals arising from the $s-$channel and
$t-$channel diagrams.  The integrals for the $u-$channel diagrams can
be obtained from the integrals of the corresponding $t-$channel
diagrams by exchanging $q_1\leftrightarrow q_2$, i.e. by exchanging
$\tau_1\leftrightarrow\tau_3$ and $\tau_2\leftrightarrow\tau_4$.  The
IR finite scalar integrals are evaluated by implementing the method
described in Ref.~\cite{Denner:1993kt} and are cross checked against
the FF package~\cite{vanOldenborgh:1990wn}.
\subsection{Box integrals}
\label{subsec:box_integrals}

The scalar and tensor box integrals arising in the computation of box diagram
$D_{i,j}^{(k)}$ are of the following form:
\begin{equation}
\label{eq:box_tensor}
D0_{D_{i,j}^{(k)}},D1_{D_{i,j}^{(k)}}^\mu,D2_{D_{i,j}^{(k)}}^{\mu\nu},D3_{D_{i,j}^{(k)}}^{\mu\nu\rho}=
\mu^{4-d}\int\frac{d^dk}{(2\pi)^d}
\frac{1,k^\mu,k^\mu k^\nu,k^\mu k^\nu k^\rho}{N_1 N_2 N_3 N_4} \,\,\,,
\end{equation}
where
\begin{eqnarray}
\label{eq:box_denominators} 
N_1= (k^2-m_0^2)\,\,\,&,& \, \, \, 
N_2= (k+p_1)^2-m_1^2\,\,\,,\nonumber \\ 
N_3= (k+p_1+p_2)^2-m_2^2\,\,\,&,& \, \, \, 
N_4= (k+p_1+p_2+p_3)^2-m_3^2\,\,\,,
\end{eqnarray} 
$p_1$, $p_2$, $p_3$, and $p_4\!=\!-p_1-p_2-p_3$ are the external (incoming)
momenta connected to the box topology, and $m_0$, $m_1$, $m_2$, and
$m_3$ are the masses of the propagators in the box loop.
We write the tensor integrals as a linear combination of the linearly
independent tensor structures built of the independent external
momenta $p_1^\mu$, $p_2^\mu$, and $p_3^\mu$ plus the metric tensor
$g^{\mu\nu}$. Our notation for the box tensor integrals is as
follows:
\begin{eqnarray}
\label{eq:d1d2d3}
D1^\mu&=&D_1^{(1)}p_1^\mu+D_1^{(2)}p_2^\mu+D_1^{(3)}p_3^\mu\,\,\,,
\nonumber\\
D2^{\mu\nu}&=& 
    D_2^{(0)}g^{\mu\nu}+D_2^{(11)}p_1^\mu p_1^\nu+D_2^{(22)}p_2^\mu p_2^\nu+
    D_2^{(33)}p_3^\mu p_3^\nu\nonumber\\
&+& D_2^{(12)}(p_1^\mu p_2^\nu+p_1^\nu p_2^\mu)+
    D_2^{(13)}(p_1^\mu p_3^\nu+p_1^\nu p_3^\mu)+
    D_2^{(23)}(p_2^\mu p_3^\nu+p_2^\nu p_3^\mu)\,\,\,,\nonumber\\
D3^{\mu\nu\rho}&=& 
    D_3^{(01)}(g^{\mu,\nu}p_1^\rho + \mbox{perm})+
    D_3^{(02)}(g^{\mu,\nu}p_2^\rho + \mbox{perm})+
    D_3^{(03)}(g^{\mu,\nu}p_3^\rho + \mbox{perm})\nonumber\\
&+& D_3^{(111)}p_1^\mu p_1^\nu p_1^\rho+
    D_3^{(222)}p_2^\mu p_2^\nu p_2^\rho+
    D_3^{(333)}p_3^\mu p_3^\nu p_3^\rho\nonumber\\
&+& D_3^{(112)}(p_1^\mu p_1^\nu p_2^\rho + \mbox{perm})+
    D_3^{(113)}(p_1^\mu p_1^\nu p_3^\rho + \mbox{perm})\nonumber\\
&+& D_3^{(221)}(p_2^\mu p_2^\nu p_1^\rho + \mbox{perm})+
    D_3^{(223)}(p_2^\mu p_2^\nu p_3^\rho + \mbox{perm})\nonumber\\
&+& D_3^{(331)}(p_3^\mu p_3^\nu p_1^\rho + \mbox{perm})+
    D_3^{(332)}(p_3^\mu p_3^\nu p_2^\rho + \mbox{perm})+
    D_3^{(123)}(p_1^\mu p_2^\nu p_3^\rho + \mbox{perm})\,\,\,,
\nonumber\\
\end{eqnarray}
where ``$+\mbox{perm}$'' indicates that the sum over all possible
permutations of the tensor indices is understood. In the following we
will give the full structure of the scalar box integrals, including
both pole and finite parts, while for the corresponding tensor
integrals we will only give the IR pole parts, since they can be of
interest in checking the IR structure of the virtual cross section.
We will write the pole part of each tensor integral coefficient as
\begin{eqnarray}
\label{eq:delta_ir}
D_i^{(j)}|_{IR-pole}&=&\frac{i}{16\pi^2}{\cal N}_t
  \Delta_{IR}(D_i^{(j)})\,\,\,,\nonumber\\
D_i^{(jk)}|_{IR-pole}&=&\frac{i}{16\pi^2}{\cal N}_t
  \Delta_{IR}(D_i^{(jk)})\,\,\,,\nonumber\\
D_i^{(jkl)}|_{IR-pole}&=&\frac{i}{16\pi^2}{\cal N}_t
\Delta_{IR}(D_i^{(jkl)})\,\,\,,
\end{eqnarray}
where ${\cal N}_t$ is defined in Eq.~(\ref{eq:nsnt}), and give for
each box integral the non zero $\Delta_{IR}(D_i^{(j)})$,
$\Delta_{IR}(D_i^{(jk)})$, and $\Delta_{IR}(D_i^{(jkl)})$
coefficients.

\subsubsection{Box scalar integral $D0_{B_{2,s}^{(1,2)}}$}
\label{subsubsec:d0_b2}

The scalar integral appearing in diagram $B_{2,s}^{(1)}$ can be
parameterized according to Eq.~(\ref{eq:box_tensor}) with:
\begin{eqnarray}
\label{eq:d0_b2_1_denominators} 
N_1= k^2\,\,\,&,& \, \, \, 
N_2= (k+p_t)^2-m_t^2\,\,\,,\nonumber \\ 
N_3= (k+p_t+p_h)^2-m_t^2\,\,\,&,& \, \, \, 
N_4= (k-p_t^\prime)^2-m_t^2\,\,\,.
\end{eqnarray} 
$D0_{B_{2,s}^{(2)}}$ is obtained from $D0_{B_{2,s}^{(1)}}$ by
exchanging $p_t\leftrightarrow p_t^\prime$.  This integral also
appears in the $q\bar{q}\to t\bar{t}h$ calculation and has already
been presented in Ref.~\cite{Reina:2001bc}.  We repeat it here for
completeness.  

The part of $D0_{B_{2,s}^{(1)}}$ which contributes to the amplitude
squared is of the form:
\begin{eqnarray} 
\label{eq:d0_b2_1_result}
D0_{B_{2,s}^{(1)}}&=&\frac{i}{16\pi^2}{\cal N}_t\,
\frac{1}{\omega_1(\sigma-\omega_1-\omega_2+M_h^2)}
\left(\frac{X_{-1}}{\epsilon}+X_0\right)\,\,\,,
\end{eqnarray}
where ${\cal N}_t$ is defined in Eq.~(\ref{eq:nsnt}). The pole
part $X_{-1}$ is:
\begin{equation}
X_{-1}=-\frac{1}{\beta_{t\bar{t}}}
\ln\left(\frac{1+\beta_{t\bar{t}}}{1-\beta_{t\bar{t}}}\right)\,\,\,,
\end{equation}
where $\beta_{t\bar{t}}$ is given in Eq.~(\ref{eq:stt_betatt}).  The
finite part $X_0$ can be calculated using
Ref.~\cite{Beenakker:1990jr}.  All tensor box integrals associated to
$B_{2,s}^{(1)}$ and $B_{2,s}^{(2)}$ are IR finite.

\subsubsection{Box scalar integral $D0_{B_{7,t}^{(1,2)}}$}
\label{subsubsec:d0_b7}

The scalar integral appearing in diagram $B_{7,t}^{(1)}$,
$D0_{B_{7,t}^{(1)}}$, can be parameterized according to
Eq.~(\ref{eq:box_tensor}) with:
\begin{eqnarray}
\label{eq:d0_b7t_1_denominators}
N_1=k^2\,\,\,&,& \, \, \, 
N_2= (k+q_1)^2\,\,\,,\nonumber \\
N_3= (k+q_1-p_t)^2-m_t^2\,\,\, &,& \, \, \, 
N_4= (k+q_1-p_t-p_h)^2-m_t^2\,\,\, .
\end{eqnarray}
The part of $D0_{B_{7,t}^{(1)}}$ which contributes to the virtual
amplitude squared is of the form:
\begin{equation} 
\label{eq:d0_b7t_1_all}
D0_{B_{7,t}^{(1)}}=\frac{i}{16\pi^2}{\cal N}_t\left(-\frac{1}
{\omega_1\tau_1}\right)
\left(\frac{X_{-2}}{\epsilon^2}+\frac{X_{-1}}{\epsilon}+X_0\right)\,\,\,,
\end{equation}
where the coefficients $X_{-2}$, $X_{-1}$, and $X_{0}$ are given by:
\begin{eqnarray}
\label{eq:d0_b7t_1_coefficients}
X_{-2}&=&\frac{1}{2}\,\,\,, \nonumber \\
X_{-1}&=& \ln\left(\frac{\tau_2 m_t^2}{\omega_1\tau_1}\right)\,\,\,,
\nonumber \\
X_0&=& Re\left\{-\frac{5}{6}\pi^2+
 \ln^2\left(\frac{\omega_1}{m_t^2}\right)+
 \ln^2\left(\frac{\tau_1}{m_t^2}\right)-
 \ln^2\left(\frac{\tau_2}{m_t^2}\right)\right.\nonumber\\
 &+&2\,\ln\left(\frac{\omega_1+\tau_2}{\tau_1}\right)
    \ln\left(\frac{\tau_2}{\omega_1}\right)+
 2\,\ln\left(\frac{\tau_1-\tau_2}{\omega_1}\right)
    \ln\left(\frac{\tau_2}{\tau_1}\right)\nonumber\\
 &-&2\left.\,\mbox{Li}_2\left(\frac{\tau_1-\tau_2-\omega_1}{\tau_1}\right)-
 2\,\mbox{Li}_2\left(\frac{\omega_1+\tau_2-\tau_1}{\omega_1}\right)+
 2\,\mbox{Li}_2\left(\frac{\tau_2(\omega_1+\tau_2-\tau_1)}{\omega_1\tau_1}
\right)-I_0\right\}\,\,\,,\nonumber\\
\end{eqnarray}
with
\begin{eqnarray}
\label{eq:d0_b7t_1_i0}
I_0&=& 
  \ln\left(\frac{\tau_1}{\tau_2}\right)\ln\left(\frac{M_h^2}{m_t^2}\right)+
\left\{-\mbox{Li}_2\left(\frac{1}{\lambda_+}\right)+
\ln\left(\frac{\tau_1}{\tau_2}\right)
\ln\left(\frac{-\tau_2-\lambda_+(\tau_1-\tau_2)}{\tau_1-\tau_2}\right)\right.
\nonumber\\
&-&\left.
\mbox{Li}_2\left(\frac{\tau_1}{\lambda_+(\tau_1-\tau_2)+\tau_2}\right)+
\mbox{Li}_2\left(\frac{\tau_2}{\lambda_+(\tau_1-\tau_2)+\tau_2}\right)+
(\lambda_+\leftrightarrow\lambda_-)\right\}\,\,\,,\nonumber\\
\end{eqnarray}
and
\begin{equation}
\label{eq:d0_b7t_1_lambdapm}
\lambda_\pm=\frac{1}{2}\left(1\pm \sqrt{1-\frac{4m_t^2}{M_h^2}}\right)
\,\,\,.
\end{equation}
The tensor integrals associated with $B_{7,t}^{(1)}$ also contain IR
divergences. Using the notation introduced in Eqs.~(\ref{eq:d1d2d3})
and (\ref{eq:delta_ir}), only the following coefficients of
$D1_{B_{7,t}^{(1)}}^{\mu}$:
\begin{eqnarray}
\label{eq:d1_b7t_1_irpoles}
\Delta_{IR}(D_1^{(1)})&=&
\frac{1}{2}\frac{1}{\tau_1\omega_1}\frac{1}{\epsilon^2}+
\frac{1}{\tau_1\omega_1}\left[
-\Lambda_{\tau_1}+
\frac{\tau_2}{\tau_2+\omega_1}\left(\Lambda_{\tau_2}-\Lambda_{\omega_1}\right)\right]
\frac{1}{\epsilon} \; ,
\end{eqnarray}
and of $D2_{B_{7,t}^{(1)}}^{\mu\nu}$:
\begin{eqnarray}
\label{eq:d2_b7t_1_irpoles}
\Delta_{IR}(D_2^{(11)})&=&
-\frac{1}{2}\frac{1}{\tau_1\omega_1}\frac{1}{\epsilon^2}
+\frac{1}{\tau_1\omega_1}\left[
\Lambda_{\tau_1}
-\frac{\tau_2^2}{(\tau_2+\omega_1)^2}\left(\Lambda_{\tau_2}-\Lambda_{\omega_1}\right)
-\frac{\omega_1}{\tau_2+\omega_1}\right]\frac{1}{\epsilon} \; ,\nonumber\\
\end{eqnarray}
are IR divergent.

$D0_{B_{7,t}^{(2)}}$ and the corresponding tensor integrals are
obtained from $D0_{B_{7,t}^{(1)}}$ by exchanging $q_1\leftrightarrow
q_2$ and $p_t\leftrightarrow p_t^\prime$, i.e. by exchanging
$\tau_1\leftrightarrow\tau_2$ and $\omega_1\leftrightarrow\omega_2$ in
Eqs.~(\ref{eq:d0_b7t_1_all})-(\ref{eq:d2_b7t_1_irpoles}).

\subsubsection{Box scalar integral $D0_{B_{8,t}^{(1,2)}}$}
\label{subsubsec:d0_b8}

The scalar box integral appearing in diagram $B_{8,t}^{(1)}$,
$D0_{B_{8,t}^{(1)}}$, can be parameterized according to
Eq.~(\ref{eq:box_tensor}) with:
\begin{eqnarray}
\label{eq:i_b8t_1_denominators}
N_1=k^2\,\,\,&,& \, \, \, 
N_2= (k+q_1)^2\,\,\,,\nonumber \\
N_3= (k+q_1+q_2)^2\,\,\,&,& \, \, \,
N_4= (k+q_1+q_2-p_t^\prime)^2-m_t^2\,\,\, .
\end{eqnarray}
The part of $D0_{B_{8,t}^{(1)}}$ which contributes to the virtual
amplitude squared is given by:
\begin{eqnarray}
\label{eq:d0_b8t_1_all}
D0_{B_{8,t}^{(1)}}&=&\frac{i}{16\pi^2}{\cal N}_t
\left(-\frac{1}{\sigma\tau_2}\right)
\left(\frac{X_{-2}}{\epsilon^2}+\frac{X_{-1}}{\epsilon}+X_0\right)\,\,\,,
\end{eqnarray}
where ${\cal N}_t$ is defined in Eq.~(\ref{eq:nsnt}), and the
coefficients $X_{-2}$, $X_{-1}$, and $X_{0}$ are given by:
\begin{eqnarray}
\label{eq:d0_b8t_1_coefficients}
X_{-2}&=&\frac{3}{2}\,\,\,, \nonumber \\
X_{-1}&=&\ln\left(\frac{\omega_1 m_t^4}{\sigma \tau_2^2}\right)\,\,\,, \nonumber \\
X_0&=&
2\ln\left(\frac{\tau_2}{m_t^2}\right)
\ln\left(\frac{\sigma}{m_t^2}\right)
-\ln^2\left(\frac{\omega_1}{m_t^2}\right)
 -2\,\mbox{Li}_2\left(1+\frac{\omega_1}{\tau_2}\right)+\frac{\pi^2}{3}
\,\,\,.
\end{eqnarray}
The tensor integrals associated with $B_{8,t}^{(1)}$ also contain IR
divergences. Using the notation introduced in Eqs.~(\ref{eq:d1d2d3})
and (\ref{eq:delta_ir}), only the following tensor coefficients of
$D1_{B_{8,t}^{(1)}}^{\mu}$:
\begin{eqnarray}
\label{eq:d1_b8t_1_irpoles}
\Delta_{IR}(D_1^{(1)})&=&\frac{3}{2}\frac{1}{\sigma\tau_2}\frac{1}{\epsilon^2}
-\frac{1}{\sigma\tau_2}\left[\Lambda_\sigma+\Lambda_{\tau_2}+
\frac{\omega_1}{\tau_2+\omega_1}\left(\Lambda_{\tau_2}-\Lambda_{\omega_1}\right)
\right]\frac{1}{\epsilon}\,\,\,,\nonumber\\
\Delta_{IR}(D_1^{(2)})&=&\frac{1}{2}\frac{1}{\sigma\tau_2}\frac{1}{\epsilon^2}
-\frac{1}{\sigma\tau_2}\Lambda_{\tau_2}\frac{1}{\epsilon}\,\,\,,
\end{eqnarray}
of $D2_{B_{8,t}^{(1)}}^{\mu\nu}$:
\begin{eqnarray}
\label{eq:d2_b8t_1_irpoles}
\Delta_{IR}(D_2^{(11)})&=&-\frac{3}{2}\frac{1}{\sigma\tau_2}\frac{1}{\epsilon^2}
+\frac{1}{\sigma\tau_2}\left[
-\frac{\tau_2}{\tau_2+\omega_1}+\Lambda_\sigma+\Lambda_{\tau_2}+
\frac{\omega_1^2}{(\tau_2+\omega_1)^2}
\left(\Lambda_{\tau_2}-\Lambda_{\omega_1}\right)\right]\frac{1}{\epsilon}
\,\,\,,\nonumber \\
\Delta_{IR}(D_2^{(12)})&=&-\frac{1}{2}\frac{1}{\sigma\tau_2}\frac{1}{\epsilon^2}
+\frac{1}{\sigma\tau_2}\Lambda_{\tau_2}\frac{1}{\epsilon}\,\,\,,\nonumber\\
\Delta_{IR}(D_2^{(22)})&=&-\frac{1}{2}\frac{1}{\sigma\tau_2}\frac{1}{\epsilon^2}
+\frac{1}{\sigma\tau_2}(-1+\Lambda_{\tau_2})\frac{1}{\epsilon}\; ,
\end{eqnarray}
and of
$D3_{B_{8,t}^{(1)}}^{\mu\nu\rho}$:
\begin{eqnarray}
\label{eq:d3_b8t_1_irpoles}
\Delta_{IR}(D_3^{(111)})&=&\frac{3}{2}\frac{1}{\sigma\tau_2}\frac{1}{\epsilon^2}
+\frac{1}{2\sigma\tau_2}\left[
\frac{3\tau_2}{\tau_2+\omega_1}+\frac{2\tau_2\omega_1}{(\tau_2+\omega_1)^2}
-2\Lambda_\sigma-2\Lambda_{\tau_2}\right.\nonumber\\
&&\left.-\frac{2\omega_1^2}{(\tau_2+\omega_1)^3}
\left(\Lambda_{\tau_2}-\Lambda_{\omega_1}\right)\right]\frac{1}{\epsilon}
\,\,\,,\nonumber\\
\Delta_{IR}(D_3^{(112)})&=&\frac{1}{2}\frac{1}{\sigma\tau_2}\frac{1}{\epsilon^2}
-\frac{1}{\sigma\tau_2}\Lambda_{\tau_2}\frac{1}{\epsilon}\,\,\,,\nonumber\\
\Delta_{IR}(D_3^{(221)})&=&\frac{1}{2}\frac{1}{\sigma\tau_2}\frac{1}{\epsilon^2}
+\frac{1}{\sigma\tau_2}\left(1-\Lambda_{\tau_2}\right)\frac{1}{\epsilon}
\,\,\,, \nonumber \\
\Delta_{IR}(D_3^{(222)})&=&\frac{1}{2}\frac{1}{\sigma\tau_2}\frac{1}{\epsilon^2}
+\frac{1}{\sigma\tau_2}\left(\frac{3}{2}-\Lambda_{\tau_2}\right)\frac{1}{\epsilon}
\; ,
\end{eqnarray}
are IR divergent.

$D0_{B_{8,t}^{(2)}}$ as well as the corresponding tensor integrals can
be obtained from $D0_{B_{8,t}^{(1)}}$ by exchanging
$q_1\leftrightarrow q_2$ and $p_t^\prime\leftrightarrow p_t$, i.e. by
exchanging $\tau_1\leftrightarrow\tau_2$ and
$\omega_1\leftrightarrow\omega_2$ in
Eqs.~(\ref{eq:d0_b8t_1_all})-(\ref{eq:d3_b8t_1_irpoles}).

\subsubsection{Box scalar integral $D0_{B_{10,t}^{(1,2)}}$}
\label{subsubsec:d0_b10}

The scalar box integral appearing in diagram $B_{10,t}^{(1)}$,
$D0_{B_{10,t}}^{(1)}$, can be parameterized according to
Eq.~\ref{eq:box_tensor} with:
\begin{eqnarray}
\label{eq:i_b10t_1_denominators}
N_1=k^2\,\,\,&,& \,\,\,
N_2= (k+q_1)^2\,\,\,,\nonumber \\
N_3= (k+q_1-p_t^\prime)^2-m_t^2\,\,\,&,& \,\,\,
N_4= (k+q_1+q_2-p_t^\prime)^2-m_t^2\,\,\,.
\end{eqnarray}
The part of $D0_{B_{10,t}^{(1)}}$ which contributes to the virtual
amplitude squared is given by:
\begin{eqnarray}
\label{eq:d0_b10t_1_all}
D0_{B_{10,t}}^{(1)}&=&\frac{i}{16\pi^2}{\cal N}_t
\left(\frac{1}{\tau_2\tau_4}\right)
\left(\frac{X_{-2}}{\epsilon^2}+\frac{X_{-1}}{\epsilon}+X_0\right)\,\,\,,
\end{eqnarray}
where the coefficients $X_{-2}$, $X_{-1}$, and $X_{0}$ are given by:
\begin{eqnarray}
\label{eq:d0_b10t_1_coefficients}
X_{-2}&=&\frac{1}{2}\,\,\,, \nonumber \\
X_{-1}&=&\ln\left(\frac{\omega_1}{\tau_4}\right)-
         \ln\left(\frac{\tau_2}{m_t^2}\right)\,\,\,, \nonumber \\
X_0&=& {\cal R}e\left\{
 \ln^2\left(\frac{\tau_2}{m_t^2}\right)
+\ln^2\left(\frac{\tau_4}{m_t^2}\right)
-\ln^2\left(\frac{\omega_1}{m_t^2}\right)+\frac{3}{2}\pi^2\right.\nonumber\\
&+&2\ln\left(\frac{\tau_2+\omega_1}{\tau_4}\right)
  \ln\left(\frac{\tau_4}{\tau_2+\tau_4+\omega_1}\right)
+2\ln\left(\frac{\tau_4+\omega_1}{\tau_2}\right)
  \ln\left(\frac{\tau_2}{\tau_2+\tau_4+\omega_1}\right)\nonumber\\
&-&2\left.\,\mbox{Li}_2\left(\frac{\tau_2+\tau_4+\omega_1}{\tau_4}\right)
-2\,\mbox{Li}_2\left(\frac{\tau_2+\tau_4+\omega_1}{\tau_2}\right)
-2\,\mbox{Li}_2\left(\frac{(\tau_2+\omega_1)(\tau_4+\omega_1)}
                        {\tau_2\tau_4}\right)\right\} \,\,\,.\nonumber\\ 
\end{eqnarray}
The tensor integrals associated with $B_{10,t}^{(1)}$ also contain IR
divergences. Using the notation introduced in Eqs.~(\ref{eq:d1d2d3})
and (\ref{eq:delta_ir}), the only IR divergent tensor coefficients of
$D1_{B_{10,t}^{(1)}}^{\mu}$:
\begin{eqnarray}
\label{eq:d1_b10t_1_irpoles}
\Delta_{IR}(D_1^{(1)})&=&-\frac{1}{2}\frac{1}{\tau_2\tau_4}\frac{1}{\epsilon^2}
+\frac{1}{\tau_2\tau_4(\tau_2+\omega_1)}\left[
(\tau_2+\omega_1)\Lambda_{\tau_4}
+\omega_1\left(\Lambda_{\tau_2}-\Lambda_{\omega_1}\right)\right]\frac{1}{\epsilon}
\,\,\,,\nonumber\\
\end{eqnarray}
of $D2_{B_{10,t}^{(1)}}^{\mu\nu}$:
\begin{eqnarray}
\label{eq:d2_b10t_1_irpoles}
\Delta_{IR}(D_2^{(11)})&=&\frac{1}{2}\frac{1}{\tau_2\tau_4}\frac{1}{\epsilon^2}
+\frac{1}{\tau_2\tau_4(\tau_2+\omega_1)^2}\left[
\tau_2(\tau_2+\omega_1)-(\tau_2+\omega_1)^2\Lambda_{\tau_4}
-\omega_1^2\left(\Lambda_{\tau_2}-\Lambda_{\omega_1}\right)\right]
\frac{1}{\epsilon}\; ,
\nonumber \\ 
\end{eqnarray}
and of  $D3_{B_{10,t}^{(1)}}^{\mu\nu\rho}$:
\begin{eqnarray}
\label{eq:d3_b10t_1_irpoles}
\Delta_{IR}(D_3^{(111)})&=&-\frac{1}{2}\frac{1}{\tau_2\tau_4}\frac{1}{\epsilon^2}
-\frac{1}{2}\frac{1}{\tau_2\tau_4(\tau_2+\omega_1)^3}\left[
-2\omega_1^2\left(\Lambda_{\tau_2}-\Lambda_{\omega_1}\right)
-2(\tau_2+\omega_1)^3\Lambda_{\tau_4}\right.\nonumber\\
&&+\left.3\tau_2(\tau_2+\omega_1)^2
+2\tau_2\omega_1(\tau_2+\omega_1)\right]\frac{1}{\epsilon} \;,
\end{eqnarray}
are IR divergent.

$D0_{B_{10,t}^{(2)}}$ can be obtained from $D0_{B_{10,t}^{(1)}}$ by
exchanging $p_t^\prime\leftrightarrow p_t$, i.e. by exchanging
$\tau_1\leftrightarrow\tau_4$, $\tau_2\leftrightarrow\tau_3$, and
$\omega_1\leftrightarrow\omega_2$ in Eqs~(\ref{eq:d0_b10t_1_all}) and
(\ref{eq:d3_b10t_1_irpoles}). 

\subsection{Pentagon integrals}
\label{subsec:pentagon_integrals}

The scalar and tensor pentagon integrals originating from the generic
pentagon diagram $P_{i,j}$ in Fig.~\ref{fg:pentagons_gg} are of the
form:
\begin{equation}
\label{eq:pentagon_tensor}
E0_{P_{i,j}},E1_{P_{i,j}}^\mu,E2_{P_{i,j}}^{\mu\nu},
E3_{P_{i,j}}^{\mu\nu\rho}=
\mu^{4-d}\int\frac{d^dk}{(2\pi)^d}
\frac{1,k^\mu,k^\mu k^\nu,k^\mu k^\nu k^\rho}{N_1 N_2 N_3 N_4 N_5} \,\,\,,
\end{equation}
where
\begin{eqnarray}
\label{eq:pentagon_denominators} 
N_1= (k^2-m_0^2)\,\,\, &,&  \,\,\,
N_2= (k+p_1)^2-m_1^2\,\,\,,\nonumber\\
N_3= (k+p_1+p_2)^2-m_2^2\,\,\,&,&\,\,\,
N_4= (k+p_1+p_2+p_3)^2-m_3^2\,\,\,,\nonumber\\
N_5= (k+p_1+p_2+p_3+p_4)^2-m_4^2\,\,\,&,&
\end{eqnarray} 
$p_1$, $p_2$, $p_3$, $p_4$, and $p_5\!=\!-p_1-p_2-p_3-p_4$ are the
external (incoming) momenta connected to the pentagon topology, while
$m_0$, $m_1$, $m_2$, $m_3$, and $m_4$ are the masses of the
propagators in the pentagon loop.

The scalar pentagon integrals are evaluated as a linear combination of
five scalar box integrals, using the technique originally proposed in
Ref.~\cite{Bern:1993em,Bern:1994kr}. In particular, we  use:
\begin{equation}
\label{eq:pent_sum_fiveboxes}
E0_{P_{i,j}}=-\frac{1}{2}\sum_{k=1}^{5}c_kD0_{P_{i,j}}^{(k)}\,\,\,,
\end{equation}
where each scalar box integral $D0_{P_{i,j}}^{(k)}$ can be obtained from
the scalar pentagon integral $E0_{P_{i,j}}$ in
Eq.~(\ref{eq:pentagon_tensor}) by dropping one of the
internal propagators. The coefficients $c_k$ are given by:
\begin{equation}
\label{eq:pentagon_ci}
c_k=\sum_{l=1}^{5}S_{kl}^{-1}\,\,\,,
\end{equation}
where $S_{kl}$ is the symmetric matrix:
\begin{equation}
S_{kl}=\frac{1}{2}\left(M_k^2+M_l^2-p_{kl}^2\right)\,\,\,,
\end{equation}
built out of the internal propagator masses $M_k$ and $M_l$ and the
linear combination of external momenta
$p_{kl}^\mu\!=\!p_k^\mu+\ldots+p_{l-1}^\mu$ ($k,l=1,\ldots,5$). A
thorough explanation of this method is given in
Ref.~\cite{Reina:2001bc,Bern:1993em,Bern:1994kr}, to which we refer
for more details.

We write the tensor pentagon integrals as a linear combination of the
linearly independent tensor structures built of the external momenta
$p_1^\mu$, $p_2^\mu$, $p_3^\mu$, and $p_4^\mu$, which in $d=4$
constitute a complete basis. Our notation for the pentagon tensor integrals
is as follows:
\begin{eqnarray}
\label{eq:e1e2e3}
E1^\mu&=&
E_1^{(1)}p_1^\mu+E_1^{(2)}p_2^\mu+E_1^{(3)}p_3^\mu+E_1^{(4)}p_4^\mu
\nonumber\,\,\,,\\
E2^{\mu\nu}&=& 
    E_2^{(11)}p_1^\mu p_1^\nu+E_2^{(22)}p_2^\mu p_2^\nu+
    E_2^{(33)}p_3^\mu p_3^\nu+E_2^{(44)}p_4^\mu p_4^\nu\nonumber\\
&+& E_2^{(12)}(p_1^\mu p_2^\nu+p_1^\nu p_2^\mu)+
    E_2^{(13)}(p_1^\mu p_3^\nu+p_1^\nu p_3^\mu)+
    E_2^{(14)}(p_1^\mu p_4^\nu+p_1^\nu p_4^\mu)\nonumber\\
&+& E_2^{(23)}(p_2^\mu p_3^\nu+p_2^\nu p_3^\mu)+
    E_2^{(24)}(p_2^\mu p_4^\nu+p_2^\nu p_4^\mu)+
    E_2^{(34)}(p_3^\mu p_4^\nu+p_3^\nu p_4^\mu)\,\,\,,\nonumber\\
E3^{\mu\nu\rho}&=& 
    E_3^{(111)}p_1^\mu p_1^\nu p_1^\rho+E_3^{(222)}p_2^\mu p_2^\nu p_2^\rho+
    E_3^{(333)}p_3^\mu p_3^\nu p_3^\rho+E_3^{(444)}p_4^\mu p_4^\nu p_4^\rho
\nonumber\\
&+& E_3^{(112)}(p_1^\mu p_1^\nu p_2^\rho + \mbox{perm})+
    E_3^{(113)}(p_1^\mu p_1^\nu p_3^\rho + \mbox{perm})+
    E_3^{(114)}(p_1^\mu p_1^\nu p_4^\rho + \mbox{perm})\nonumber\\
&+& E_3^{(221)}(p_2^\mu p_2^\nu p_1^\rho + \mbox{perm})+
    E_3^{(223)}(p_2^\mu p_2^\nu p_3^\rho + \mbox{perm})+
    E_3^{(224)}(p_2^\mu p_2^\nu p_4^\rho + \mbox{perm})\nonumber\\
&+& E_3^{(331)}(p_3^\mu p_3^\nu p_1^\rho + \mbox{perm})+
    E_3^{(332)}(p_3^\mu p_3^\nu p_2^\rho + \mbox{perm})+
    E_3^{(334)}(p_3^\mu p_3^\nu p_4^\rho + \mbox{perm})\nonumber\\
&+& E_3^{(441)}(p_4^\mu p_4^\nu p_1^\rho + \mbox{perm})+
    E_3^{(442)}(p_4^\mu p_4^\nu p_2^\rho + \mbox{perm})+
    E_3^{(443)}(p_4^\mu p_4^\nu p_4^\rho + \mbox{perm})\nonumber\\
&+& E_3^{(123)}(p_1^\mu p_2^\nu p_3^\rho + \mbox{perm})+
    E_3^{(124)}(p_1^\mu p_2^\nu p_4^\rho + \mbox{perm})+
    E_3^{(134)}(p_1^\mu p_3^\nu p_4^\rho + \mbox{perm})\nonumber\\
&+& E_3^{(234)}(p_2^\mu p_3^\nu p_4^\rho + \mbox{perm})\nonumber\,\,\,.\\
\end{eqnarray}
The calculation of $gg\to t\bar{t}h$ involves the six pentagon
structures illustrated in Fig.~\ref{fg:pentagons_gg}. For each of them
we will give in the following the IR pole parts of the corresponding
scalar integrals, as well as the coefficient $c_k$ (in terms of the
$S_{kl}$ matrix) and the IR singular box scalar integrals
$D0_{P_{i,j}}^{(k)}$ out of which they can be calculated. We will
moreover list the IR pole parts of the corresponding tensor integral
coefficients, since they may be of interest in checking the IR
structure of the virtual cross section.  We will write the pole part
of each tensor integral coefficient as
\begin{eqnarray}
\label{eq:delta_ir_pent}
E_i^{(j)}|_{IR-pole}&=&\frac{i}{16\pi^2}{\cal N}_t
  \Delta_{IR}(E_i^{(j)})\,\,\,,\nonumber\\
E_i^{(jk)}|_{IR-pole}&=&\frac{i}{16\pi^2}{\cal N}_t
  \Delta_{IR}(E_i^{(jk)})\,\,\,,\nonumber\\
E_i^{(jkl)}|_{IR-pole}&=&\frac{i}{16\pi^2}{\cal N}_t
\Delta_{IR}(E_i^{(jkl)})\,\,\,,
\end{eqnarray}
where ${\cal N}_t$ is defined in Eq.~(\ref{eq:nsnt}), and give for
each pentagon integral the non zero $\Delta_{IR}(E_i^{(j)})$,
$\Delta_{IR}(E_i^{(jk)})$, and $\Delta_{IR}(E_i^{(jkl)})$
coefficients.

As in Section~\ref{subsec:box_integrals} we express our results in
terms of the kinematic invariants $\sigma,\tau_i,\omega_i$ of
Eqs.~(\ref{eq:kinematic_invariants})
and (\ref{eq:kinematic_invariants_all}), and $\beta_{t\bar t}$ of
Eq.~(\ref{eq:stt_betatt}).

\subsubsection{Pentagon scalar integral $E0_{P_{1,t}}$}
\label{subsubsec:e0_p1t}

The pentagon scalar integral arising from diagram $P_{1,t}$ coincides
with $E0_{P_1}$ of the $q\bar{q}\to t\bar{t}h$ calculation
of Ref.~\cite{Reina:2001bc}, 
and can be parameterized according to
Eq.~(\ref{eq:pentagon_tensor}) with:
\begin{eqnarray}
\label{eq:i_p1t_denominators}
N_1= k^2 \,\,\,\,,\,\,\,\,N_2= (k+q_1)^2\,\,\,&,&\,\,\,
N_3= (k+q_1+q_2)^2\,\,\,,\nonumber\\
N_4= (k+q_1+q_2-p_t^\prime)^2-m_t^2\,\,\,&,&\,\,\,
N_5= (k+q_1+q_2-p_t^\prime-p_h)^2-m_t^2\,\,\,.
\end{eqnarray}
We summarize here for completeness the results obtained in
Ref.~\cite{Reina:2001bc}.  The $c_k$ ($k\!=\!1,\ldots,5$) coefficients
in Eq.~(\ref{eq:pent_sum_fiveboxes}) are obtained, according to
Eq.~(\ref{eq:pentagon_ci}), as:
\begin{equation}
\label{eq:e0_p1t_coefficients}
c_k=\sum_{l=1}^{5}\left[S(P_{1,t})\right]_{kl}^{-1} \,\,\,,
\end{equation}
where
\begin{equation}
\label{eq:p1t_s_matrix}
S(P_{1,t})=\frac{1}{2}\left(
\begin{array}{c c c c c}
0 & 0 & -\sigma & -\omega_1 & 0\\
0 & 0 & 0 & \tau_2 & \tau_1 \\
-\sigma & 0 & 0 & 0 & -\omega_2 \\
-\omega_1 & \tau_2 & 0 & 2m_t^2 & (2 m_t^2-M_h^2) \\
0 & \tau_1 & -\omega_2 & (2 m_t^2-M_h^2) & 2 m_t^2
\end{array}
\right)\,\,\,.
\end{equation}
The part of $E0_{P_{1,t}}$ that contributes to the virtual amplitude
squared can be written as:
\begin{equation}
E0_{P_{1,t}}=\frac{i}{16\pi^2}{\cal N}_t\left[\frac{X_{-2}}{\epsilon^2}
+\frac{X_{-1}}{\epsilon}+X_0\right]\,\,\,,
\end{equation}
where $X_{-2}$, $X_{-1}$, and $X_0$ are obtained using
Eqs.~(\ref{eq:pent_sum_fiveboxes}), (\ref{eq:e0_p1t_coefficients}),
(\ref{eq:p1t_s_matrix}) and the results for the $D0^{(k)}_{P_{1,t}}$
integrals presented in the following. The expressions for $X_{-2}$ and
$X_{-1}$ have the following form:
\begin{eqnarray}
\label{eq:e0_p1t_poles}
X_{-2} &=& \frac{1}{2\sigma}\left(-\frac{1}{\omega_1\tau_1}-
\frac{1}{\omega_2\tau_2}+\frac{2}{\tau_1\tau_2}\right)\,\,\,,\nonumber\\
X_{-1} &=& \frac{1}{\sigma\tau_1\tau_2}\left(-\Lambda_\sigma+
\Lambda_{\omega_1}+\Lambda_{\omega_2}-\Lambda_{\tau_1}-\Lambda_{\tau_2}\right)
+\frac{1}{\sigma\tau_2\omega_2}\left(\Lambda_{\tau_2}-\Lambda_{\tau_1}+
\Lambda_{\omega_2}\right)+\nonumber\\
&+&\frac{1}{\sigma\tau_1\omega_1}\left(\Lambda_{\tau_1}-\Lambda_{\tau_2}+
\Lambda_{\omega_1}\right)\,\,\,.
\end{eqnarray} 
The tensor integrals associated with $P_{1,t}$ contain IR divergences.
Using the notation introduce in Eqs.~(\ref{eq:e1e2e3}) and
(\ref{eq:delta_ir_pent}), only the following coefficients of
$E1_{P_{1,t}}^{\mu}$:
\begin{eqnarray}
\label{eq:e1_p1t_irpoles}
\Delta_{IR}(E_1^{(1)})&=&
\frac{1}{2\sigma\tau_2}\left(\frac{1}{\omega_2}-\frac{2}{\tau_1}\right)
\frac{1}{\epsilon^2}
+\frac{1}{\sigma}\left[
\frac{1}{\tau_1\tau_2}\left(\Lambda_\sigma+\Lambda_{\tau_1}+\Lambda_{\tau_2}
-\Lambda_{\omega_1}-\Lambda_{\omega_2}\right)\right.\nonumber\\
&&\left.+\frac{1}{\omega_2\tau_2}\left(\Lambda_{\tau_1}-\Lambda_{\tau_2}-
\Lambda_{\omega_2}\right)\right]\frac{1}{\epsilon}\,\,\,,\nonumber\\
\Delta_{IR}(E_1^{(2)})&=&\frac{1}{2\sigma\tau_2\omega_2}\frac{1}{\epsilon^2}
+\frac{1}{\sigma\tau_2\omega_2}\left(\Lambda_{\tau_1}-\Lambda_{\tau_2}
-\Lambda_{\omega_2}\right)\frac{1}{\epsilon}\,\,\,, \nonumber\\
\end{eqnarray}
of $E2_{P_{1,t}}^{\mu\nu}$:
\begin{eqnarray}
\label{eq:e2_p1t_irpoles}
\Delta_{IR}(E_2^{(11)})&=&-\frac{1}{2\sigma\tau_2}\left(\frac{1}{\omega_2}
-\frac{2}{\tau_1}\right)\frac{1}{\epsilon^2}
+\frac{1}{\sigma}\left[\frac{1}{\tau_1\tau_2}\left(\Lambda_{\omega_2}-
\Lambda_{\tau_1}-\Lambda_{\sigma}\right)
+\frac{1}{\tau_2\omega_2}\left(\Lambda_{\tau_2}+\Lambda_{\omega_2}-
\Lambda_{\tau_1}\right)\right.\nonumber\\
&&\left.+\frac{\omega_1}{\tau_1\tau_2(\tau_2+\omega_1)}\left(\Lambda_{\omega_1}-
\Lambda_{\tau_2}\right)\right]\frac{1}{\epsilon}\,\,\,,\nonumber\\
\Delta_{IR}(E_2^{(12)})&=&-\frac{1}{2\sigma\tau_2\omega_2}\frac{1}{\epsilon^2} 
-\frac{1}{\sigma\tau_2\omega_2}\left(\Lambda_{\tau_1}-\Lambda_{\tau_2}
-\Lambda_{\omega_2}\right)\frac{1}{\epsilon}\,\,\,,\nonumber\\
\Delta_{IR}(E_2^{(22)})&=&-\frac{1}{2\sigma\tau_2\omega_2}\frac{1}{\epsilon^2}
+\frac{1}{\sigma\tau_2\omega_2}\left[\Lambda_{\tau_2}+
\frac{\tau_1}{(\tau_2+\omega_1)}\left(\Lambda_{\omega_2}-\Lambda_{\tau_1}\right)
\right]\frac{1}{\epsilon}\,\,\,,\nonumber\\
\end{eqnarray}
and of  $E3_{P_{1,t}}^{\mu\nu\rho}$:
\begin{eqnarray}
\label{eq:e3_p1t_irpoles}
\Delta_{IR}(E_3^{(111)})&=&\frac{1}{2\sigma\tau_2}\left(\frac{1}{\omega_2}
-\frac{2}{\tau_1}\right)\frac{1}{\epsilon^2}
-\left[\frac{1}{\tau_1\tau_2}\left(\Lambda_{\omega_2}-\Lambda_{\tau_1}
-\Lambda_\sigma\right)
+\frac{1}{\tau_2\omega_2}\left(\Lambda_{\tau_2}+\Lambda_{\omega_2}
-\Lambda_{\tau_1}\right)\right.\nonumber\\
&&\left.
+\frac{\omega_1^2}{\tau_1\tau_2(\tau_2+\omega_1)^2}\left(\Lambda_{\omega_1}
-\Lambda_{\tau_2}\right)+\frac{1}{\tau_1(\tau_2+\omega_1)}\right]
\frac{1}{\epsilon}\,\,\,,\nonumber\\
\Delta_{IR}(E_3^{(112)})&=&\frac{1}{2\sigma\tau_2\omega_2}\frac{1}{\epsilon^2}
+\frac{1}{\sigma\tau_2\omega_2}\left(\Lambda_{\tau_1}-\Lambda_{\tau_2}
-\Lambda_{\omega_2}\right)\frac{1}{\epsilon}\,\,\,,\nonumber\\
\Delta_{IR}(E_3^{(221)})&=&\frac{1}{2\sigma\tau_2\omega_2}\frac{1}{\epsilon^2}
-\frac{1}{\sigma\tau_2\omega_2}\left[\Lambda_{\tau_2}+
\frac{\tau_1}{(\tau_2+\omega_1)}\left(\Lambda_{\omega_2}-\Lambda_{\tau_1}\right)
\right]\frac{1}{\epsilon}\,\,\,,\nonumber\\
\Delta_{IR}(E_3^{(222)})&=&\frac{1}{2\sigma\tau_2\omega_2}\frac{1}{\epsilon^2}
-\frac{1}{\sigma}\left[\frac{1}{\tau_2\omega_2}\Lambda_{\tau_2}
+\frac{\tau_1^2}{\tau_2\omega_2(\tau_1+\omega_2)^2}\left(\Lambda_{\omega_2}
-\Lambda_{\tau_1}\right)-\frac{1}{\tau_2(\tau_1+\omega_2)}\right]
\frac{1}{\epsilon}\,\,\,,\nonumber\\
\end{eqnarray}
are IR divergent.

We present in the following the IR singular box scalar integrals
$D0_{P_{1,t}}^{(k)}$, which are used in
Eq.~(\ref{eq:pent_sum_fiveboxes}) to calculate $E0_{P_{1,t}}$.
$D0_{P_{1,t}}^{(2)}$ is finite and we will not discuss it further.

\underline{Box scalar integral \boldmath$D0_{P_{1,t}}^{(1)}$\unboldmath}\\
$D0_{P_{1,t}}^{(1)}$ can be parameterized according to
Eq.~(\ref{eq:box_tensor}) with:
\begin{eqnarray}
\label{eq:d0_1_p1t_denominators}
N_1=k^2 \,\,\,&,& \,\,\,
N_2=(k+q_2)^2\,\,\,, \nonumber \\
N_3=(k+q_2-p_t^\prime)^2-m_t^2 \,\,\,&,& \,\,\, 
N_4=(k+q_2-p_t^\prime-p_h)^2-m_t^2 \,\,\,.
\end{eqnarray}
and can be obtained from $D0_{B_{7,t}^{(1)}}$ in
Section~\ref{subsubsec:d0_b7} by exchanging $q_1\leftrightarrow q_2$
and $p_t\leftrightarrow p_t^\prime$, i.e. by exchanging
$\tau_1\leftrightarrow\tau_2$, and $\omega_1\leftrightarrow\omega_2$.

\underline{Box scalar integral \boldmath$D0_{P_{1,t}}^{(3)}$\unboldmath}

$D0_{P_{1,t}}^{(3)}$ can be parameterized according to
Eq.~(\ref{eq:box_tensor}) with:
\begin{eqnarray}
\label{eq:d0_3_p1t_denominators}
N_1=k^2 \,\,\,&,& \,\,\,
N_2=(k+q_1)^2 \,\,\, , \nonumber \\
N_3=(k+q_1-p_t)^2-m_t^2 \,\,\,&,& \,\,\,
N_4=(k+q_1-p_t-p_h)^2-m_t^2 \,\,\,.
\end{eqnarray}
and is equal to $D0_{B_{7,t}^{(1)}}$ in Section~\ref{subsubsec:d0_b7}.

\underline{Box scalar integral \boldmath $D0_{P_{1,t}}^{(4)}$\unboldmath}

$D0_{P_{1,t}}^{(4)}$ can be parameterized according to
Eq.~(\ref{eq:box_tensor}) with:
\begin{eqnarray}
\label{eq:d0_4_p1t_denominators}
N_1=k^2 \,\,\, &,& \,\,\,
N_2=(k+q_2)^2\,\,\,, \nonumber\\
N_3=(k+q_1+q_2)^2 \,\,\, &,& \,\,\, 
N_4=(k+q_1+q_2-p_t)^2-m_t^2\,\,\,.
\end{eqnarray}
and is equal to $D0_{B_{8,t}}^{(2)}$ in Section~\ref{subsubsec:d0_b8}.

\underline{Box scalar integral \boldmath $D0_{P_{1,t}}^{(5)}$\unboldmath}

$D0_{P_{1,t}}^{(5)}$ can be parameterized according to
Eq.~(\ref{eq:box_tensor}) with:
\begin{eqnarray}
\label{eq:d0_5_p1t_denominators}
N_1=k^2 \,\,\, &,& \,\,\,
N_2=(k+q_1)^2\,\,\,,\nonumber\\
N_3=(k+q_1+q_2)^2 \,\,\, &,&\,\,\,
N_4=(k+q_1+q_2-p_t^\prime)^2-m_t^2\,\,\,,
\end{eqnarray}
and coincides with $D0_{B_{8,t}^{(1)}}$ in Section~\ref{subsubsec:d0_b8}.

\subsubsection{Pentagon scalar integral $E0_{P_{2,t}}$}
\label{subsubsec:e0_p2t}

The pentagon scalar integral arising from diagram $P_{2,t}$ can be
parameterized according to Eq.~(\ref{eq:pentagon_tensor}) with:
\begin{eqnarray}
\label{eq:i_p2t_denominators}
N_1=k^2 \,\,\,,\,\,\,
N_2= (k-p_t^\prime)^2-m_t^2\,\,\,&,& \,\,\,
N_3= (k-p_t^\prime+q_2)^2-m_t^2\,\,\,,\nonumber \\
N_4= (k-p_t^\prime+q_1+q_2)^2-m_t^2\,\,\,&,& \,\,\,
N_5=(k-p_t^\prime+q_1+q_2-p_h)^2-m_t^2\,\,\,.
\end{eqnarray}
The $c_k$ ($k\!=\!1,\ldots,5$) coefficients of
Eq.~(\ref{eq:pent_sum_fiveboxes}) are obtained, according to
Eq.~(\ref{eq:pentagon_ci}), as:
\begin{equation}
\label{eq:e0_p2t_coefficients}
c_k=\sum_{l=1}^{5}\left[S(P_{2,t})\right]_{kl}^{-1} \,\,\,,
\end{equation}
where
\begin{equation}
\label{eq:p2t_s_matrix}
S(P_{2,t})=\frac{1}{2}\left(
\begin{array}{c c c c c}
0 & 0 & \tau_2 & -\omega_1 & 0\\
0 & 2 m_t^2 & 2 m_t^2 & 2m_t^2-\sigma & a_1 \\
\tau_2 & 2 m_t^2 & 2 m_t^2& 2 m_t^2 & a_2 \\
-\omega_1 & 2 m_t^2-\sigma & 2 m_t^2& 2m_t^2 & 2 m_t^2-M_h^2 \\
0 & a_1 & a_2 & 2 m_t^2-M_h^2 & 2 m_t^2
\end{array}
\right)\,\,\,,
\end{equation}
and we have defined
\begin{eqnarray}
\label{eq:p2t_a1a2}
a_1 &=& 2m_t^2-(p_t+p_t^\prime)^2 = 2m_t^2-\sigma+\omega_1+\omega_2-M_h^2
\,\,\,,\nonumber\\
a_2 &=& 2m_t^2-(q_1-p_h)^2 = 2 m_t^2+\omega_1-\tau_1+\tau_2-M_h^2
\,\,\,.
\end{eqnarray}
The part of $E0_{P_{2,t}}$ that contributes to the virtual amplitude
squared can be written as:
\begin{equation}
E0_{P_{2,t}}=\frac{i}{16\pi^2}{\cal N}_t\left[
\frac{X_{-1}}{\epsilon}+X_0\right]\,\,\,,
\end{equation}
where $X_{-1}$ and $X_0$ are obtained using
Eqs.~(\ref{eq:pent_sum_fiveboxes}),
(\ref{eq:e0_p2t_coefficients})-(\ref{eq:p2t_a1a2}), and the results
for the $D0^{(k)}_{P_{2,t}}$ integrals presented in the following. The
expression for $X_{-1}$ has the following form:
\begin{equation}
\label{eq:e0_p2t_poles}
X_{-1} = \frac{1}{\tau_2\omega_1(\sigma-\omega_1-\omega_2+M_h^2)}
\frac{1}{\beta_{t\bar{t}}}\ln\left(\frac{1+\beta_{t\bar{t}}}
{1-\beta_{t\bar{t}}}\right)\,\,\, .
\end{equation}
All tensor pentagon integrals associated with $P_{2,t}$ are IR finite.

We present in the following the IR singular box scalar integrals
$D0_{P_{2,t}}^{(k)}$ which are used in
Eq.~(\ref{eq:pent_sum_fiveboxes}) to calculate $E0_{P_{2,t}}$.
$D0_{P_{2,t}}^{(1)}$, $D0_{P_{2,t}}^{(2)}$, and $D0_{P_{2,t}}^{(5)}$
are finite and we will not discuss them further.

\underline{Box scalar integral \boldmath $D0_{P_{2,t}}^{(3)}$ \unboldmath}

$D0_{P_{2,t}}^{(3)}$ can be parameterized according to
Eq.~(\ref{eq:box_tensor}) with:
\begin{eqnarray}
\label{eq:d0_3_p2t_denominators}
N_1=k^2 \,\,\, &,&\,\,\,
N_2=(k-p_t^\prime)^2-m_t^2\,\,\,,\nonumber\\
N_3=(k-p_t^\prime+q_1+q_2)^2-m_t^2\,\,\, &,&\,\,\,
N_4=(k+p_t)^2-m_t^2\,\,\,,
\end{eqnarray}
and is equal to $D0_{B_{2,s}^{(1)}}$ in Section~\ref{subsubsec:d0_b2}.

\underline{Box scalar integral \boldmath $D0_{P_{2,t}}^{(4)}$\unboldmath.}

$D0_{P_{2,t}}^{(4)}$ can be parameterized according to
Eq.~(\ref{eq:box_tensor}) with:
\begin{eqnarray}
\label{eq:d0_4_p2t_denominators}
N_1=k^2\,\,\, &,&\,\,\,
N_2=(k-p_t^\prime)^2-m_t^2\,\,\,,\nonumber\\
N_3=(k-p_t^\prime+q_2)^2-m_t^2\,\,\, &,&\,\,\,
N_4=(k+p_t)^2-m_t^2\,\,\,,
\end{eqnarray}
and can be written as 
\begin{eqnarray} 
\label{eq:i_d0_4_p2t_all}
D0_{P_{2,t}}^{(4)}&=&\frac{i}{16\pi^2}{\cal N}_t
\left(\frac{X_{-1}}{\epsilon}+X_0\right)\,\,\,,
\end{eqnarray}
where the pole part $X_{-1}$ is given by:
\begin{equation}
\label{eq:i_d0_4_p2t_pole}
X_{-1}=\frac{1}{\tau_2(\sigma-\omega_1-\omega_2+M_h^2)}
\frac{1}{\beta_{t\bar t}}\ln\left(\frac{1+\beta_{t\bar t}}{1-\beta_{t\bar t}}\right)\,\,\,,
\end{equation}
while the finite part $X_0$ can be found from Eq.~(2.9) of
Ref.~\cite{Beenakker:1990jr} with the identifications:
\begin{eqnarray}
m_0^2=m_1^2=m_4^2 &\rightarrow & m_t^2\,\,\,,\\
s &\rightarrow & (p_t+p_t^\prime)^2
=\sigma-\omega_1-\omega_2+M_h^2 \,\,\,,\nonumber \\
t &\rightarrow & (q_2-p_t^\prime)^2=m_t^2-\tau_2\,\,\,.\nonumber
\end{eqnarray}

\subsubsection{Pentagon scalar integral $E0_{P_{3,t}}$}
\label{subsubsec:e0_p3t}

The pentagon scalar integral arising from diagram $P_{3,t}$ can be
parameterized according to Eq.~(\ref{eq:pentagon_tensor}) with:
\begin{eqnarray}
\label{eq:i_p3t_denominators}
N_1=k^2 \,\,\,, \,\,\,
N_2= (k-p_t^\prime)^2-m_t^2\,\,\, &,& \,\,\,
N_3= (k-p_t^\prime+q_2)^2-m_t^2\,\,\,,\nonumber \\
N_4= (k-p_t^\prime+q_2-p_h)^2-m_t^2\,\,\, &,& \,\,\,
N_5= (k+p_t)^2-m_t^2\,\,\,.
\end{eqnarray}
The $c_k$ ($k\!=\!1,\ldots,5$) coefficients of
Eq.~(\ref{eq:pent_sum_fiveboxes}) are obtained, according to
Eq.~(\ref{eq:pentagon_ci}), as:
\begin{equation}
\label{eq:e0_p3t_coefficients}
c_k=\sum_{l=1}^{5}\left[S(P_{3,t})\right]_{kl}^{-1} \,\,\,,
\end{equation}
where
\begin{equation}
\label{eq:p3t_s_matrix}
S(P_{3,t})=\frac{1}{2}\left(
\begin{array}{c c c c c}
0 & 0 & \tau_2 & \tau_1 & 0\\
0 & 2 m_t^2 & 2 m_t^2 & a_3 & a_1 \\
\tau_2 & 2 m_t^2 & 2 m_t^2 & 2 m_t^2- M_h^2 & a_2 \\
\tau_1 & a_3 & 2 m_t^2-M_h^2 & 2m_t^2 & 2 m_t^2 \\
0 & a_1 & a_2 & 2 m_t^2& 2 m_t^2
\end{array}
\right)\,\,\,,
\end{equation}
and we have defined 
\begin{equation}
\label{eq:p3t_a3}
a_3 = 2m_t^2-(q_2-p_h)^2 = 2 m_t^2-M_h^2+\omega_2+\tau_1-\tau_2\,\,\,,
\end{equation}
while $a_1$ and $a_2$ are given in Eq.~(\ref{eq:p2t_a1a2}).

The part of $E0_{P_{3,t}}$ that contributes to the virtual amplitude
squared can be written as:
\begin{equation}
E0_{P_{3,t}}=\frac{i}{16\pi^2}{\cal N}_t\left[
\frac{X_{-1}}{\epsilon}+X_0\right]\,\,\,,
\end{equation}
where $X_{-1}$ and $X_0$ are obtained using
Eqs.~(\ref{eq:pent_sum_fiveboxes}),
(\ref{eq:e0_p3t_coefficients})-(\ref{eq:p3t_a3}), and the results for
$D0_{P_{3,t}}^{(k)}$ given in the following. The expression for
$X_{-1}$ has the following form:
\begin{equation}
\label{eq:e0_p3t_poles}
X_{-1} = -\frac{1}{\tau_1\tau_2(\sigma-\omega_1-\omega_2+M_h^2)}
\frac{1}{\beta_{t\bar{t}}}
\ln\left(\frac{1+\beta_{t\bar{t}}}{1-\beta_{t\bar{t}}}\right)\,\,\, .
\end{equation}
All tensor pentagon integrals associated with $P_{3,t}$ are IR finite.

We present in the following the box scalar integrals
$D0_{P_{3,t}}^{(k)}$, which are used in
Eq.~(\ref{eq:pent_sum_fiveboxes}) to calculate $E0_{P_{3,t}}$.
$D0_{P_{3,t}}^{(1)}$, $D0_{P_{3,t}}^{(2)}$, and $D0_{P_{3,t}}^{(5)}$
are finite and we will not discuss them further.

\underline{Box scalar integral \boldmath $D0_{P_{3,t}}^{(3)}$ \unboldmath}

$D0_{P_{3,t}}^{(3)}$ can be parameterized according to
Eq.~(\ref{eq:box_tensor}) with:
\begin{eqnarray}
\label{eq:d0_3_p3t_denominators}
N_1=k^2 \,\,\, &,& \,\,\,
N_2=(k-p_t^\prime)^2-m_t^2\,\,\,,\nonumber\\
N_3=(k-p_t^\prime+q_2-p_h)^2-m_t^2\,\,\, &,& \,\,\,
N_4=(k+p_t)^2-m_t^2\,\,\,,
\end{eqnarray}
and can be written as 
\begin{eqnarray} 
\label{eq:i_d0_3_p3t_all}
D0_{P_{3,t}}^{(3)}&=&\frac{i}{16\pi^2}{\cal N}_t
\left(\frac{X_{-1}}{\epsilon}+X_0\right)\,\,\,,
\end{eqnarray}
where the pole part $X_{-1}$ is:
\begin{equation}
\label{eq:i_d0_3_p3t_pole}
X_{-1}=\frac{1}{\tau_1(\sigma-\omega_1-\omega_2+M_h^2)}
\frac{1}{\beta_{t\bar{t}}}
\ln\left(\frac{1+\beta_{t\bar{t}}}{1-\beta_{t\bar{t}}}\right)\,\,\,,
\end{equation}
while the finite part $X_0$ can be found from Eq.~(2.9) of
Ref.~\cite{Beenakker:1990jr} with the identifications:
\begin{eqnarray}
m_0^2=m_1^2=m_4^2 &\rightarrow & m_t^2\,\,\,,\nonumber\\
s &\rightarrow & (p_t+p_t^\prime)^2 =
\sigma+m_h^2-\omega_1-\omega_2\,\,\,,\nonumber\\
t &\rightarrow &(q_1-p_t)^2= m_t^2-\tau_1\,\,\,. 
\end{eqnarray}

\underline{Box scalar integral \boldmath $D0_{P_{3,t}}^{(4)}$\unboldmath.}

$D0_{P_{3,t}}^{(4)}$ can be parameterized according to
Eq.~(\ref{eq:box_tensor}) with:
\begin{eqnarray}
\label{eq:d0_4_p3t_denominators}
N_1=k^2 \,\,\,&,&\,\,\,
N_2=(k-p_t^\prime)^2-m_t^2\,\,\,,\nonumber\\
N_3=(k-p_t^\prime+q_2)^2-m_t^2\,\,\,&,&\,\,\,
N_4=(k+p_t)^2-m_t^2\,\,\,,
\end{eqnarray}
and is equal to $D0_{P_{2,t}}^{(4)}$ in
Section~\ref{subsubsec:e0_p2t}.

\subsubsection{Pentagon scalar integral $E0_{P_{4,t}}$}
\label{subsubsec:e0_p4t}

The pentagon scalar and tensor integrals arising from diagram
$P_{4,t}$ can be found from the corresponding integrals for diagram
$P_{2,t}$ by exchanging $q_1\leftrightarrow q_2$ and
$p_t\leftrightarrow p_t^\prime$, i.e. by exchanging
$\tau_1\leftrightarrow\tau_2$, $\tau_3\leftrightarrow\tau_4$, and
$\omega_1\leftrightarrow\omega_2$.

\subsubsection{Pentagon scalar integral $E0_{P_{5,t}}$}
\label{subsubsec:e0_p5t}

The pentagon scalar integral arising from diagram $P_{5,t}$ can be
parameterized according to Eq.~(\ref{eq:pentagon_tensor}) with:
\begin{eqnarray}
\label{eq:i_p5t_denominators}
N_1=k^2 \,\,\,,\,\,\,
N_2= (k+q_1)^2\,\,\,&,&\,\,\,
N_3= (k+q_1-p_t^\prime)^2-m_t^2\,\,\,,\nonumber \\
N_4= (k+q_1+q_2-p_t^\prime)^2-m_t^2\,\,\,&,&\,\,\,
N_5= (k+q_1+q_2-p_t^\prime-p_h)^2-m_t^2\,\,\,.
\end{eqnarray}
The $c_k$ ($k\!=\!1,\ldots,5$) coefficients of
Eq.~(\ref{eq:pent_sum_fiveboxes}) are obtained, according to
Eq.~(\ref{eq:pentagon_ci}), as:
\begin{equation}
\label{eq:e0_p5t_coefficients}
c_k=\sum_{l=1}^{5}\left[S(P_{5,t})\right]_{kl}^{-1} \,\,\,,
\end{equation}
where
\begin{equation}
\label{eq:p5t_s_matrix}
S(P_{5,t})=\frac{1}{2}\left(
\begin{array}{c c c c c}
0 & 0 & \tau_4 & -\omega_1 & 0\\
0 & 0 & 0 & \tau_2 & \tau_1 \\
\tau_4 & 0 & 2 m_t^2 & 2 m_t^2 & a_3 \\
-\omega_1 & \tau_2 & 2 m_t^2 & 2 m_t^2 & 2m_t^2-M_h^2  \\
0 & \tau_1 & a_3 & 2 m_t^2-M_h^2 & 2 m_t^2
\end{array}
\right)\,\,\,,
\end{equation}
with $a_3$ as defined in Eq.~(\ref{eq:p3t_a3}).

The part of $E0_{P_{5,t}}$ that contributes to the virtual amplitude
squared can be written as:
\begin{equation}
E0_{P_{5,t}}=\frac{i}{16\pi^2}{\cal N}_t\left[\frac{X_{-2}}{\epsilon^2}
+\frac{X_{-1}}{\epsilon}+X_0\right]\,\,\,,
\end{equation}
where $X_{-2}$, $X_{-1}$ and $X_0$ are obtained using
Eqs.~(\ref{eq:pent_sum_fiveboxes}), (\ref{eq:e0_p5t_coefficients}),
(\ref{eq:p5t_s_matrix}), and the results for $D0_{P_{5,t}}^{(k)}$
given below. The expressions for $X_{-2}$ and $X_{-1}$
have the following form:
\begin{eqnarray}
\label{eq:e0_p5_poles}
X_{-2} &=& \frac{1}{2\tau_1\tau_4}
\left(\frac{1}{\omega_1}-\frac{1}{\tau_2}\right)\,\,\,,\nonumber\\
X_{-1} &=& \frac{1}{\tau_1\tau_4}\left[
\frac{1}{\omega_1}\left(\Lambda_{\tau_2}-\Lambda_{\tau_1}
-\Lambda_{\omega_1}\right)+
\frac{1}{\tau_2}\left(\Lambda_{\tau_2}+
\Lambda_{\tau_4}-\Lambda_{\omega_1}\right)\right]\,\,\,.
\end{eqnarray} 
The tensor integrals associated with $P_{5,t}$ also contain IR
divergences. Only the following tensor coefficients of
$E1_{P_{5,t}}^{\mu}$:
\begin{eqnarray}
\label{eq:e1_p5t_irpoles}
\Delta_{IR}(E_1^{(1)})&=&\frac{1}{2\tau_1\tau_2\tau_4}\frac{1}{\epsilon^2}
+\frac{1}{\tau_1\tau_2\tau_4}\left(
\Lambda_{\omega_1}-\Lambda_{\tau_2}-\Lambda_{\tau_4}\right)\frac{1}{\epsilon}
\,\,\,,\nonumber\\
\end{eqnarray}
of $E2_{P_{5,t}}^{\mu\nu}$:
\begin{eqnarray}
\label{eq:e2_p5t_irpoles}
\Delta_{IR}(E_2^{(11)})&=&-\frac{1}{2\tau_1\tau_2\tau_4}\frac{1}{\epsilon^2}
+\frac{1}{\tau_1\tau_2\tau_4(\tau_2+\omega_1)}\left[
(\tau_2+\omega_1)\Lambda_{\tau_4}
-\omega_1\left(\Lambda_{\omega_1}-\Lambda_{\tau_2}\right)\right]
\frac{1}{\epsilon}\,\,\,,\nonumber\\
\end{eqnarray}
and of $E3_{P_{5,t}}^{\mu\nu\rho}$:
\begin{eqnarray}
\label{eq:e3_p5t_irpoles}
\Delta_{IR}(E_3^{(111)})&=&\frac{1}{2\tau_1\tau_2\tau_4}\frac{1}{\epsilon^2}
-\frac{1}{\tau_1\tau_2\tau_4(\tau_2+\omega_1)^2}\left[
-\tau_2(\tau_2+\omega_1)+(\tau_2+\omega_1)^2\Lambda_{\tau_4}\right.\\
&&\left.+\omega_1^2\left(\Lambda_{\tau_2}-\Lambda_{\omega_1}\right)\right]
\frac{1}{\epsilon}\,\,\,\nonumber\\
\end{eqnarray}
are IR divergent.

We present in the following the IR singular box scalar integrals
$D0_{P_{5,t}}^{(k)}$, which are used in
Eq.~(\ref{eq:pent_sum_fiveboxes}) to calculate $E0_{P_{5,t}}$.
$D0_{P_{5,t}}^{(1)}$ and $D0_{P_{5,t}}^{(2)}$ are finite and we will
not discuss them further.

\underline{Box scalar integral \boldmath $D0_{P_{5,t}}^{(3)}$ \unboldmath}

$D0_{P_{5,t}}^{(3)}$ can be parameterized according to
Eq.~(\ref{eq:box_tensor}) with:
\begin{eqnarray}
\label{eq:d0_3_p5t_denominators}
N_1=k^2 \;,\,\, &,& \,\,\,
N_2=(k+q_1)^2\,\,\,,\nonumber\\
N_3=(k+q_1+q_2-p_t^\prime)^2-m_t^2 \,\,\, &,&\,\,\,
N_4=(k+p_t)^2-m_t^2\,\,\,,
\end{eqnarray}
and coincides with $D0_{P_{1,t}}^{(3)}$ in
Section~\ref{subsubsec:e0_p1t}, after shifting $k\rightarrow -k-q_1$.

\underline{Box scalar integral \boldmath $D0_{P_{5,t}}^{(4)}$\unboldmath.}

$D0_{P_{5,t}}^{(4)}$ can be parameterized according to
Eq.~(\ref{eq:box_tensor}) with:
\begin{eqnarray}
\label{eq:d0_4_p5t_denominators}
N_1=k^2 \,\,\,&,&\,\,\,
N_2=(k+q_1)^2\,\,\,,\nonumber\\
N_3=(k+q_1-p_t^\prime)^2-m_t^2\,\,\,&,&\,\,\,
N_4=(k+p_t)^2-m_t^2\,\,\,.
\end{eqnarray}
The part of $D0_{P_{5,t}}^{(4)}$ which contributes to the virtual amplitude
squared is given by:
\begin{eqnarray}
\label{eq:d0_4_p5t_all}
D0_{P_{5,t}}^{(4)}&=&\frac{i}{16\pi^2}{\cal N}_t
\left(\frac{1}{\tau_1\tau_4}\right)
\left(\frac{X_{-2}}{\epsilon^2}+\frac{X_{-1}}{\epsilon}+X_0\right)\,\,\,,
\end{eqnarray}
where the coefficients $X_{-2}$, $X_{-1}$, and $X_{0}$ are given by:
\begin{eqnarray}
X_{-2}&=& 1\,\,\,, \nonumber\\
X_{-1}&=&
-\ln\left(\frac{\tau_1}{m_t^2}\right)
-\ln\left(\frac{\tau_4}{m_t^2}\right)\,\,\,,\nonumber\\
X_0&=&{\cal R}e\left\{
  \ln^2\left(\frac{\tau_1}{m_t^2}\right)
 +\ln^2\left(\frac{\tau_4}{m_t^2}\right)
 -\ln^2\left(\frac{\tau_4}{\tau_1}\right)-\frac{2}{3}\pi^2
 +2\mbox{Li}_2\left(\frac{1}{z_+}\right)
 +2\mbox{Li}_2\left(\frac{1}{z_-}\right)\right\}\,\,\,,\nonumber\\
\end{eqnarray}
with
\begin{equation}
z_\pm=\frac{1}{2}\left(1\pm\Delta\right)\,\,\,,\,\,\,\,
\Delta=\sqrt{1-\frac{4m_t^2}{2m_t^2-a_3}}\,\,\,,
\end{equation}
and $a_3$ defined in Eq.~(\ref{eq:p3t_a3}).

\underline{Box integral \boldmath $D0_{P_{5,t}}^{(5)}$\unboldmath.}
\label{par:d0_5_p5t}

$D0_{P_{5,t}}^{(5)}$ can be parameterized according to
Eq.~(\ref{eq:box_tensor}) with:
\begin{eqnarray}
\label{eq:d0_5_p5t_denominators}
N_1=k^2\,\,\,&,&\,\,\,
N_2=(k+q_1)^2\,\,\,,\nonumber\\
N_3=(k+q_1-p_t^\prime)^2-m_t^2\,\,\,&,&\,\,\,
N_4=(k+q_1+q_2-p_t^\prime)^2-m_t^2\,\,\,,
\end{eqnarray}
and is equal to $D0_{B_{10,t}^{(1)}}$ in
Section~\ref{subsubsec:d0_b10}.

\subsubsection{Pentagon scalar integral $E0_{P_{6,t}}$}
\label{subsubsec:e0_p6t}

The pentagon scalar integral arising from diagram $P_{6,t}$ can be
parameterized according to Eq.~(\ref{eq:pentagon_tensor}) with:
\begin{eqnarray}
\label{eq:i_p6t_denominators}
N_1=k^2 \,\,\,,\,\,\, 
N_2= (k+q_1)^2\,\,\, &,& \,\,\, 
N_3= (k+q_1-p_t^\prime)^2-m_t^2\,\,\,,\nonumber \\
N_4= (k+q_1-p_t^\prime-p_h)^2-m_t^2\,\,\,&,&\,\,\, 
N_5= (k+q_1-p_t^\prime-p_h+q_2)^2-m_t^2\,\,\,.
\end{eqnarray}
We note that $E0_{P_{6,t}}$ can be obtained from $E0_{P_{5,t}}$ by
shifting $k\rightarrow -k-q_1$ and exchanging $p_t\leftrightarrow
p_t^\prime$, or equivalently by exchanging
$\tau_1\leftrightarrow\tau_4$, $\tau_2\leftrightarrow\tau_3$, and
$\omega_1\leftrightarrow\omega_2$. The same applies to the tensor
pentagon integrals $E1_{P_{6,t}}^{\mu}$, $E2_{P_{6,t}}^{\mu\nu}$, and
$E3_{P_{6,t}}^{\mu\nu\rho}$.
\section{Phase Space Integrals for the emission of a soft gluon in the
  two cutoff PSS method.}
\label{sec:app_soft_integrals}

In this Appendix we collect the phase space integrals for a final
state soft gluon
that are used in calculating the results reported in
Eq.~(\ref{eq:soft_a2_poles}). We parameterize the soft gluon
$d$-momentum in the $gg$ rest frame as:
\begin{equation}
\label{eq:gluon_param}
k=E_g(1,\ldots,\sin\theta_1\sin\theta_2, \sin\theta_1\cos\theta_2, 
\cos\theta_1)\,\,\,,
\end{equation} 
such that the phase space of the soft gluon in $d\!=\!4-2\epsilon$
dimensions can be written as:
\begin{eqnarray}
\label{eq:gluon_ps}
d(PS_g)_{soft}&=&\frac{\Gamma(1-\epsilon)}{\Gamma(1-2\epsilon)}
\frac{\pi^\epsilon}{(2\pi)^3} \int_0^{\delta_s \sqrt{s}/2} dE_g
E_g^{1-2\epsilon}\times\nonumber\\
&&\int_0^{\pi} d \theta_1
\sin^{1-2\epsilon}\theta_1
\int_0^\pi d\theta_2 \sin^{-2\epsilon}\theta_2\,\,\,.
\end{eqnarray}
Then all the integrals we need are the following four:
\begin{eqnarray}
\int d(PS_g)_{soft}\frac{(q_1\!\cdot\!q_2)}{(q_1\!\cdot\!k)(q_2\!\cdot\!k)}&=&
\frac{1}{(4\pi)^2}{\cal N}_t\,2\left[\frac{1}{\epsilon^2}
-\frac{2}{\epsilon}\ln(\delta_s)-\frac{1}{\epsilon}
\Lambda_{\sigma}\right.\nonumber\\
&&\left.-\frac{\pi^2}{3}+\frac{1}{2}\left(\Lambda_{\sigma}^2+4\Lambda_{\sigma}
\ln(\delta_s)+4\ln^2(\delta_s)\right)\right]\,\,\,,\nonumber\\
\int d(PS_g)_{soft}\frac{(q_1\!\cdot\!p_t)}{(q_1\!\cdot\!k)(p_t\!\cdot\!k)}&=&
\frac{1}{(4\pi)^2}{\cal N}_t \left[
\frac{1}{\epsilon^2}-\frac{2}{\epsilon}\Lambda_{\tau_1}
-\frac{2}{\epsilon}\ln(\delta_s)-\frac{\pi^2}{3}\right.\nonumber\\
&&\left.-\frac{1}{2}\Lambda_\sigma^2+2\Lambda_{\tau_1}\Lambda_\sigma
+2\ln^2(\delta_s)+4\Lambda_{\tau_1}\ln(\delta_s)+F(q_1,p_t)\right]\,\,\,,
\nonumber\\
\int d(PS_g)_{soft}\frac{(p_t\!\cdot\!p_t^\prime)}{(p_t\!\cdot\!k)
(p_t^\prime\!\cdot\!k)} &=& 
\frac{1}{(4\pi)^2}{\cal N}_t
\left(\frac{\bar s_{t\bar{t}}-2m_t^2}{\bar s_{t\bar{t}}}\right)
\left[\left(-\frac{2}{\epsilon}
+2\Lambda_\sigma+4\ln(\delta_s)\right)\frac{1}{\beta_{t\bar{t}}}
\Lambda_{t\bar t} \right.\nonumber \\
&&\left.-\frac{1}{\beta_{t\bar{t}}}\Lambda_{t\bar{t}}^2-\frac{4}{\beta_{t\bar{t}}}
\mbox{Li}_2\left(\frac{2\beta_{t\bar{t}}}{1+\beta_{t\bar{t}}}\right)\right]
\,\,\,,\nonumber\\
\int d(PS_g)_{soft}\frac{p_t^2}{(p_t\!\cdot\!k)^2}&=& 
\frac{1}{(4\pi)^2}{\cal N}_t
\left[-\frac{2}{\epsilon}+2\Lambda_\sigma+4\ln(\delta_s)-
2\frac{1}{\beta_{t\bar{t}}}\Lambda_{t\bar{t}}\right]\,\,\,,
\end{eqnarray}
where $\bar s_{t\bar t}$, $\beta_{t\bar t}$ and $\Lambda_{t\bar t}$ 
are defined in Eq.~(\ref{eq:stt_betatt}).
Moreover we have denoted by $F(p_i,p_f)$ the function:
\begin{eqnarray}
\label{eq:f_if}
F(p_i,p_f)&=&\ln^2\left(\frac{1-\beta_f}{1-\beta_f\cos\theta_{if}}\right)-
\frac{1}{2}\ln^2\left(\frac{1+\beta_f}{1-\beta_f}\right)\nonumber\\
&&+2\mbox{Li}_2\left(-\frac{\beta_f(1-\cos\theta_{if})}{1-\beta_f}\right)
-2\mbox{Li}_2\left(-\frac{\beta_f(1+\cos\theta_{if})}
{1-\beta_f\cos\theta_{if}}\right)\,\,\,,
\end{eqnarray}
where $\cos\theta_{if}$ is the angle between partons $i$ and $f$ in
the center-of-mass frame of the initial state partons, and
\begin{equation}
\beta_f=\sqrt{1-\frac{m_t^2}{(p^0_f)^2}}\,\,\,\,,\,\,\,\,
1-\beta_f\cos\theta_{if}=\frac{s_{if}}{p^0_f\sqrt{s}}\,\,\,.
\end{equation}
All the quantities in Eq.~(\ref{eq:f_if}) can be expressed in terms of
kinematical invariants, once we use $s_{if}\!=\!2p_i\!\cdot\!p_f$ and:
\begin{equation}
p_t^0=\frac{s-{\bar s}_{{\bar t}h}+m_t^2}{2\sqrt{s}}
\,\,\,\,\,\mbox{and}\,\,\,\,\,
p_{\bar{t}}^0=\frac{s-{\bar s}_{th}+m_t^2}{2\sqrt{s}}\,\,\,,
\end{equation}
with ${\bar s}_{fh}\!=\!(p_f+p_h)^2$.
\boldmath
\section{Integrated soft functions for the one-cutoff Phase
  Space Slicing method.}
\label{sec:app_sab}
\unboldmath

In this appendix we give the explicit form of the integrated soft
functions $S_{ab}$ in the three possible cases in which: both partons
$(a,b)$ are massless, one is massless and the other is massive, and
when both are massive.  These expressions have been originally
presented in Refs.~\cite{Giele:1992vf,Keller:1998tf}, and used in the
calculation of the soft part of the real cross section for $h\to
q\bar{q}t\bar{t}+g$ in Ref.~\cite{Reina:2001bc}.

When both partons $a$ and $b$ are massless $S_{ab}$ is simply given by
\cite{Giele:1992vf}:
\begin{equation}
S_{ab}=\frac{\alpha_s}{2\pi}N\frac{1}{\Gamma(1-\epsilon)}
\left(\frac{4\pi\mu^2}{s_{min}}\right)^{\epsilon}
\left(\frac{{s}_{ab}}{s_{min}}\right)^\epsilon\frac{1}{\epsilon^2}\,\,\,.
\end{equation}
For $h\to ggt\bar{t}+(g,q,\bar{q})$, this occurs when $a$ and $b$
correspond to the two hard gluons $g^A$ and $g^B$ of
Eq.(\ref{eq:h_ggttbg}), in which case $s_{ab}\!=\!s$ is the partonic
center of mass energy.

When one parton is massive and the other is massless, the function
$S_{ab}$ has the form \cite{Keller:1998tf}:
\begin{eqnarray}
S_{ab}&=&
\label{eq:soft_function_ab}
\frac{\alpha_s}{2\pi}N\frac{1}{\Gamma(1-\epsilon)}
\left(\frac{4\pi\mu^2}{s_{min}}\right)^\epsilon
\left(\frac{s_{ab}}{s_{min}}\right)^\epsilon\times\nonumber \\
&&
\left\{
\frac{1}{\epsilon^2}\left[1-\frac{1}{2} 
\left(\frac{s_{ab}}{m_t^2}\right)^\epsilon\right]
+\frac{1}{2\epsilon}\left(\frac{s_{ab}}{m_t^2}\right)^\epsilon
-\frac{1}{2}\zeta(2)+\frac{m_t^2}{s_{ab}}\right\}
\nonumber \\ 
&=&
\frac{\alpha_s}{2\pi}N\frac{1}{\Gamma(1-\epsilon)}
\left(\frac{4\pi\mu^2}{s_{min}}\right)^\epsilon\times\nonumber\\
&&\left\{ \frac{1}{2\epsilon^2}+\frac{1}{2\epsilon}+\frac{1}{2\epsilon}
\ln\left(\frac{m_t^2}{s_{min}}\right)\right.\nonumber\\
&+& \left.
\frac{1}{4}\ln^2\left(\frac{m_t^2}{s_{min}}\right)
-\frac{1}{2}\ln^2\left(\frac{s_{ab}}{m_t^2}\right)
+\frac{1}{2}\ln\left(\frac{s_{ab}}{m_t^2}\right)
+\frac{1}{2}\ln\left(\frac{s_{ab}}{s_{min}}\right)
-\frac{1}{2}\zeta(2)+\frac{m_t^2}{s_{ab}}\right\}\,\,\,.\nonumber\\
\end{eqnarray}
For $h\to ggt\bar{t}+(g,q,\bar{q})$, this occurs when $a\!=\!1,2$
(where $1,2$ denote the initial gluons $g^A,g^B$) and
$b\!=\!t,\bar{t}$, and we therefore have four possible integrated soft
functions of this type: $S_{1t}$, $S_{1\bar{t}}$, $S_{2t}$, and
$S_{2\bar{t}}$.

Finally, when both partons are massive, i.e. when $a\!=\!t$ and
$b\!=\!\bar{t}$, the integrated soft function $S_{t\bar{t}}$ is
\cite{Keller:1998tf}:
\begin{equation}
\label{eq:soft_function_ttbar}
S_{t\bar t}=\frac{\alpha_s}{2\pi}N
\frac{1}{\Gamma(1-\epsilon)}
\left(\frac{4\pi\mu^2}{s_{min}}\right)^\epsilon 
\frac{m_t^2}{\sqrt{\lambda_{t\bar{t}}}}
\left(J_{s}\frac{1}{\epsilon}+J_a+J_b\right)\,\,\,,
\end{equation}
where we have defined:
\begin{eqnarray}
\label{eq:ttbar_jsjajb}
\frac{m_t^2}{\sqrt{\lambda_{t\bar{t}}}}J_{s}&=&
1-\frac{s_{t\bar{t}}}{(2 m_t^2+s_{t\bar{t}})\beta_{t\bar{t}}}
\Lambda_{t\bar{t}}\,\,\,,\nonumber\\
J_a&=& J_{s} 
\ln\left(\frac{\tau_+^2\lambda_{t\bar{t}}}{s_{min}m_t^2}\right)
\,\,\,,\nonumber \\
J_b&=&\bigl(\tau_+-\tau_-\bigr)
\bigl[1-2\ln(\tau_+-\tau_-)-\ln(\tau_+)\bigr]\nonumber\\
&+&\left(\frac{\tau_++\tau_-}{2}\right)
\left[\ln\left(\frac{\tau_+}{\tau_-}\right)
\bigl(1+2\ln(\tau_+-\tau_-)\bigr)\right.\nonumber\\
&+&\left.\mbox{Li}_2\left(1-\frac{\tau_+}{\tau_-}\right)
-\mbox{Li}_2\left(1-\frac{\tau_-}{\tau_+}\right)\right]
+1+\tau_-\tau_+\nonumber\\
&+&(\tau_-+\tau_+)\left[
-1-\ln(\tau_+)\ln(\tau_-)+\frac{1}{2}\ln^2(\tau_+)
\right]\,\,\,,
\end{eqnarray}
$\beta_{t\bar{t}}$ and $\Lambda_{t\bar{t}}$ are defined in
Eq.~(\ref{eq:stt_betatt}) while $\lambda_{t\bar{t}}$ and $\tau_{\pm}$ are
given by:
\begin{eqnarray}
\label{eq:lambdatt_taupm}
\lambda_{t\bar{t}} &\equiv& s_{t\bar{t}}^2-4 m_t^4\,\,\,,\nonumber\\
\tau_{\pm}&=&\frac{s_{t\bar{t}}}{2m_t^2}\pm
\sqrt{\left(\frac{s_{t\bar{t}}}{2m_t^2}\right)^2-1}\,\,\,.
\end{eqnarray}
\bibliography{tth_lhc}
\end{document}